\let\kernellabel\label
\documentclass[10pt,aps,prd,twocolumn,showpacs,superscriptaddress,nofootinbib,nobibnotes,longbibliography,floatfix]{revtex4-2}
\let\label\kernellabel

\usepackage[utf8]{inputenc}
\usepackage[T1]{fontenc}
\usepackage{fix-cm}
\usepackage{comment}
\usepackage{amsmath, amssymb, amsfonts, mathtools}
\usepackage{bm}
\usepackage{commath}
\usepackage{tensor}
\usepackage{leftindex}
\usepackage{cancel}
\usepackage{siunitx}
\usepackage{graphicx}
\usepackage{array}
\usepackage{booktabs}
\usepackage{hhline}
\usepackage{tabularray}
\usepackage{enumitem}
\usepackage{csquotes}
\usepackage{accents}
\usepackage{chngcntr}
\usepackage{pifont}
\usepackage[normalem]{ulem}
\usepackage{anyfontsize}
\usepackage{mathrsfs}
\usepackage[dvipsnames]{xcolor}
\usepackage{hyperref}
\hypersetup{
    colorlinks=true, 
    citecolor=MidnightBlue,
    linkcolor=MidnightBlue, 
    urlcolor=MidnightBlue, 
    linktocpage=true
}
\usepackage{orcidlink}

\DeclareMathAlphabet{\mathpzc}{OT1}{pzc}{m}{it}
\def\sinh{{\mathrm{sinh}}}
\def\cosh{{\mathrm{cosh}}}

\def\a{{\alpha}}

\def\Re{{\rm Re}}

\def\d{{\mathrm{d}}}
\def\be{\begin{equation}}
\def\ee{\end{equation}}
\def\ba{\begin{eqnarray}}
\def\ea{\end{eqnarray}}

\newcommand{\centra}{CENTRA, Departamento de Física, Instituto Superior Técnico -- IST,
Universidade de Lisboa -- UL, Avenida Rovisco Pais 1, 1049-001 Lisboa, Portugal}

\begin{document}

\title{When the Ringing Stops: Purely Imaginary Modes in the Ringdown Spectrum of Dynamical Black Holes}

\author{Lodovico Capuano\,\textsuperscript{\,$\sharp$}~\orcidlink{0009-0001-0369-6635}}
\email{lodovico.capuano@uniroma1.it}
\affiliation{Dipartimento di Fisica, Sapienza Università di Roma, Piazzale Aldo Moro 5, 00185, Roma, Italy}
\affiliation{INFN, Sezione di Roma, Piazzale Aldo Moro 2, 00185, Roma, Italy}

\author{Thomas Lovo\,\textsuperscript{\,$\sharp$}\,\orcidlink{0009-0000-6719-0847}}
\email{tlovo@sissa.it}
\affiliation{SISSA, Via Bonomea 265, 34136 Trieste, Italy \& INFN Sezione di Trieste, Trieste, Italy}
\affiliation{IFPU -- Institute for Fundamental Physics of the Universe, via Beirut 2, 34014 Trieste, Italy}

\author{Gorka Prieto-Varela\,\textsuperscript{\,$\sharp$}\,\orcidlink{0009-0006-6295-7694}}
\email{gprietov@sissa.it}
\affiliation{SISSA, Via Bonomea 265, 34136 Trieste, Italy \& INFN Sezione di Trieste, Trieste, Italy}
\affiliation{IGAP -- Institute for Geometry and Physics, via Beirut 2, 34151 Trieste, Italy}

\author{Subhodeep Sarkar\,\textsuperscript{\,$\sharp$}\,\orcidlink{0000-0001-9015-8837}}
\email{subhodeep.sarkar1@gmail.com}
\affiliation{CSGC -- Centre for Strings, Gravitation and Cosmology, Department of Physics, Indian Institute of Technology Madras, Chennai 600 036, India}

\author{Adrien Kuntz~\orcidlink{0000-0002-4803-2998}}
\email{adrien.kuntz@tecnico.ulisboa.pt}
\affiliation{\centra}

\author{Enrico Barausse~\orcidlink{0000-0001-6499-6263}}
\email{barausse@sissa.it}
\affiliation{SISSA, Via Bonomea 265, 34136 Trieste, Italy \& INFN Sezione di Trieste, Trieste, Italy}
\affiliation{IFPU -- Institute for Fundamental Physics of the Universe, via Beirut 2, 34014 Trieste, Italy}

\author{Dawood Kothawala\,\orcidlink{0000-0001-6540-6405}~}
\email{dawood@iitm.ac.in}
\affiliation{CSGC -- Centre for Strings, Gravitation and Cosmology, Department of Physics, Indian Institute of Technology Madras, Chennai 600 036, India}

\begin{abstract}
	We extend the frequency-domain analysis of quasinormal modes in a dynamical, spherically symmetric black hole spacetime undergoing constant-rate mass evolution. In particular, we report a novel feature of the spectrum: the presence of purely imaginary eigenvalues in addition to the usual light-ring modes.
	We study the frequencies of these modes both analytically and numerically. The analytical calculation uses a novel formalism based on recent advances in connection coefficients of Heun functions. We then compute the frequencies numerically using a spectral method on hyperboloidal slices and find excellent agreement between the two approaches.
	Finally, we validate the frequency-domain results against an independent set of time-domain simulations. Our analysis shows that the purely imaginary modes govern the late-time signal through exponentially decaying tails.
	In the Schwarzschild limit, both frequency- and time-domain studies consistently show that the purely imaginary modes give rise to the familiar Schwarzschild power-law tail.
\end{abstract}

\maketitle

\begingroup
\renewcommand\thefootnote{$\sharp$}
\footnotetext[1]{LC, TL, GP-V, and SS contributed equally to this work and should be considered joint first authors.}
\endgroup

\section{Introduction}\label{introduction}

The gravitational wave (GW) signal emitted by a binary black hole system, such as those detected by the LIGO-Virgo-KAGRA (LVK) collaboration~\cite{KAGRA:2021vkt,LIGOScientific:2021sio}, is divided into three main phases: inspiral, merger, and ringdown. During the inspiral, the binary gradually loses orbital energy and angular momentum through GW emission, causing the two black holes (BHs) to spiral toward each other until they merge. Immediately after coalescence, the remnant BH relaxes toward a final stationary configuration, in the ringdown stage.

The ringdown signal is well understood within BH perturbation theory. In General Relativity (GR), linear perturbations of a stationary and spherically-symmetric BH are described by the Regge-Wheeler and Zerilli equations~\cite{Regge:1957td,Zerilli:1970se}. After imposing the appropriate boundary conditions, the solution to the perturbation equation can be written as a linear superposition of quasinormal modes (QNMs), namely exponentially damped sinusoids~\cite{1970Natur.227..936V,Chandra1975,Anninos_1993,Nollert:1999ji,Kokkotas_1999,Berti:2009kk,Konoplya_2011,Berti:2025hly,Cardoso:2025npr}. Each QNM is characterized by a complex frequency and a complex amplitude, with the imaginary part of the frequency determining the damping timescale.

A linear superposition of QNMs provides an accurate description of the GW signal at intermediate times after the merger. At later times, however, the signal is dominated by a non-oscillatory power-law tail~\cite{Price_tails,DeWitt:1960fc,Ma:2024hzq,DeAmicis:2024eoy}, while at earlier times, closer to the merger, non-linear effects become increasingly important~\cite{Bhagwat:2017tkm,Okounkova:2020vwu,Ma:2022wpv,London:2014cma,Cheung:2022rbm,Mitman:2022qdl,Khera:2023oyf,Zhu:2024rej,Redondo-Yuste:2023seq,Bucciotti:2023ets,Bucciotti:2024jrv,Bucciotti:2024zyp,Bucciotti:2025rxa,Ma:2024qcv,Khera:2024yrk}.
Although ringdown GWs are often modeled as perturbations of a stationary BH background, realistic BHs are generically non-stationary, the simplest example being the spherically symmetric Vaidya BH~\cite{Vaidya1953,Vaidya1943,Vaidya1951,Vaidya1999a,Vaidya1999b,PhysRev.83.10,PhysRev.137.B1364}. The geometry of such dynamical BHs~\cite{Kothawala:2004fy} may evolve over time due to GW self-absorption or interactions with the environment~\cite{Pringle:1981ds,Hawking:1974rv,Hawking:1975vcx,Page:1976df,Brito:2015oca,Herdeiro:2021znw,Penrose:1971uk,Bardeen:1973gs,Bekenstein:1973ur,Kothawala:2004fy,deFreitasPacheco:2007ysm,Ghosh:2008zza,Lima_2010,Cardoso:2025npr,Redondo-Yuste:2023ipg}. Modeling a time-dependent background is therefore important for achieving a more complete and physically realistic description of the post-merger dynamics. The impact of the evolution of the BH mass on the QNM spectrum has gained attention in recent times~\cite{PhysRevD.105.064046,PhysRevD.109.044048,Capuano:2024qhv}. These studies highlight time-dependent deformation or excitation of standard QNMs, whose frequencies are related to the null geodesics at the light-ring (LR)~\footnote{The evolution of the LR of dynamical spacetimes is interesting in its own right and has been explored in the literature as well~\cite{Mishra:2019trb,Sarkar:2021djs,Solanki:2022glc,Koga:2022dsu}.} according to the well known correspondence~\cite{Cardoso:2008bp, Konoplya_2011}. However, the possibility that other families of modes, not related to the LR, could be activated in a dynamical BH spacetime has not yet been explored.

As shown in~\cite{Capuano:2024qhv}, a dynamical spherically-symmetric BH background with constant mass accretion/radiation rate can be related by a conformal transformation to a static Schwarzschild-Rindler (SR)~\cite{Grumiller:2002nm,Grumiller:2010bz,Grumiller:2019fmp,Culetu:2011tz,Grumiller:2013mxa} spacetime, characterized by a term growing linearly with the distance from the BH, and hence displaying two separate horizons.
Therefore, this spacetime resembles the Schwarzschild-de Sitter (SdS) BH, which is known to possess a family of purely imaginary (PI) non-LR QNMs~\cite{Jansen:2017oag,Cardoso:2017soq,Konoplya:2022xid}~\footnote{Note that, in the limit where the event horizon and the cosmological horizon coincide, the scattering potential of the SdS black hole effectively reduces to the P\"{o}schl-Teller potential, making it possible to obtain analytical approximations for the LR QNM frequencies~\cite{Cardoso:2003sw,Zhidenko:2003wq}. We study the corresponding limit for the SR metric for completeness in Appendix~\ref{appendix:next_limit}, and focus exclusively on the PI modes in the main text.}. These modes are called de Sitter modes since they are smooth deformations of the QNMs of pure de Sitter spacetime~\cite{Cardoso:2004up,Lopez-Ortega:2006aal,Lopez-Ortega:2012xvr,Cardoso:2017soq}. Moreover, it has been recently shown in~\cite{Arnaudo:2025kit} that in the Schwarzschild limit, such modes collapse into the well-known branch-cut appearing along the imaginary axis in the Laplace-transformed Green function of the Regge-Wheeler/Zerilli equation~\cite{Leaver:1986gd,Casals:2011aa,Casals:2012ng,Su:2026fvj,Aruquipa:2026tga}. It is worth mentioning that the Green function approach also shows the existence of purely imaginary Matsubara modes, distinct from the de Sitter ones, that contribute to the prompt response of the Green function~\cite{Arnaudo:2025uos,Kuntz:2025gdq}.
It is also well known that the branch-cut gives rise to the late-time power-law tail appearing in the ringdown signal of stationary BHs~\cite{Leaver:1986gd,DeAmicis:2024eoy,Rosato:2026moe}. Recently, motivated by a Weyl law for BH QNMs~\cite{Jaramillo:2022zvf}, the relation between the Schwarzschild branch-cut and an infinite accumulation of de Sitter modes was numerically demonstrated in~\cite{Zhou:2025xta}. The analogy between dynamical BHs with constant mass accretion/radiation and SdS spacetimes motivates both the frequency-domain search for similar PI modes spectrum of Vaidya BHs, and, if present, a time-domain analysis to clarify their physical interpretation.

In this paper, we build on the formalism developed in~\cite{Capuano:2024qhv} for a dynamical BH, described by a Vaidya metric~\cite{Vaidya1953,Vaidya1943,Vaidya1951,Vaidya1999a,Vaidya1999b,PhysRev.83.10,PhysRev.137.B1364} with constant mass derivative, that we will denote as linear-mass Vaidya (LMV) spacetime.
The advantage of this last assumption resides in the highest degree of spacetime symmetry, which allows for a frequency-domain treatment of the QNM problem.

The master perturbation equation in the frequency domain typically exhibits four regular singular points. We show that the master equation derived in~\cite{Capuano:2024qhv} can be mapped into an Heun equation, without any approximation. Using recent developments on connection coefficients of Heun functions~\cite{Bonelli:2022ten}, we obtain a quantization condition for the QNMs analogous to the condition derived in~\cite{Arnaudo:2025kit,Aminov_2023} for the SdS spacetime. Due to this analogy, we are able to find analytically the set of PI QNMs as series expansion in the mass-evolution parameter. We also show that in suitable approximate limiting cases, one of the regular singular points of the master equation can be removed, allowing for a solution in terms of hypergeometric functions. This latter approach, although rudimentary compared to the Heun formalism, provides a clear physical picture supporting the existence of PI modes in the LMV spacetime.

We then move on to test our analytic findings against a numerical computation in the frequency-domain using hyperboloidal slices.
In~\cite{Capuano:2024qhv}, some of the present authors used the method of continued fractions~\cite{leaver} to compute the QNM frequencies. Usually, this approach involves finding the roots of an algebraic equation and hence requires us to specify a guess value. As a result, this method may skip modes while scanning the complex frequency plane. To overcome this problem, we write the SR metric in hyperboloidal coordinates by constructing the so-called height function, closely following the method employed for SdS BHs~\cite{Sarkar:2023rhp}. We then use the hyperboloidal slicing approach to find QNMs~\cite{Zenginoglu:2011jz,PanossoMacedo:2018hab,PanossoMacedo:2024nkw} and compute the eigenvalues numerically using the Chebyshev spectral method~\cite{trefethen2000spectral} with mesh refinement~\cite{Pynn:2016mtw,PanossoMacedo:2022fdi,Zhou:2025xta}. This method presents some clear advantages: we are able to incorporate the QNM boundary conditions geometrically~\cite{Zenginoglu:2011jz,PanossoMacedo:2018hab,PanossoMacedo:2024nkw} and formulate an eigenvalue problem that is capable of revealing the entire spectrum without needing any initial seed value. The hyperboloidal approach also provides insight into the limiting geometries of the spacetime and its QNM spectrum as well, as recently shown for the SdS geometry~\cite{Zhou:2025xta}.

All of these frequency-domain methods confirm, with very good agreement, the presence of PI QNMs. In light of the aforementioned analogy between SdS and SR spacetime, we shall often refer to these  PI modes as Rindler modes.
We also show that as we approach the stationary limit, these PI modes accumulate near the origin of the imaginary axis, giving rise to the characteristic branch-cut of the Schwarzschild spacetime.

Finally, we numerically evolve the time-domain perturbation equation in the SR spacetime and show that, when the mass derivative does not exceed a certain threshold, PI modes dominate over the LR ones at late times, producing exponential tails. As we approach the Schwarzschild limit, the sum of several exponential tails reconstructs the well-known power-law one.

The paper is organized as follows. In Sec.~\ref{sec:Vaidya_geometry}, we set up the spacetime and review the frequency-domain formalism for the LMV spacetime, and the main theoretical findings of~\cite{Capuano:2024qhv} regarding the LR mode structure. In Sec.~\ref{sec:Hyperboloidal_coord}, we construct the hyperboloidal coordinates for the SR metric, highlighting some relevant limiting geometries of the spacetime. In Sec.~\ref{sec:QNMviaHeun}, we show that the master perturbation equation can be cast in the form of a Heun equation, and present a rigorous computation of the QNM frequencies. In Sec.~\ref{sec:freq_domain_numerical}, we write the master equation in hyperboloidal coordinates and compute the complete QNM spectrum numerically, clearly showing the presence of a new family of purely imaginary modes, in addition to the usual LR modes. We also study the behavior of the QNM spectrum under different limits and show how the spectrum scales with the parameters of the spacetime. We also compare frequency-domain analytic and numerical results. Finally, in Sec.~\ref{sec:time_domain}, we solve the wave equation in the time-domain using a finite-difference scheme, and discuss the potential role of PI QNM excitation in the ringdown signal. We summarize our findings and discuss the possible impact for BH spectroscopy in Sec.~\ref{sec:conclusions}.
We also include appendices that supplement the discussion in the text; in particular, Appendix~\ref{appendix:next_limit} contains a detailed description of QNMs of the SR metric in the nearly-extremal limit, and Appendix \ref{appendix:convergence} contains extensive tests demonstrating convergence of our frequency- and time-domain codes.

\emph{Notations and conventions:} We will work in geometric units, i.e., we set $c = G = 1$. Throughout this paper, we will use the mostly positive signature convention, such that the Minkowski spacetime will have the metric $ \eta_{\mu \nu} =\mathrm{diag}(-1,1,1,1)$.

\section{Basic Setup}
Here, we introduce the geometric setup and the resulting perturbation equations for our dynamical black hole spacetime. We define the Linear Mass Vaidya (LMV) geometry and the conformally related Schwarzschild-Rindler (SR) black hole. We discuss the conformal diagram, continuous symmetries and degenerate limits, and introduce the master perturbation equation. Finally, we define quasinormal modes (QNMs) and summarize the known results about the QNM spectrum.
\label{sec:Vaidya_geometry}
\subsection{The Vaidya Spacetime}

A spherically symmetric, dynamical black hole can be described by the Vaidya metric~\cite{Vaidya1953,Vaidya1943,Vaidya1951,Vaidya1999a,Vaidya1999b,PhysRev.83.10,PhysRev.137.B1364}. In Eddington--Finkelstein coordinates $(w,r,\theta,\phi)$, the Vaidya metric takes the form of the Schwarzschild solution with the constant mass parameter promoted to a function of the null coordinate $w$. The line element is given by
\begin{equation}
	{\rm d}s^2=-\left(1-\dfrac{2 M(w)}{r}\right){\rm d}w^2+ 2 \mathcal{S}\,{\rm d}w \, {\rm d}r+r^2{\rm d}\Omega_{S^2}^2\,,
	\label{eq:Vaidya}
\end{equation}
with $\mathcal{S}$ being the sign of the
derivative of the mass function\footnote{The mass function $M(w)$ is typically assumed to be monotonic, so that $M'(w)$ has definite sign.}, and with ${\rm d}\Omega_{S^2}^2={\rm d}\theta^2+\sin^2\theta \, {\rm d}\phi^2$.

The metric given by Eq.~\eqref{eq:Vaidya} describes an accreting BH for $M'(w)>0$, with $w$ being the advanced time (ingoing null) coordinate, while it describes a radiating BH for $M'(w)<0$, with $w$ being the retarded time (outgoing null) coordinate. Note that the Vaidya metric of Eq.~\eqref{eq:Vaidya} is a nonvacuum solution of the Einstein field equations,
\begin{equation}
	R_{\mu\nu}-\frac{1}{2}R\,g_{\mu\nu}= 8\pi T_{\mu\nu }\,,
	\label{eq:Einstein}
\end{equation}
where $R_{\mu\nu}$ and $R$ are the Ricci tensor and Ricci scalar, respectively, while the energy content of the spacetime is represented by the stress-energy tensor $T_{\mu\nu}$,
\begin{equation}
	T_{\mu\nu}=\frac{| M'(w)|}{ 4 \pi r^2}\partial_\mu w  \partial_\nu w\,,
	\label{Vaidya_SET}
\end{equation}
which describes a null radiation field in the geometrical optics limit.
Hereafter, we focus on the case where the black hole mass changes at a constant rate, namely, $M(w) = M_0 + M'\,(w-w_0)$, with $M' = \text{const}$. This metric, which we refer to as the Linear Mass Vaidya (LMV) spacetime, exhibits peculiar geometric features. These features are more apparent in the coordinate system given by
\begin{figure}[t]
	\includegraphics[width=1.0\columnwidth]{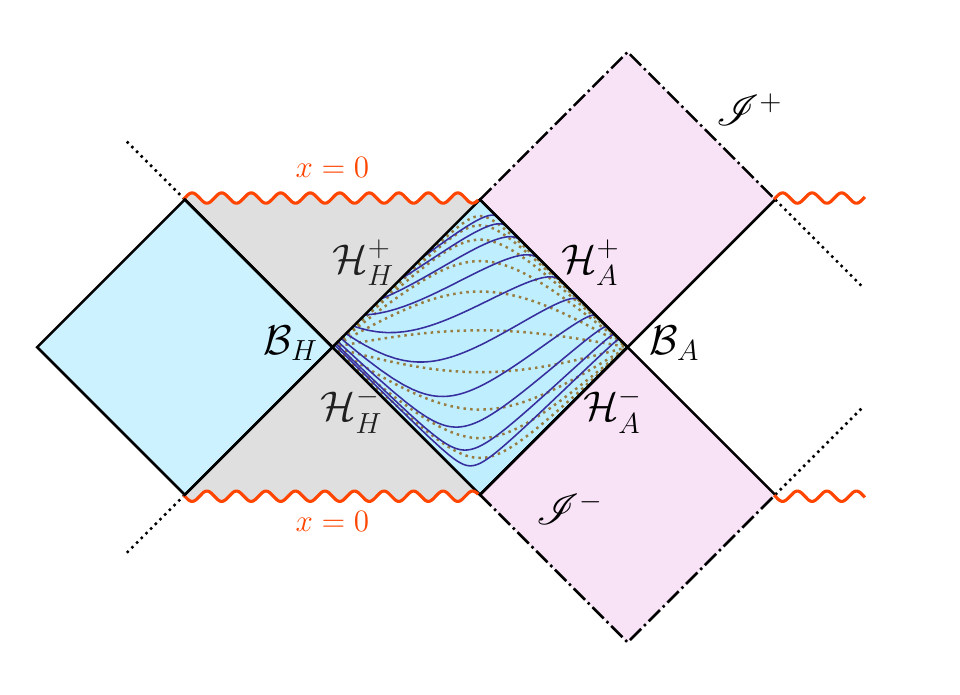}
	\caption{\textbf{Conformal diagram of SR spacetime} representing the static region (light blue), the black hole interior (gray), and the region beyond the acceleration horizon (pink). The additional patches illustrate the repeating lattice structure of the maximal analytic extension of the spacetime (indicated by dotted lines). The (future and past) event and acceleration horizons, $\mathcal{H}_H^{\pm}$ and $\mathcal{H}_A^{\pm}$, are indicated by diagonal lines, while the spacelike singularities at $x = 0$ are represented by red wavy lines. The future and past null infinities, $\mathscr{I}^{\pm}$, are indicated by dashed lines. Slices of constant Schwarzschild time $T$ (orange dotted lines) pile up and terminate at the bifurcation spheres $\mathcal{B}_H$ and $\mathcal{B}_A$, i.e., at the intersections of the future and past event and acceleration horizons. In contrast, constant hyperboloidal time $\tau$ hypersurfaces (solid purple lines) foliate the static region, smoothly penetrating the future event and acceleration horizons. The constant time slices were generated using the coordinate transformations involved in obtaining the conformal compactification of the SR metric with $|M'|=0.05$.}
	\label{fig:penrose_diagram}
\end{figure}
\begin{align}
	W\equiv \int \frac{{\rm d}w}{2M(w)},\quad \mathrm{and} \quad
	T= W-\mathcal{S} \, x_*\,,
	\label{coordinate_transformation}
\end{align}
where we introduce a dimensionless radial coordinate $x\equiv {r}/{2M(w)}$ and the tortoise coordinate
\begin{equation}
	x_*=\int  \frac{{\rm d}x}{f(x)} \,,
	\label{eq:tortoisex}
\end{equation}
with
\begin{equation}
	f(x)=1-\frac{1}{x}-4|M'| x\,.
\end{equation}
In these coordinates, the line element in Eq.~\eqref{eq:Vaidya} becomes
\begin{equation}
	{\rm d}s^2=4 M(w)^2{\rm d}\tilde s^2.
	\label{conformal_metric}
\end{equation}
with
\begin{equation}
	{\rm d}\tilde s^2=-f(x){\rm d}T^2+\frac{1}{f(x)}{\rm d}x^2+x^2{\rm d}\Omega_{S^2}^2 \,,
	\label{eq:vaidya_metric_static}
\end{equation}
which is the Schwarzschild--Rindler metric. Notice that this geometry has two horizons, determined by the roots of $g^{xx}$, viz.,
\begin{align}
	 & x_H = \dfrac{r_H}{2M(w)} =\frac{1-\sqrt{1- 16|M'|}}{8| M'|}\,, \nonumber \\
	 & x_A = \dfrac{r_A}{2M(w)} =\frac{1+\sqrt{1-16 |M'|}}{8 |M'|}\,.
	\label{eq:horizons}
\end{align}
Here $x_H$ is the location of the event horizon, while $x_A>x_H$ is the acceleration or Rindler horizon.
The condition for the two horizons to exist and remain distinct is $|M'|<1/16$, which we shall assume holds throughout. The metric function $f(x)$ and the tortoise coordinate $x_*$ can be expressed in terms of the horizons as,
\begin{align}
	f(x)=  -\frac{4 |M'|}{x}(x-x_H)(x-x_A)\,,  \label{eq:lapse_vaidya_static} \\
	x_* = x_H \eta \ln(x-x_H) - x_A \eta\ln|x-x_A|\,,\label{eq:tortoisecoordinate}
\end{align}
where
\begin{equation}
	\eta = \dfrac{1}{4 |M'|(x_A-x_H)}\,. \label{eq:defTortoise_eta}
\end{equation}
The conformal diagram for this spacetime is presented in Fig.~\ref{fig:penrose_diagram}. Our primary domain of interest is the static exterior diamond (shaded blue), defined by $x_H < x < x_A$, which is bounded by the past and future black hole event horizons ($\mathcal{H}_H^{\pm}$) and the acceleration horizons ($\mathcal{H}_A^{\pm}$). The past and future branches of each horizon intersect at the corresponding bifurcation spheres, namely the black hole bifurcation sphere $\mathcal{B}_H$ and the acceleration bifurcation sphere $\mathcal{B}_A$. To the left, crossing $\mathcal{H}_H^{+}$, lies the black hole interior (shaded gray). This region is bounded by $\mathcal{H}_H^{+}$ and terminates at the future spacelike curvature singularity at $x=0$ (indicated by a red wavy line). Symmetrically, traversing downward across $\mathcal{H}_H^{-}$ leads into the white hole interior, which is bounded by a past spacelike singularity. To the right, crossing $\mathcal{H}_A^{+}$, we enter a future asymptotic patch (shaded pink), where $f<0$, bounded by future null infinity ($\mathscr{I}^+$). Extending from the lower right is a corresponding past asymptotic patch bounded by past null infinity ($\mathscr{I}^-$). Finally, a maximal analytic extension of the spacetime reveals a global causal structure that forms an infinite, repeating lattice continuing across these horizons.

\subsection{Continuous Symmetries and Degenerate Limits}
We see that the Vaidya solution given by Eq.~\eqref{eq:Vaidya} forms a one-parameter family of solutions characterized by $|M'| \in (0, 1/16)$, where $M_0$ sets the characteristic length scale, spanning the weak ($|M'| \to 0$) and strong ($|M'| \to 1/16$) accretion/radiation limits. For most of this work, we restrict our analysis to the conformally related Schwarzschild-Rindler (SR) spacetime of Eq.~\eqref{eq:vaidya_metric_static}, specifying how our results translate back to the full physical spacetime when needed.
The SR metric (Eq.~\eqref{eq:vaidya_metric_static}) admits the timelike Killing vector (KV) field
\[
	\xi^T = \partial_T \, ,
\]
associated with stationarity and satisfying the Killing equation
\begin{equation}
	\tilde{\nabla}_\mu \xi^T_\nu + \tilde{\nabla}_\nu \xi^T_\mu = 0 \, ,
\end{equation}
where $\tilde{\nabla}$ denotes the covariant derivative with respect to the static metric. The KV $\xi^T$ is timelike in the blue region of Fig.~\ref{fig:penrose_diagram}, while it becomes spacelike in the pink and gray regions.
Since the full LMV geometry is conformally related to this static background, the same vector field $\xi^T$ obeys
\begin{align}
	\nabla_\mu \xi^T_\nu + \nabla_\nu \xi^T_\mu
	= & 2 g_{\mu\nu}\,\xi^{T\alpha}\partial_\alpha \ln M(w)\,, \nonumber \\
	= & 2 g_{\mu\nu} M'\,,
	\label{eq:conformal_killing}
\end{align}
where $g_{\mu\nu}$ is the Vaidya metric and $\nabla$ is the covariant derivative associated with it. This is precisely the conformal Killing equation, showing that $\xi^T$ is not an exact KV of the full spacetime, but rather a conformal KV \cite{galaxies2010062,Nielsen:2017hxt}. Similar considerations also apply for the charged \cite{Tarafdar:2022rzz,Koh:2020hta} and rotating Vaidya geometries \cite{Ghoshal:2026pdn} as well.

It is well known that one can define a suitable charge associated with a Killing symmetry that is conserved along geodesics. In the conformal case, this conservation law survives only along null geodesics, consistently with the conformal invariance of causal structure. In the limit $|M'| \to 0$, where the spacetime reduces to Schwarzschild, the conformal factor becomes constant and $\xi^T$ reduces to an ordinary KV of the full metric. On the other hand, in the $M_0\to 0$ limit one approaches the self-similar Vaidya case ($M_0=0$)~\cite{PhysRevD.34.2978,WAUGH1986154,Berezin:2016ubu,Nolan:2005is,Nolan:2006pz}, in which $\xi^T$ coincides with the homothetic KV $\xi^H=w\partial_w+r\partial_r$, associated with the symmetry under the rescaling $(w,r)\to (\lambda w,\lambda r)$, for a generic real scalar $\lambda$. The self-similar Vaidya spacetime, while displaying the same generalized Killing equation (Eq.~\eqref{eq:conformal_killing}), is more symmetric than the general LMV spacetime, since the additional scale introduced by $M_0$ is absent and the metric components depend only on the ratio $w/r$. However, while mathematically interesting, this spacetime exhibits physical pathologies, such as a naked singularity located at $(w,r)=(0,0)$~\cite{Nolan:2005is,Nolan:2006pz}, which make it less appealing from an astrophysical perspective.
\begin{figure*}    \includegraphics[width=0.75\textwidth]{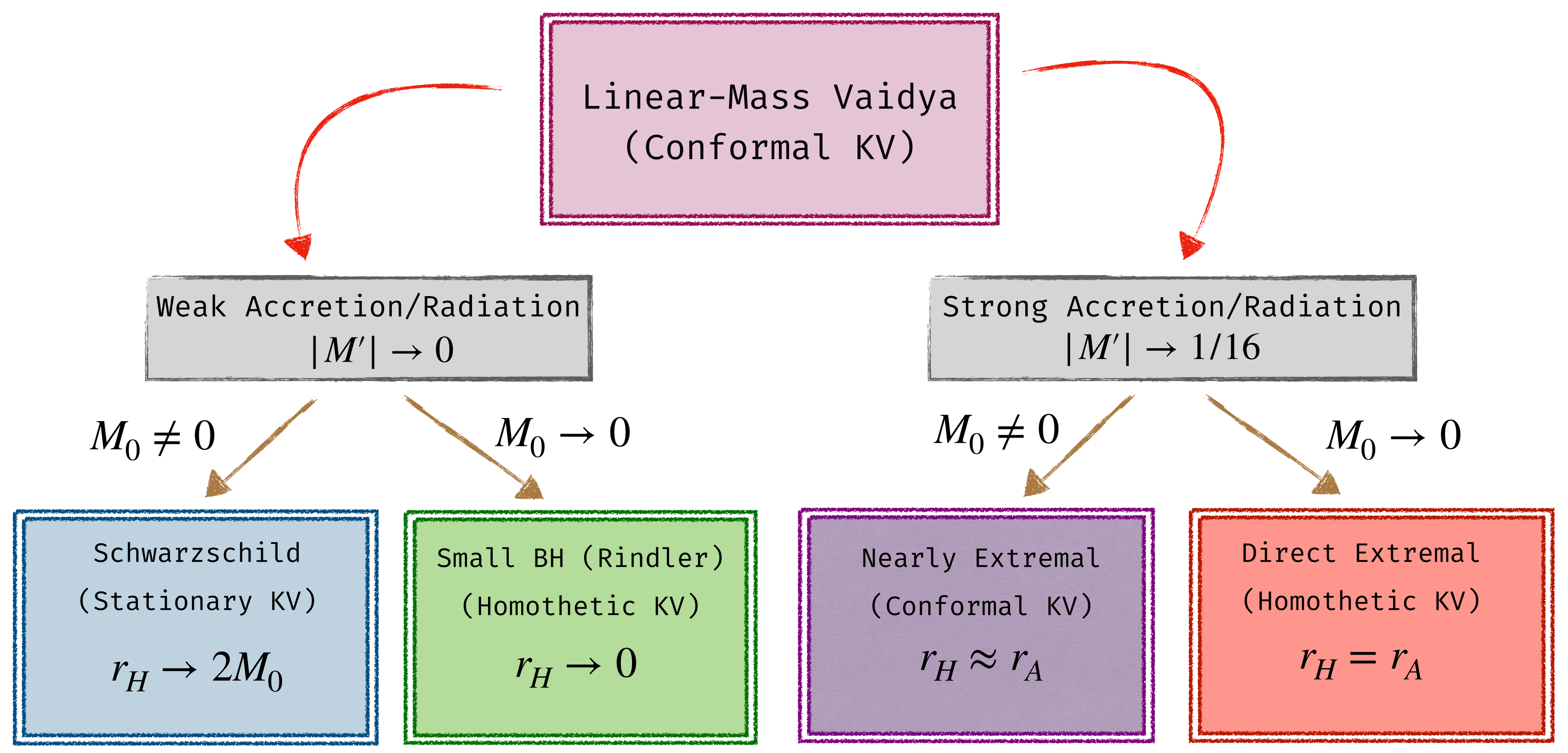}
	\caption{\textbf{Symmetries and degenerate limits of the LMV spacetime:} The one-parameter family of LMV solutions, having two horizons, spans weak ($|M'| \to 0$) and strong ($|M'| \to 1/16$) accretion/radiation regimes, each with two degenerate limits depending on $M_0$. When $M_0 \neq 0$ the spacetime carries a conformal Killing vector (KV) $\xi^T$, which asymptotes to an ordinary stationary KV in the Schwarzschild limit ($M' \to 0$), where the event horizon is located at $r = 2M_0$. In the nearly-extremal limit ($|M'| \to 1/16$, $M_0 \neq 0$) the two horizons approach one another but remain strictly distinct. When $M_0 \to 0$ the additional scale is absent and $\xi^T$ asymptotes to the homothetic KV $\xi^H = w\partial_w + r\partial_r$, tending to the self-similar Vaidya spacetime in both the small BH ($M' \to 0$) and direct extremal ($|M'| \to 1/16$) limits. In the direct extremal case the two horizons merge into a single degenerate horizon, whereas in the small BH case the event horizon shrinks to almost zero while the outer horizon remains finite. In the conformal SR geometry, the second horizon is interpreted as an acceleration horizon, and the small BH limit is also called the Rindler limit.}
	\label{fig:Vaidya_Killing_Vectors}
\end{figure*}

While the preceding discussion summarizes how the parameters $M'$ and $M_0$ govern the continuous symmetries of the spacetime, these same parameters also dictate its global causal structure. In this regard, it is important to note that the spacetime has degenerate limits, similar to the SdS geometry~\cite{Zhou:2025xta}. The existence of these degenerate geometric limits is perhaps better understood within the conformal SR geometry by introducing the parameter
\begin{equation}
	t = \dfrac{x_H}{x_A}.
\end{equation}
Because the horizon locations $x_H$ and $x_A$ are determined entirely by $|M'|$, the parameter $t$ naturally maps the limits of the LMV spacetime to those of the SR geometry: the weak accretion/radiation limit ($|M'| \to 0$) maps to $t \to 0$, while the strong accretion/radiation limit ($|M'| \to 1/16$) maps to $t \to 1$. However, this mapping is degenerate in the sense that a single limiting value of $t$ can admit multiple distinct limiting geometries. The conformally related SR spacetime does not explicitly retain the initial length scale $M_0$, whose dependence enters only implicitly through the relation $x_{H,A}=r_{H,A}/2M(w)$. Consequently, distinct LMV spacetimes with different physical length scales may correspond to the same limiting SR geometry.

Using this parameterization, the $t \to 0$ (weak accretion/radiation) limit encompasses two distinct limiting geometries. The first corresponds to the Schwarzschild limit, where $x_H = 1$ and $x_A$ merges with null infinity. The second corresponds to the Rindler (or small BH) limit, in which the acceleration horizon remains finite at $x_A \gg x_H$, while the BH horizon shrinks toward $x_H \to 0$. In the LMV spacetime, fixing $M_0 \neq 0$ while taking $|M'| \to 0$ recovers the Schwarzschild spacetime with $r_H = 2M_0$, whereas the simultaneous scaling $M_0 \to 0$ and $|M'| \to 0$ approaches the self-similar Vaidya geometry. For $0 < M_0 \ll 1$, this geometry describes a small BH undergoing weak accretion/radiation. The latter is analogous to the de Sitter limit of the SdS metric describing a small BH embedded in de Sitter spacetime \cite{Zhou:2025xta,Hintz:2021vfl}.

On the other hand, the $t \to 1$ (strong accretion/radiation) limit governs geometries where the two horizons of the SR metric approach one another. It is crucial to distinguish between two distinct limiting geometries in this branch. The first corresponds to the nearly-extremal geometry, characterized by two horizons that become arbitrarily close while remaining distinct. The second corresponds to the direct extremal geometry, where the two boundaries coincide exactly into a single degenerate horizon.  This structure is analogous to the degenerate branch of the SdS spacetime that admits both the Nariai and the direct extreme SdS limiting geometries~\cite{Zhou:2025xta,Bini:2025wef,Bousso:2002fq,Podolsky:1999ts}. In the LMV geometry, $M_0$ sets the length scale of the extremal configuration, with $M_0 \to 0$ yielding the self-similar realization of the direct extremal geometry.

We summarize the KV structure and degenerate limits of the LMV spacetime in Fig.~\ref{fig:Vaidya_Killing_Vectors}. Consequently, the primary focus of this work will be on the $t \to 0$ regimes of the SR metric, viz., the Schwarzschild and Rindler limits. We provide further details regarding these non-trivial limiting geometries in Sec.~\ref{sec:limitVaidyaHyperboloidal}, where we introduce hyperboloidal coordinates to facilitate their discussion\footnote{We explicitly exclude the direct extremal case from our analysis, as the Killing vector $\partial_T$ becomes globally spacelike; therefore, the spacetime does not contain an \emph{exterior region} where static geodesic observers exist, making a Fourier study of the wave equation impossible~\cite{Zhou:2025xta}.}. For completeness, the analysis of the nearly-extremal limit has been included in Appendix~\ref{appendix:next_limit}.

\subsection{Wave Equation and Quasinormal Modes}
As shown in~\cite{Capuano:2024qhv}, the conformal Killing symmetry discussed in the previous subsection allows us to factor out the time dependence from equations of motion for scalar, electromagnetic and axial gravitational perturbations \footnote{The polar gravitational perturbations of the Vaidya BH have only been treated in~\cite{Nolan:2005is}, with the assumption of self-similarity. However, a Zerilli equation for the general case, as far as we are aware, has not been derived.}. In particular, all the aforementioned equations can be cast into the single master perturbation equation, for a suitable master variable $\tilde\phi(T,x)$
\begin{equation}
	\left[\frac{\partial^2}{\partial T^2}-\frac{\partial^2}{\partial x_*^2}+\, f(x_*)V^{\rm eff}_{\ell,s} (x_*)\right]\tilde\phi(T,x_*)=0\,,
	\label{eq:wave-equation}
\end{equation}
where we have defined the effective potential

\begin{equation}
	V_{\ell,s}^{\rm eff}(x)=\left(\frac{\ell(\ell +1)}{x^2}+\frac{1-s^2}{x^3}\right)\,,
\end{equation}
with the spin parameter being $s = 0,1,2$ for scalar, electromagnetic and (axial) gravitational perturbations, respectively. Notice that the angular dependence has been factored out from Eq.~\eqref{eq:wave-equation} due to spherical symmetry.
Furthermore, the fact that the effective potential carries no time dependence allows for the separation of variable $\tilde\phi\sim \exp(-i\Omega T) R(x_*)$ for a suitable radial master variable $R(x_*)$.
Hence, Eq.~\eqref{eq:wave-equation} can be rewritten as
\begin{equation}
	\left[\frac{{\rm d}^2}{{\rm d}x_*^2}+\tilde{\Omega}^2- \, f(x_*) V^{\rm eff}_{\ell,s}(x_*)\right]R(x_*)=0\,,
	\label{eq:VaidyaFourierDomain}
\end{equation}
with $\Tilde{\Omega}=\Omega$ in the scalar and electromagnetic case, and $\Tilde{\Omega} = \Omega+2iM'$ in the axial gravitational case. Note that if one sets $\abs{M'}=0$, $f$ reduces to $f=1-{1}/{x}$ and Eq.~\ref{eq:VaidyaFourierDomain} becomes the Regge-Wheeler equation under the identification $x={r}/{2M_0}$, with $M_0$ being the Schwarzschild mass parameter.

Now Eq.~\eqref{eq:VaidyaFourierDomain} can be solved as an eigenvalue problem for $\tilde\Omega$, given a proper set of boundary conditions. Since we are interested in QNMs, we require that
\begin{equation}
	\begin{split}
		&R(x_*)\xrightarrow{x \rightarrow x_A}{\rm e}^{i\tilde{\Omega} x_*} , \\
		&R(x_*)\xrightarrow{x \rightarrow x_H}{\rm e}^{-i\tilde{\Omega} x_*}\,,
	\end{split}
	\label{eq:Rbcs}
\end{equation}
which means imposing that the solution to Eq.~\eqref{eq:VaidyaFourierDomain} behaves as a purely outgoing plane-wave at $x = x_A$ and as a purely ingoing plane-wave at $x=x_H$.
It is important to mention that the physical scalar, electromagnetic, and axial gravitational fields are related to the master variable appearing in Eq.~\eqref{eq:VaidyaFourierDomain} through a factor $(2 M(w))^\chi$, where $\chi$ is a conformal weight~\cite{Capuano:2024qhv}. In particular, for the aforementioned fields, one has $\chi = s(s+1)/2-1$. Hence, the physical fields oscillate with the shifted frequency
\begin{equation}
	\Omega_* = \Omega+2i\chi M'\,.
	\label{eq:conformal_shift}
\end{equation}
Remarkably, the shifts in Eq.~\eqref{eq:conformal_shift} and in the definition of $\tilde\Omega$ for axial gravitational perturbations are purely imaginary. Hence, they do not affect the boundary conditions of Eq.~\eqref{eq:Rbcs}, which can be safely imposed on $\tilde\Omega$~\cite{Capuano:2024qhv}.

As already noted in~\cite{Capuano:2024qhv}, the imaginary shift of Eq.~\ref{eq:conformal_shift} can potentially push the imaginary part of the physical QNM frequencies into the positive-half complex plane, producing an exponential growth of the perturbation in time, i.e., an instability. In more detail, this phenomenon can only occur if the product $\chi M'$ is positive and exceeds a given threshold, while not violating the condition $|M'|<1/16 = 0.0625$. In the case of gravitational LR QNMs, this occurs in the mass-accreting case for $0.0625>M'>0.0420$. On the other hand, for scalar perturbations, an instability can occur in the mass-radiating case, for $-0.0625<M'<-0.0436$, while electromagnetic perturbations are never affected by instabilities of this kind~\cite{Capuano:2024qhv}.

Finally, once the original dimensionful coordinates $(w,r)$ are restored, one obtains the actual time-dependent spectrum, given by
\begin{equation}
	\omega = \frac{\Omega_*}{2 M(w)}\,.
	\label{eq:time_dependent_omega}
\end{equation}
Therefore, Eq.~\eqref{eq:time_dependent_omega} states that, in the LMV spacetime, the time-evolution of physical modes is completely encoded in the evolution of the BH mass. Note that we will often refer to the QNM spectrum of the SR metric as the \emph{static} QNM spectrum of the LMV BH, omitting this distinction whenever it is clear from context.

Based on the discussion so far, we note that when we compute QNMs in the frequency domain, we traditionally use a Cauchy foliation. First, we write down the wave equation using the usual Schwarzschild time coordinate $T$, where the constant-$T$ hypersurfaces are spacelike and extend from $\mathcal{B}_H$ to $\mathcal{B}_A$ (or to spatial infinity $i^0$ for asymptotically flat spacetimes), as shown in Fig.~\ref{fig:penrose_diagram} by orange dotted lines. The next step involves imposing the QNM boundary conditions at the two bifurcation spheres (viz., Eq.~\eqref{eq:Rbcs}). However, this approach has a problematic consequence: the QNM eigenfunctions exhibit exponential growth near the two boundaries as $x \to \pm \infty$ \cite{PanossoMacedo:2024nkw}. To avoid this issue, we can select a spacetime foliation that respects the QNM boundary conditions. We choose a new time coordinate $\tau$ that not only penetrates the future event horizon $\mathcal{H}^+_H$ but also intersects the future acceleration horizon $\mathcal{H}^+_A$ ($\mathscr{I}^+$ for asymptotically flat geometries). When we rewrite the wave equation in these so-called hyperboloidal slices (shown in Fig.~\ref{fig:penrose_diagram} by solid purple lines), the eigenfunctions do not display any pathologies. Along with an appropriate radial coordinate, we can then recast the problem of finding QNMs as a generalized eigenvalue problem. For massless perturbations, it reduces to a quadratic eigenvalue problem. In the next section, we will construct such a foliation, which we will later use for our numerical computations.

\section{Hyperboloidal coordinates for the Vaidya black hole}
\label{sec:Hyperboloidal_coord}

The use of hyperboloidal foliations in the study of QNMs is relatively new; the idea was first introduced in \cite{Zenginoglu:2011jz} and later implemented in a particularly convenient gauge in \cite{Ansorg:2016ztf,PanossoMacedo:2018hab}. In recent times, this framework has proved pivotal in addressing questions related to the stability of the QNM spectrum through the notion of the pseudospectrum \cite{Jaramillo:2020tuu,Jaramillo:2021tmt,Ripley:2022ypi}. An extension of these ideas to asymptotically de Sitter spacetimes was given in \cite{Sarkar:2023rhp}, and is directly relevant for the present work. Since the hyperboloidal foliations allow the inclusion of future null infinity $\mathscr{I}^+$, (or more generally, the asymptotic wave zone) within the computational domain, their utility extends beyond the frequency-domain computations of QNM spectra \cite{Zenginoglu:2007jw,Zenginoglu:2008pw,Zenginoglu:2008uc,Rinne:2009qx,PanossoMacedo:2014dnr}, and they have found applications in time-domain studies of the wave equation on fixed backgrounds, extreme-mass-ratio inspirals, mathematical and numerical relativity, and theoretical high-energy physics~\cite{Zenginoglu:2011zz,Cardoso:2022whc,Rahman:2025mip,Friedrich1986OnTE,Schinkel:2013zm,Peterson:2023bha,Peterson:2024bxk,Peterson:2025csg,Alvares:2025pbi,Gregory:2017sor,Gregory:2018ghc,Anderson:2020dim,Anderson:2022zuc}. We refer the reader to the topical collections \cite{10.3389/fphy.2025.1753507,Hilditch:2025ind}, and the references therein, for further information.

We shall use the height function approach~\cite{Zenginoglu:2007jw,PanossoMacedo:2018hab,PanossoMacedo:2023qzp} to construct the required hyperboloidal coordinates for the SR metric, following a strategy used in~\cite{Sarkar:2023rhp}. Since the constant $T$ hypersurfaces converge at the bifurcation spheres, $\mathcal{B}_H$ and $\mathcal{B}_A$, the basic idea is to deform the $T$ coordinate using a so-called \textit{height function} $h(x)$ such that it behaves like a null coordinate at the boundaries but remains spacelike in the bulk, viz.,
\begin{equation}
	\tau = T + h(x), \label{eq:height_func_transformation}
\end{equation}

The functional form of $h(x)$ can be deduced by simple geometric arguments: Starting from Eq.~\eqref{eq:vaidya_metric_static}, introduce an ingoing null coordinate $V$ so that the metric reduces to a form that penetrates the event horizon, and write the metric in terms of $\tau$ using Eq.~\eqref{eq:height_func_transformation}. Then consider the ingoing and outgoing null vectors ($k^\alpha$ and $l^\alpha$ respectively) at the asymptotic boundary, and impose $k^\alpha \partial_\alpha \tau = 1$ (making $\tau$ a regular parameter at the boundary), while ensuring that the components of $l^\alpha$ remain well-behaved. This gives $h(x)$ up to a gauge choice. We also introduce a new radial coordinate $\sigma$ that linearly rescales the region between the event and acceleration horizons to a suitable range \cite{Sarkar:2023rhp}. One can also compactify the radial coordinate \cite{PanossoMacedo:2018hab, Zhou:2025xta}, but a linear rescaling is conceptually simple and adequate for our purpose (also see the discussion in \cite{PanossoMacedo:2023qzp}).

\subsection{Construction of Hyperboloidal Slices}
\label{sec:Hyperboloidal_coord_geometric_approach}
We first introduce an ingoing null coordinate for the static line element $\d \tilde{s}^2$ such that the constant time slices become ingoing at the horizon,
\begin{equation}
	T = V - x_*,
\end{equation}
where $x_*$ is the tortoise coordinate given by Eq.~\eqref{eq:tortoisecoordinate}. We shall also rescale the radial coordinate using the relation
\begin{equation}
	x = \frac{\rho(\sigma)}{\sigma} \,,
\end{equation}
such that $x_H=\sigma_H$ and $x_A=\sigma_A$. We can always choose $\sigma_H=1$ and $\sigma_A=0$ for convenience.
Note that we can also define an outgoing null coordinate, viz., $U = T - x_*$.
We can then write the static metric in horizon-penetrating coordinates as
\begin{align}
	\d \tilde{s}^2 = & ~ {\sigma^{-2}}
	\left[- \sigma^2 F(\sigma) \d V^2 - 2 \beta(\sigma)\d V \d\sigma + \rho(\sigma)^2
	\d\Omega^2_{S^2}\right]\, \nonumber         \\=~
	                 & \sigma^{-2} \d\bar{s}^2,
\end{align}
where
\begin{equation}
	F(\sigma) = f\bigl(x(\sigma)\bigr),
	\qquad
	\beta(\sigma) = \rho(\sigma) - \sigma \rho'(\sigma),
\end{equation}
and $\d \bar{s}^2$ is another conformal line element. Now, we can define the ingoing and outgoing null vectors, ${k}^\a$ and ${l}^\a$ respectively, using the relations
\begin{equation}
	{k}_\a = - B \nabla_\a V \,, \qquad  {l}_\a = - A \nabla_\a U \,,
\end{equation}
where $A$ and $B$ are constants. A straightforward calculation~\cite{Sarkar:2023rhp} shows that
\begin{align}
	{k}^\a & = \frac{B}{\beta} \delta^\a_\sigma,                                              \\
	{l}^\a & = B^{-1}\left( \delta^\a_V - \frac{F \sigma^2}{2 \beta} \delta^\a_\sigma\right),
\end{align}
where we have used ${k}^\alpha {l}_\alpha = 1$.

Let us now deform the ingoing null slices of constant $V$ into spacelike slices using a height function ${h_0}(\sigma)$,
\begin{equation}
	\tau = V + {h_0}(\sigma).
\end{equation}
To ensure that the spacelike hypersurfaces foliate the acceleration horizon, we must have
\begin{equation}
	{k}^\a \partial_\a \tau = 1. \label{tau_condition}
\end{equation}
In this new coordinate system, the line element becomes
\begin{align}
	\d\bar{s}^2
	= & - \sigma^2 F \d\tau^2
	+  h_0'\left( 2\beta-\sigma^2 F h_0'\right)\d\sigma^2 \nonumber \\&
	- 2\left(\beta-\sigma^2 F h_0' \right)\d\tau \d\sigma
	+ \rho^2 \d\Omega^2_{S^2}. \label{eqn:hyperboloidal_metric}
\end{align}
We can now write our null vectors in the $(\tau, \sigma, \theta,\varphi)$ coordinates, and we get
\begin{equation}
	{k}^\a= \delta^\a_\tau+\frac{1}{h_0'}\delta^\a_\sigma,
\end{equation}
\begin{equation}
	l^\a = \frac{h_0'}{2\beta^2}\left(2\beta - F\sigma^2 h_0'\right)\delta^\a_\tau
	- \frac{h_0'F\sigma^2}{2\beta^2}\delta^\a_\sigma,
\end{equation}
where we have used Eq.~\eqref{tau_condition} to fix $B=\beta/h_0'$.

Now, in order to ensure that $\sigma=\sigma_A$ (along constant $\tau$) is a null surface corresponding to the future acceleration horizon, we must have
\begin{equation}
	\lim_{\sigma \to \sigma_A} {k}^\a = \delta^\a_\tau \Rightarrow \lim_{\sigma \to \sigma_A} \frac{1}{h'_0}=0.
\end{equation}
However, we must also ensure that in this limit, the components of ${l}^\a$ remain finite. In other words,
\begin{equation}
	\frac{h_0'}{2\beta^2}\left(2\beta - F\sigma^2 h_0'\right) = C(\sigma),
\end{equation}
such that $C(\sigma)\to C_0$, a constant, as $\sigma \to \sigma_A$. In particular, in the limit $\sigma \to \sigma_A=0$, we can solve this quadratic equation to write
\begin{equation}
	h_0' = \frac{2 \beta}{F \sigma^2} - \beta C_0.
\end{equation}
The above expression for $h'_0$ ensures that the outgoing null vector is finite at $\sigma_A$.
Let us now make things more concrete by imposing the following relation between $x$ and $\sigma$, viz.,
\begin{equation}
	x(\sigma) = x_H \sigma + x_A (1-\sigma), \label{eqn:x_sigma_linear}
\end{equation}
which fixes $\rho$. Then, we get $\beta = (x_A-x_H)\sigma^2$, and we can write
\begin{equation}
	h_0' \simeq \frac{2 \beta}{F \sigma^2} = 2(x_A-x_H)\frac{1}{F(\sigma)}, \label{h_0_prime}
\end{equation}
where we have neglected $\mathcal{O}(\sigma^2)$ terms \cite{PanossoMacedo:2023qzp}. We can integrate the above equation to obtain
\begin{equation}
	h_0 = {2x_A}\eta \ln{\sigma}, \label{eq:h0_formula}
\end{equation}
where $\eta$ is given by Eq.~\eqref{eq:defTortoise_eta} and we have used the following expansion near $\sigma_A=0$,

\begin{equation}
	F(x(\sigma))^{-1} \simeq \frac{x_A}{4 |M'|(x_A-x_H)^2 \sigma}+ \mathcal{A}(\sigma)\,.
\end{equation}
The height function thus consists of a singular logarithmic term and a regular term ($\sim \int \d \sigma\mathcal{A}(\sigma)$). We can omit the regular term from the height function by exploiting gauge freedom; including this term will merely change how the hyperboloidal slices foliate the acceleration horizon.

Let us now establish the relation between the time coordinate $T$ and the hyperboloidal coordinate $\tau$, viz.,
\begin{equation}
	\tau = V + h_0(\sigma) = T + x_* + h_0(\sigma) = T + h(\sigma). \label{eq:T_tau_relation}
\end{equation}
Thus, we obtain the final expression for the height function,
\begin{equation}
	h(\sigma) = x_*(\sigma) + h_0(\sigma) = x_H \eta \ln(1-\sigma) + x_A \eta \ln{\sigma}, \label{eqn:height_func}
\end{equation}
where $\eta$ is given by Eq.~\eqref{eq:defTortoise_eta}. The constant $\tau$ hypersurfaces have been shown in the conformal diagram in Fig.~\ref{fig:penrose_diagram} by solid purple lines. We also write the expression for the tortoise coordinate in terms of $\sigma$,
\begin{align}
	g(\sigma) & \equiv  x_* = \int \frac{\d x}{f(x)} =\int\frac{\d \sigma}{F(\sigma)}\frac{\d x}{\d \sigma} \nonumber \\
	          & = x_H \eta \ln{(1-\sigma)} - x_A \eta \ln \sigma.
	\label{compact_func_tortoise}
\end{align}

We shall use Eqs.~\eqref{eq:T_tau_relation},~\eqref{eqn:height_func}, and \eqref{compact_func_tortoise} to rewrite the wave equation in hyperboloidal coordinates before finding QNMs.

Finally, note that the height function obtained here using geometric arguments can also be motivated using a purely algebraic approach. In fact, when evaluating QNMs via the standard Cauchy foliation using techniques such as the method of continued fractions, we factor out the singular behavior~\cite{PanossoMacedo:2024nkw} of the wave function at the boundaries by using a suitable ansatz motivated by a Frobenius analysis. The choice of the ansatz is equivalent to fixing a hyperboloidal foliation. We demonstrate this briefly in Appendix~\ref{appendix:height_function_algebraic_approach}, and we refer the reader to~\cite{PanossoMacedo:2018hab,Shen:2025nbq} for a more detailed discussion.

\subsection{Limits of the Vaidya Spacetime in Hyperboloidal Coordinates} \label{sec:limitVaidyaHyperboloidal}

Before evaluating the QNMs numerically using hyperboloidal coordinates, it is instructive to understand the various limiting geometries of the spacetime under consideration. A clear understanding of these limits will naturally help us anticipate how the QNM spectrum will change as we vary the parameters of the Vaidya geometry, especially if the limit corresponds to a spacetime whose QNMs are either known analytically or which reduces to a well-known example.

We will now demonstrate the existence of the geometric limit of the SR spacetime, focusing on the weak accretion/radiation regime. We will do so by defining a suitable parameter-dependent diffeomorphism that maps the points of the manifold to a specific limiting manifold \cite{Geroch:1969ca,Bengtsson:2014fha,Bugden:2018ekg}~(also see \cite{Paiva:1993bv}). We will then show that the hyperboloidal metric (Eq.~\eqref{eqn:hyperboloidal_metric}) also has the corresponding limit. Specifically, we shall demonstrate that the particular hyperboloidal slicing used to establish the limiting geometry is well defined in these limits. We also show that in the nearly-extremal (or strong accretion/radiation) limit, the spacetime is Nariai-like in Appendix~\ref{appsec:nearly_ext_limit_of_metric}.

\subsubsection{Schwarzschild Limit}
For the LMV spacetime, the Schwarzschild limit corresponds to the case where $|M'| \to 0$ with $M_0 \neq 0$. In this limit, a simple Taylor expansion of Eq.~\eqref{eq:horizons} shows that $x_H \to 1$ and $x_A \to \infty$ in the SR geometry, corresponding to $t \to 0$ as mentioned earlier. Consequently, the horizons of the LMV BH approach the values $r_H \to 2M_0$ and $r_A \to \infty$. It further reveals that the radial rescaling employed in the construction of the hyperboloidal coordinates in the previous sections is inadequate for theoretically studying this limit, since the Jacobian corresponding to Eq.~\eqref{eqn:x_sigma_linear} diverges, viz.,
\begin{equation}
	\frac{\d x}{\d\sigma} = x_H - x_A \to -\infty.
\end{equation}
In particular, the $g_{\sigma \sigma}$ and $g_{\sigma \tau}$ components of the hyperboloidal metric Eq.~\eqref{eqn:hyperboloidal_metric} diverge. We therefore introduce a new coordinate
\begin{equation}
	\sigma = 1 - t y \quad \implies y = \frac{1-\sigma}{t},
\end{equation}
where $t = x_H/x_A$ as before. In this coordinate system, the event horizon is fixed at $y_H=0$, whereas the location of the acceleration horizon varies with the spacetime parameter $t$ as $y_A=1/t$. This construction is similar in spirit to the compactified radial coordinate employed in~\cite{Zhou:2025xta} to probe the Schwarzschild limit of the Schwarzschild-de Sitter geometry. We can now write
\begin{equation}
	x = x_H \left(1 + y(1 - t)\right).
\end{equation}
Now in the limit, $t \to 0$, and we can write
\begin{equation}
	x \to 1+y , \quad \mathrm{and} \quad \frac{\d x}{\d y} \to 1,
\end{equation}
since $x_H \to 1$. Thus, the Jacobian is now finite. From Eq.~\eqref{eq:lapse_vaidya_static}, we can write
\begin{equation}
	f(y) \sim 1 - \frac{1}{1+{y}},
\end{equation}
as $t \to 0$ and we see that the SR metric (Eq.~\eqref{eq:vaidya_metric_static}) indeed reduces to the Schwarzschild metric with the event horizon at $y=0$ in this limit. Now, let us examine the behavior of the hyperboloidal metric in $(\tau,y)$ coordinates: using $\beta = (x_A-x_H) \sigma^2$, a careful evaluation of the Schwarzschild limit ($t \to 0$) of Eq.~\eqref{eqn:hyperboloidal_metric} gives us
\begin{equation}
	\begin{split}
		\d s^2 \sim\;&
		-f\,\d\tau^2
		+2(1-2f)\,\d\tau\,\d y \\
		&+4(1-f)\,\d y^2
		+(1+y)^2 \d\Omega_{S^2}^2\,.\label{eq:metric_Sch_limit}
	\end{split}
\end{equation}
where we used $h_0'(\sigma)\sim 2x_H/(t \sigma)$.
Note that, the above metric is regular and ingoing at the event horizon (since $\d y/\d \tau = 0$ for outgoing radial null rays at $y=0$, whereas $\d y/\d \tau=-1/2$ for ingoing null rays). One can put the above metric in standard ingoing Eddington-Finkelstein form using $\d \tau \sim \d V - 2 \d y$.
We can write Eq.~\eqref{eq:metric_Sch_limit} in the standard Schwarzschild form by using ${x} = 1 + y$ and
\begin{equation}
	\d \tau \sim \d T + \frac{1-y}{y} \d y.
\end{equation}
The above coordinate transformation is the Schwarzschild limit of the height function given by Eq.~\eqref{eqn:height_func}. Note that in the Schwarzschild limit, the height function behaves like
\begin{equation}
	h(y) \sim \ln t + \ln y - y.
\end{equation}
The appearance of $\ln t$ in the height function in the limit $t \to 0$ might appear worrying, but we can cure this by adding a constant of integration $-\ln t$ to $h_0$ in Eq.~\eqref{eq:h0_formula}. Since the metric, the coordinate transformations, and the wave equation involve only derivatives of the height function, the addition of a constant to $h_0$ does not affect our calculations. Finally, note that this irksome term  in $h$ comes from the tortoise coordinate $x_*$, since $h_0 \sim -2y$ in the Schwarzschild limit; hence we can get rid of it by adding a constant to $x_*$ as well.

\subsubsection{Rindler or Small Black Hole Limit}
Consider the case when $0<M_0\ll 1$ and $|M'| \to 0$: in this small BH limit, $r_H \to 0$ but $r_A \gg r_H$ is finite and the LMV solution describes the threshold of BH formation. We shall now show that the SR metric given by Eq.~\eqref{eq:vaidya_metric_static} has the Rindler spacetime as the limiting geometry. This situation is different from the Schwarzschild limit since we consider a regime in which the acceleration effect dominates over the contribution from the static BH. We want to isolate the growing and the constant term in $f (x)$. This means considering $x \gg 1$, which can still be compatible with $x < x_A \sim 1/|M'|$. In this regime, we have
\begin{equation}
	f \simeq 1-4 |M'| x\,. \label{eq:small_lapse}
\end{equation}
Using $t = x_H/x_A$, we now write Eq.~\eqref{eqn:x_sigma_linear} as,
\begin{equation}
	x  = x_A(t \sigma + 1 - \sigma).
\end{equation}
Using this, we can write,
\begin{equation}
	f = \dfrac{(1-t)^2 \sigma(1-\sigma)}{(1+t)(t \sigma +1 - \sigma)} \sim \sigma,
\end{equation}
where we have used $t \to 0$ to obtain the last approximation. We could have also obtained this result from Eq.~\eqref{eq:small_lapse}. We now define $\bar \rho = 2 x_A \sqrt{\sigma}$ and $\tilde T = T/2x_A$, and write Eq.~\eqref{eq:vaidya_metric_static} as
\begin{equation}
	ds^2 \sim -\bar \rho^2 d\tilde T^2 + d \bar \rho^2 + x_A^2\left(1-\dfrac{\bar \rho^2}{4x_A^2}\right)^2 d\Omega^2_{S{^2}}. \label{eq:rindler}
\end{equation}
The above metric gives us the usual Rindler metric in the limit $\sigma \to 0$, near $x_A$. Now, if we look at the metric in hyperboloidal coordinates Eq.~\eqref{eqn:hyperboloidal_metric}, we can write the following in the small BH limit,
\begin{equation}
	\begin{aligned}
		\d s^2 \sim \sigma^{2}\Big(
		 & - \sigma \, \d \tau^2
		+ 2x_A  \, \d \tau \, \d \sigma
		+ x_A^2 (1-\sigma)^2 \, \d \Omega^2_{S{^2}}
		\Big),
	\end{aligned}
\end{equation}
where we have used the fact that $h'_0 \to 2x_A/\sigma$ in this limit.  Note that, as $t \to 0$ with $x_H \to 0$, $x_* \sim -x_A \ln \sigma$. So $\tau \sim T +x_A \ln \sigma = T - x_* = U$ and $\d x \sim -x_A \d \sigma$. Therefore the hyperboloidal metric resembles the outgoing null metric near $x_A$. The total height function $h \sim x_A \ln \sigma \sim - x_*$ is also regular in this limit, and we can also recover the Rindler metric Eq.~\eqref{eq:rindler} from the hyperboloidal metric using our earlier definition of $\bar \rho$ and $\tilde T$.

\section{Computing QNMs analytically using the Heun Equation}\label{sec:QNMviaHeun}

In this Section, we provide an alternative way of computing QNM frequencies in Vaidya spacetime, which makes use of the fact that Eq.~\eqref{eq:VaidyaFourierDomain} can be mapped to a Heun equation~\cite{Heun1889,Ronveaux1995HeunsDE}. Recent results on the connection formulae for the Heun differential equation developed in~\cite{Bonelli:2022ten} will allow us to explicitly write the algebraic equation satisfied by the QNMs, which we will analytically solve in different limits. The method developed in~\cite{Bonelli:2022ten} exploits the Alday-Gaiotto-Tachikawa (AGT) correspondence~\cite{Alday_2010,Le_Floch_2022}, which relates the partition function of $\mathcal{N}=2$ $SU(2)$ gauge theories with correlation functions in Liouville conformal field theory (CFT)~\cite{ZamolodchikovZamolodchikov1991,Teschner_2001}. Here we carry out the computation for the weak accretion/radiation limit, $|M'| \to 0$, since we want to focus on demonstrating the existence of the PI modes. We treat the nearly-extremal (or the strong accretion/radiation) limit separately in Appendix~\ref{sec:HeunNariai}.

The first step that we need to carry out is to transform the differential equation for the radial function~\eqref{eq:VaidyaFourierDomain} into the normal form of the Heun equation.
This is achieved by introducing the variable $z=x/x_A$, which maps the points $\{0,x_H,x_A,\infty\}$ to $\{0,t,1,\infty\}$, where we have introduced the parameter $t=x_H/x_A$. We also rescale the dependent variable as $R(x)=\psi(x)/\sqrt{f(x)}$. The resulting differential equation is
\begin{equation} \label{eq:HeunNormalForm}
	(\partial_z^2+\bar V)\psi=0,
\end{equation}
with
\begin{equation}
	\bar V=\frac{1}{z^2(z-1)^2(z-t)^2}\sum_{i=0}^4\bar V_iz^i,
\end{equation}
where we have again used the definition $t=x_H/x_A$. \\
The explicit form of the coefficients $\bar V_i$ of the potential can be found in Appendix~\ref{appendix:Vaidya_potential}. Comparing with the Heun equation written in normal form, see Appendix~\ref{appendix:NSHeun}, we find the dictionaries between the coefficients describing the Liouville conformal field theory $a_0, a_1, a_\infty, a_t, u$ (see Appendix~\ref{appendix:NSHeun}) and the parameters of Vaidya spacetime:
\begin{align}
	 & a_0 =\theta^{(0)}\sqrt{1-\frac{1-s^2}{4|M'| \,x_H\, x_A}}, \nonumber   \\
	 & a_1 =\theta^{(1)}\frac{i\tilde{\Omega}}{4|M'|\,(1-x_H/x_A)}, \nonumber \\
	 & a_\infty =\theta^{(\infty)}\frac{i\tilde{\Omega}}{4|M'|}, \nonumber    \\
	 & a_t =\theta^{(t)}\frac{i\,\tilde{\Omega}x_H/x_A}{4|M'|\,(1-x_H/x_A)},
	\label{eq:a_iparameters}
\end{align}
and
\begin{align}
	u= & \frac{1}{2(1-x_H/x_A)}+ \frac{\ell(\ell+1)\, x_H +1-s^2}{4|M'|(x_H/x_A-1)\,x_H\, x_A} \nonumber \\
	   & -\frac{\tilde{\Omega}^2(x_H/x_A)^2}{8|M'|^2(1-x_H/x_A)^3}.
	\label{eq:uparameter}
\end{align}
The $\theta^{(i)}$ are independent sign choices that we will take to be $+1$. The behavior of the solutions around the singular point $z\sim z_i$ is
\begin{equation}
	\psi_\theta^{(z_i)}\sim(z-z_i)^{\frac{1}{2}+\theta a_i}(1+\mathcal{O}(z-z_i)).
\end{equation}
The connection formula for the solutions around $z=t$ (event horizon) and $z=1$ (acceleration horizon) is
\begin{equation}
	\psi^{(t)}_\theta=\sum_{\theta'=\pm}C_{\theta\theta'}\psi^{(1)}_{\theta'},
	\label{eq:connectionformula}
\end{equation}
with the connection coefficients given by the expression~\cite{Aminov_2023}
\begin{widetext}
	\begin{equation}
		C_{\theta\theta'}=t^{\frac{1}{2}-a_0+\theta a_t} (1-t)^{a_t-\theta a_1}e^{i\pi(a_t-\theta'a_1)}\sum_{\sigma=\pm}\frac{\Gamma(1-2\sigma a)\Gamma(-2\sigma a)\Gamma(1+2\theta a_t)\Gamma(-2\theta'a_1)t^{\sigma a}e^{-(\frac{\sigma}{2}\partial_a+\frac{\theta}{2}\partial_{a_t}-\frac{\theta'}{2}\partial_{a_1})F}}{\prod_{\sigma'=\pm}\Gamma(\frac{1}{2}+\theta a_t+\sigma'a_0-\sigma a)\Gamma(\frac{1}{2}-\theta'a_1+\sigma'a_\infty-\sigma a)}.
	\end{equation}
\end{widetext}
In this equation we have introduced $F$, the instanton part of the Nekrasov-Satashvili (NS) free energy \cite{NEKRASOV_2010}, which can be computed perturbatively as an instanton expansion in the parameter $t$ \cite{FLUME_2003}. Solving the Matone relation \cite{Matone_1995}, one gets a similar expansion for the parameter $a$ as a series in $t$. See Appendix~\ref{appendix:NSHeun} for more details.\\
To find the QNM condition, we have to impose ingoing waves at the event horizon, which behave as $e^{-i \tilde \Omega x_*}$, and outgoing waves at the acceleration horizon, which behave as $e^{i \tilde \Omega x_*}$. This amounts to choosing $\psi^{(t)}_-$ as our solution around $z\sim t$. Since the outgoing waves at $z\sim1$ are encoded in $\psi^{(1)}_-$ (see the definition of $x_*$ in Eq.~\eqref{eq:tortoisecoordinate}, which contains a minus sign in front of the $\ln |x-x_A|$ term), the QNM condition is $C_{-+}=0$. In particular, this condition implies
\begin{widetext}
	\begin{equation}
		\sum_{\sigma=\pm}
		\frac{
			\Gamma(1-2\sigma a)\Gamma(-2\sigma a)\Gamma(1-2a_t)\Gamma(-2a_1)
			\, t^{\sigma a} e^{-\frac{\sigma}{2}\partial_a F}
		}{
			\prod_{\sigma'=\pm}
			\Gamma\!\left(\tfrac{1}{2}- a_t+\sigma' a_0-\sigma a\right)
			\Gamma\!\left(\tfrac{1}{2}- a_1+\sigma' a_\infty-\sigma a\right)
		}
		=0
	\end{equation}
\end{widetext}
A family of solutions to this condition is given by the values that make a $\Gamma$ function in the denominator of each of the summands hit a pole.
In particular, in one of the summands we look at $\Gamma( \frac{1}{2}-a_1-a_\infty+a)$, and in the other at $\Gamma( \frac{1}{2}-a_1-a_\infty-a)$. Their pole spectrum is:
\begin{equation}
	\frac{1}{2}-a_1-a_\infty\pm a=-n,\quad n\in\mathbb Z_{\geq 0}.
\end{equation}
Since $t=x_H/x_A=4\abs{M'}+\mathcal{O}\left(\abs{M'}^2\right)$, the instanton expansion in $t$ is equivalent to the expansion in $\abs{M'}$. This means that if we want to compute some quantity to order $\abs{M'}^m$, we have to include $m$ instantons in the series. So, in order to get an expression for the QNMs, we expand $a_1$, $a_\infty$ and $a$ in powers of $\abs{M'}$. At leading order, we find that $a_1, a_\infty=\frac{i\tilde\Omega}{4|M'|}(1+\mathcal{O}(M'))$ and $a=(\frac{1}{2}+\ell)(1+\mathcal{O}(M'))$ (see appendix~\ref{appendix:NSHeun} for the procedure). The pole spectrum of the second $\Gamma$-function is the same as that of the first one, but shifted by $2i\abs{M'}(2\ell+1)$. Therefore, the common set of poles that satisfy the QNM condition is
\begin{equation}
	\tilde\Omega=\tilde{\Omega}^{(1)}\abs{M'}+\mathcal{O}\left(\abs{M'}^2\right), \label{eq:PI_spectrum_Heun_leading}
\end{equation}
with
\begin{equation}
	\tilde{\Omega}^{(1)}=-2i\left(n+\ell+1\right).
	\label{eq:PI_spectrum}
\end{equation}
Note also that, as already observed in the approximate regime, the leading contribution to the QNMs is independent of $s$.
Analogously, expanding $a_1$, $a_\infty$ and computing $a$ to next-to-leading order, we find that the condition to hit the poles of the $\Gamma$-functions gives the order $\abs{M'}^2$ correction to the QNMs
\begin{equation}
	\tilde{\Omega}=\tilde{\Omega}^{(1)}\abs{M'}+\tilde{\Omega}^{(2)}\abs{M'}^2+\mathcal{O}\left(\abs{M'}^3\right),
\end{equation}
with
\begin{equation}
	\tilde{\Omega}^{(2)}=-4i\frac{(\ell-n)(2\ell+1)-2s^2}{2\ell +1}.
\end{equation}
Including more instantons in our computations and expanding $a_i$ to higher orders in $\abs{M'}$ we can get the higher order corrections to the QNMs. The explicit expressions can be found in appendix~\ref{appendix:QNMsHeunSchwarzschild}.\\
As anticipated in the introduction, when reconstructing the physical QNM frequencies of the LMV metric, one should include an imaginary shift, given by Eq.~\eqref{eq:conformal_shift}. In the $M'>0$ case, this can potentially lead to an instability, which is the case for LR QNMs with $M'$ above a certain threshold. We now briefly comment about this possibility in the context of Rindler modes. The full LMV frequency (modulo the time-dependent overall factor of Eq.~\eqref{eq:time_dependent_omega}), for gravitational perturbations, and $M'>0$ reads
\begin{equation}
	\Omega_*^{(\ell,n)}=-2i(n+\ell) |M'|+\mathcal{O}(|M'|^2)\,,
	\label{eq:QNM_instability}
\end{equation}
which is always strictly negative, as $\ell\geq2$. Hence, no instability is possible in the gravitational case. On the other hand, in the scalar case, for $M'<0$, we obtain the same result as in Eq.~\eqref{eq:QNM_instability}. However, for scalar perturbations, the multipole number can be $\ell = 0$. In this case, the fundamental $n = 0$ has exactly vanishing leading-order $\Omega_*$, and using the higher-order corrections in Appendix~\ref{appendix:QNMsHeunSchwarzschild}, one obtains
\begin{equation}
	\Omega_*^{(0,0)}=-16i |M'|^3+\mathcal{O}(|M'|^4)\,.
	\label{eq:QNM_instability2}
\end{equation}
This mode, while very long-lived, is decaying in time for $M' <0$, and thus we can conclude that no instability is affecting Rindler modes in LMV.

We conclude this section showing that the order $\mathcal{O}\left(|M'|\right)$ result for the PI modes frequencies, given by Eq.~\eqref{eq:PI_spectrum}, can be derived in the small-BH/Rindler limit. In this regime, the metric function $f(x)$ is given by Eq.~\eqref{eq:small_lapse}, and the tortoise coordinate reads
\begin{equation}
	x_*=-\frac{\ln(1-4 |M'|x)}{4|M'|}\,.
\end{equation}
Notice that the physically relevant domain of this new tortoise coordinate is $(0,\infty)$, and not $(-\infty,+\infty)$ anymore. This is a consequence of the fact that the BH event horizon, which is located at $-\infty$ in a tortoise-coordinate frame, does not appear here, as in the purely de-Sitter case. Keeping only the leading term in $1/x$ in the potential, one gets the perturbation equation
\begin{equation}
	\left[\frac{{\rm d}^2}{{\rm d}x_*^2}+\left(\tilde{\Omega}^2-\mathcal{V}(x_*)\right)\right]R(x_*)=0\,,
	\label{eq:master_eq_approx}
\end{equation}
with the potential
\begin{equation}
	\mathcal{V}(x_*)=\frac{4|M'|^2\ell(\ell+1)}{\sinh(2|M'|x_*)^2}\,.
\end{equation}

Eq.~\eqref{eq:master_eq_approx} can be solved analytically in terms of hypergeometric functions.
The perturbation master variable, in the limit $x_*\rightarrow \infty$, behaves asymptotically as
\begin{equation}
	R(x_*)=C_1\,\exp\left(i\tilde\Omega x_*\right)+C_2\,\exp\left(-i\tilde\Omega x_*\right)\,,
\end{equation}
with $C_1$ and $C_2$ are integration constants.
Imposing a purely outgoing boundary condition, one gets $C_2=0$. Then, one needs regularity at $x_*=0$. The solution in this limit behaves, modulo a numerical prefactor, as
\begin{align}
	 & R(x_*)\sim -\left(2|M'|x_*\right)^{1-2\beta}\frac{2^{2(\alpha+\beta)-1}}{\Gamma\left(\tfrac{3}{2}-2\beta\right)\Gamma\left(2(\alpha+\beta)\right)} \nonumber \\
	 & +\left(2|M'|x_*\right)^{2\beta}\frac{4^{2(\alpha-\beta)}}{\Gamma\left(2(\alpha-\beta)+1\right)\Gamma\left(2(\beta+\tfrac{1}{2})\right)}\,,
	\label{eq:small_bh_near_origin}
\end{align}
with
\begin{align}
	 & \alpha = \pm \frac{i\tilde\Omega}{4|M'|}\,, \nonumber \\
	 & \beta = \frac{1}{2}\pm\frac{1+\ell}{2}\,.
	\label{eq:alpha_beta}
\end{align}
The choice of the sign for $\alpha$ and $\beta$ is a matter of convention and will not affect the QNM spectrum.
In order to obtain the final quantization condition on $\tilde{\Omega}$, we first need to determine which term in Eq.~\eqref{eq:small_bh_near_origin} can become singular in the limit $x_* \to 0$. This depends on the choice of root for $\beta$.
For $\beta > 1/2$, corresponding to the positive root $\beta = (\ell+1)/2$, the first term in Eq.~\eqref{eq:small_bh_near_origin} diverges as $x_* \to 0$, while the second term remains regular. Conversely, for the negative root, i.e. $\beta = -\ell/2$, the first term is regular and the second term becomes divergent in the same limit.

In both cases, physical regularity requires that the divergent contribution be removed. This can be achieved by imposing that the coefficient of the divergent term vanishes, which occurs when the corresponding gamma function develops a pole. This procedure leads to the same quantization condition on $\tilde{\Omega}$ independently of the choice of the root for $\beta$:
\begin{equation}
	\tilde\Omega =-2i|M'|(n+\ell+1)\,,
\end{equation}
which is the same as Eq.~\eqref{eq:PI_spectrum}.

In this section, both in the Heun equation and in the hypergeometric equation approach, we relied on the Rindler limit geometry. We will explore the opposite, nearly-extremal limit towards the end of Appendix~\ref{sec:HeunNariai}.

\section{Quasinormal Modes using the Hyperboloidal Approach}
\label{sec:freq_domain_numerical}

In this section, we derive the wave equation in hyperboloidal coordinates and describe the numerical method used for computing QNMs in the frequency domain. We then establish the existence of PI modes in the QNM spectrum numerically. We then move on to examining how the QNM spectrum behaves under the degenerate limit $t \to 0$. We therefore look at both the Schwarzschild and the Rindler limit of the QNM spectrum numerically. We shall also establish how these PI modes give rise to the branch-cut in the Schwarzschild limit. We also provide a detailed comparison of these numerical results with the analytic ones derived in Sec.~\ref{sec:QNMviaHeun}. Lastly, we demonstrate the nearly-extremal limit of the QNM spectrum in Appendix~\ref{appsec:numerical_nariai_limit_QNMs}.

\subsection{The wave equation in hyperboloidal coordinates}
\label{sec:wave_eq_hyperboloidal}
We employ the hyperboloidal coordinates derived in Section \ref{sec:Hyperboloidal_coord_geometric_approach}, viz.,
\begin{equation}
	T = \tau - h(\sigma), \quad x_* = g(\sigma)
\end{equation}
to rewrite Eq.~\eqref{eq:wave-equation} as
\begin{equation}
	\begin{split}
		\Bigg[& \left(1-\left(\dfrac{h'}{g'}\right)^2\right)\partial_\tau^2 -\dfrac{2}{g'}\left(\dfrac{h'}{g'}\right)\partial_\tau\partial_\sigma\\&-\dfrac{1}{g'}\left(\dfrac{h'}{g'}\right)'\partial_\tau -\dfrac{1}{g'}\partial_\sigma\left(\dfrac{1}{g'}\partial_\sigma\right)+ V^{\rm eff}_{\ell,s} \Bigg]\tilde\phi=0,
	\end{split}
\end{equation}
where $h(\sigma)$ and $g(\sigma)$ are given by Eqs.~\eqref{eqn:height_func}, and \eqref{compact_func_tortoise} respectively~\cite{Jaramillo:2020tuu,Sarkar:2023rhp}. We introduce $ \tilde \zeta=\partial_\tau \tilde\phi$, and write the first order reduction of the wave equation as
\begin{equation}
	\partial_\tau \tilde{u} = i L \tilde u, \label{wave_eqn_first_order}
\end{equation}
with
\begin{equation}
	\tilde u = \begin{pmatrix}
		\tilde \phi \\
		\tilde \zeta
	\end{pmatrix},
	\quad
	L = -{i}\left(\begin{array}{cc}
			0   & 1   \\
			L_1 & L_2
		\end{array}
	\right)
\end{equation}
where
\begin{align}
	\label{eqn:L_L1_L2}
	L_1 & = \dfrac{1}{w(x)}\left(\partial_x \left(p(x)\partial_x\right)-q_{\ell,s}(x)\right), \nonumber \\
	L_2 & = \dfrac{1}{w(x)}\left(2\gamma(x)\partial_x+\partial_x\gamma(x)\right),
\end{align}
and
\begin{align}
	\label{eqn:aux_func_L1_2}
	w(x)          & = \dfrac{g'^2-h'^2}{\lvert g' \rvert},\quad p(x) = \dfrac{1}{\lvert g' \rvert}, \nonumber \\
	q_{\ell,s}(x) & = \lvert g' \rvert\,V^{\rm eff}_{\ell,s}, \quad \gamma(x) = \dfrac{h'}{\lvert g' \rvert}.
\end{align}
The expression for $L$ that we obtain for our choice of height function and rescaling, that is, using Eqs.~\eqref{eqn:height_func} and \eqref{compact_func_tortoise} is given in Appendix~\ref{appendix:wave_operator_hyperboloidal_explicit_expressions}. We note that a major motivation behind adopting hyperboloidal coordinates lies in how the QNM boundary conditions are incorporated geometrically for massless perturbations. To see this, note that $L_1$ has the structure of a Sturm-Liouville operator, where the coefficient $p(x)$ vanishes at the two boundaries while all other quantities appearing in $L$ remain finite. Therefore, $L_1$ is a singular Sturm-Liouville operator. If we now impose that the eigenfunctions are sufficiently regular, then $L_1$ does not require any explicit boundary conditions, and this extends to the full operator $L$ as well \cite{Jaramillo:2020tuu, Sarkar:2023rhp}. Now, to find the QNMs, we put
\begin{equation}
	\tilde u  \sim e^{-i\tilde\Omega \tau}
\end{equation}
into Eq.~\eqref{wave_eqn_first_order} to get the following eigenvalue problem
\begin{equation}
	L \tilde u = -\tilde \Omega \tilde u. \label{eqn:qnm_eigenvalue}
\end{equation}

It is worth highlighting that so far we have been dealing with dimensionless coordinates $x$ and $T$. However, to probe the limits of the QNM spectrum and verify that they correspond to the QNMs of the limiting geometries, we have to introduce a suitable characteristic length scale $x_0$ where $x_0 \in \{ 1, x_H, x_A\}$. This means that we have to rescale our coordinates, as
\begin{equation}
	T \to T/x_0, \quad x \to x/x_0 \equiv z.
\end{equation}
Consequently, after rescaling the wave equation, our modes scale as
\begin{equation}
	\tilde \Omega \to x_0 \tilde \Omega,
\end{equation}
and the potential becomes
\begin{equation}
	f(x)V^{\rm eff}_{\ell,s} (x) \to x_0^2 f(z)\bar{V}^{\rm eff}_{\ell,s}(z)
\end{equation}
with

\begin{align}
	f(z)                          & = - \frac{(z -z_H)(z-z_A)}{z (z_H+z_A)},\nonumber                          \\
	\bar{V}^{\rm eff}_{\ell,s}(z) & = \frac{1}{x_0^2}\left[ \frac{l(l+1)}{z^2} + \frac{1-s^2}{z^3 x_0}\right].
\end{align}

We now find the eigenvalues of the matrix $L$ using the Chebyshev spectral method. The modes which pass the convergence test correspond to the static QNMs of the LMV spacetime.

\subsection{Numerical Method}

We employ the Chebyshev spectral method using a Gauss-Lobatto collocation grid to convert a calculus problem into a linear algebra problem. In other words, we approximate the differential operator governing the scattering problem with a suitable finite-dimensional matrix whose eigenvalues correspond to the QNM frequencies. This method can help us find the eigenfunctions as well, and if the eigenfunctions are analytic, the method guarantees exponential convergence. Moreover, in our implementation, the last stage of the process involves using a standard eigenvalue solver to compute the eigenspectrum. Therefore we can obtain the entire eigenspectrum of the discretized operator in a single step. This strategy gives the method a distinct advantage over iterative methods, which require an initial guess value to find a particular mode. So, this enables us to locate modes that may be missed by, say, the method of continued fractions\footnote{Indeed, the choice of our numerical method is intimately linked to the result presented in this paper, that is, the existence of PI static modes of the Vaidya BH that was previously unreported.}. However, the method is not without its disadvantages. Since we truncate an essentially infinite-dimensional operator to a finite size, the method generates many spurious eigenvalues that do not correspond to physical QNMs. Hence, to reject spurious modes, the spectrum must be subjected to a rigorous convergence test. Moreover, the choice of Chebyshev polynomials as our basis generates highly ill-conditioned matrices. In addition, we are dealing with an inherently dissipative system whose eigenspectrum is extremely sensitive to numerical noise~\cite{Jaramillo:2020tuu,Sarkar:2023rhp}. So, we often have to set the internal precision to a value much larger than machine precision to compute higher overtones. The method can be improved by using an ultraspherical polynomial basis~\cite{Ripley:2022ypi,Zhong:2022jke}, physically motivated gauge choices, and mesh refinement \cite{Hintz:2021vfl,Zhou:2025xta}. In this work, we employ the latter two techniques while probing the QNM spectrum under the limits described in Section~\ref{sec:limitVaidyaHyperboloidal}. We discuss some of the salient features of the method in Appendix \ref{app:chebyshev_spectral_method_details}.
We have carried out the numerical computations in the \texttt{Wolfram Language}~\cite{wolfram_language,wolfram_language_manual} and set the internal numerical precision to $10\times$\texttt{MachinePrecision}.

\begin{figure}[htpb]
	\centering
	\begin{minipage}{\columnwidth}
		\centering
		\includegraphics[width=0.9\linewidth]{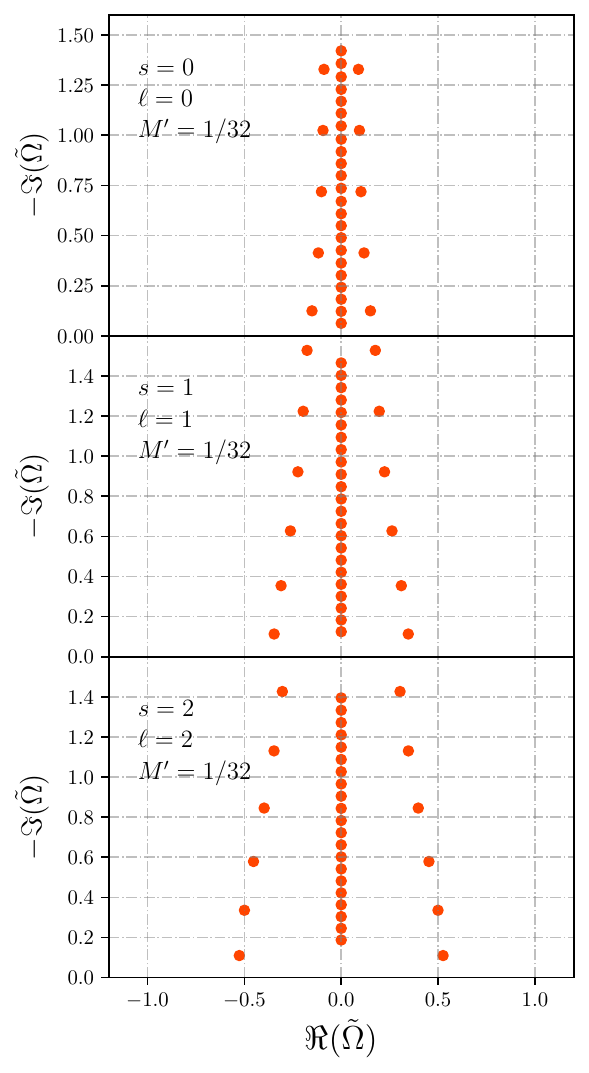}
	\end{minipage}
	\caption{\textbf{Static QNM Spectrum of the Vaidya BH} for scalar perturbations ($s=0$) with $\ell = 0$ \textbf{(top)}, electromagnetic perturbations ($s=1$) with $\ell = 1$ \textbf{(middle)}, and gravitational perturbations ($s=2$) with $\ell = 2$ \textbf{(bottom)}. We fix the accretion rate to $M' = 1/32$ and set the numerical precision to $10\times\mathrm{MachinePrecision}$. We compute QNMs using the Chebyshev spectral method with standard Gauss-Lobatto grids of sizes $N_1 = 200$ and $N_2 = 250$ in hyperboloidal coordinates, and display here only those modes that satisfy $|1 - \omega_{N_1}/\omega_{N_2}| < 10^{-8}$. The spectrum contains PI modes, termed here as Rindler modes, in addition to the usual LR modes. The characteristic length here is chosen to be $x_0=1$.}
	\label{fig:combined_qnm_spectrum}
\end{figure}

\subsection{Numerical Results}
\label{sec:numerical_results}
In this subsection, we shall discuss various aspects of the QNM spectrum that we have obtained numerically using the spectral method with hyperboloidal foliations. The discussion bears a strong parallel to the analysis carried out in~\cite{Zhou:2025xta} for the Schwarzschild-de Sitter case. A demonstration of the exponential convergence of the numerical scheme can be found in Appendix~\ref{appendix:convergence}. Since a typical convergence test involves computing the same spectrum multiple times for several grids of varying sizes, it would become a rather computationally expensive affair if we wanted to show that every spectrum is exponentially convergent for all the values of the parameter $t$ used in this work. However, given the possibility that a spectrum may be contaminated with spurious eigenvalues, we adopt the following filtering process~\cite{Jansen:2017oag,Sarkar:2023rhp}: we compute each spectrum for two grids of size $N_1$ and $N_2$, and we claim that a mode $\tilde \Omega_{n}$ is convergent if,
\begin{equation}
	\delta \tilde \Omega_n \equiv \abs{1-\dfrac{\tilde \Omega_n^{N_1}}{\tilde\Omega_n^{N_2}}} < 10^{-p}, \label{eq:numerical_filter}
\end{equation}
where $n$ denotes the mode number with $n=0$ being the lowest-lying mode, $\tilde \Omega^{N}$ indicates that the spectrum was computed with a grid of size $N$. We usually take $N_1=200, N_2=250$ and $p=8$ unless stated otherwise. Note that we label both the LR modes and Rindler modes individually with $n$ starting from zero in the figures below. Furthermore, we use the same $n$ to denote the two LR modes satisfying $\tilde \Omega = - \tilde \Omega^{\dagger}$ where the dagger denotes complex conjugation. In the figures we often just show the LR mode with positive real part.

In Fig.~\ref{fig:combined_qnm_spectrum}, we show a set of representative static QNM spectra of the Vaidya BH for a fixed accretion rate $M' = 1/32$, for scalar ($s = \ell = 0$), electromagnetic ($s = \ell = 1$), and gravitational ($s = \ell = 2$) perturbations in the top, middle, and bottom panel respectively. The spectrum shows the expected  LR modes associated with the existence of the light ring. But, importantly, it also contains a set of convergent purely imaginary modes. These modes are associated with the presence of the acceleration horizon in the static conformal Schwarzschild-Rindler spacetime. The existence of these purely imaginary Rindler modes of the Vaidya BH is the main new result, as it points to a qualitatively richer mode content in the ringdown spectrum of dynamical BHs.  One should also note that the spectrum bears a striking resemblance to that of the Schwarzschild-de Sitter BH~\cite{Sarkar:2023rhp}. The computation was carried out using the foliation given by Eqs.~\eqref{eqn:height_func} and \eqref{eqn:x_sigma_linear} on a standard CGL grid, the characteristic length scale being $x_0=1$.

\subsubsection{QNMs in the Schwarzschild Limit}
\begin{figure}[t]
	\centering

	\begin{minipage}{0.95\columnwidth}
		\centering
		\includegraphics[width=\linewidth]{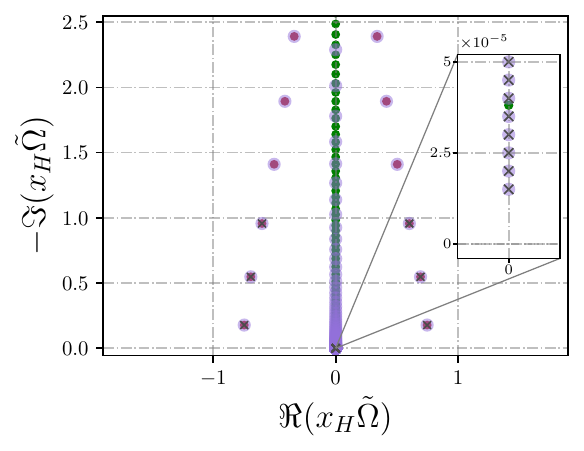}
	\end{minipage}
	\caption{\textbf{Static QNM Spectrum of the Vaidya BH in the Schwarzschild Limit} for gravitational perturbations ($s=2$) with $\ell = 2$. We choose $t\equiv x_H/x_A = 10^{-5}$ and set $x_0 =x_H$, resulting in a very small accretion rate ($M' \sim 2.499 \times10^{-6}$), and show the spectrum for $N= 250$ (translucent purple circles) superimposed on the QNM spectrum of the Schwarzschild BH (filled red circles). The green circles are a numerical artifact due to the existence of a branch-cut for Schwarzschild BHs (see main text). The Vaidya BH modes marked by crosses have a relative difference $\delta \tilde \Omega_n<10^{-8}$ when compared to the spectrum obtained for $N=200$. The Vaidya spectrum in the Schwarzschild limit bears a striking resemblance to that of the Schwarzschild BH with the lowest-lying LR mode being $\tilde \Omega M= \pm 0.373672 - i 0.0889615$ which is very close to the fundamental mode of the Schwarzschild BH. In the \textbf{Inset}, we zoom into the region around $\tilde \Omega=0$ and find several convergent Rindler modes with $\mathfrak{I}(\tilde\Omega)<0$ of the order $ \sim10^{-5}$, indicating a rapid accumulation of modes. These results were obtained using mesh refinement. }

	\label{fig:schwarzschild_limit_spectrum_vs_sch}
\end{figure}
We now probe the Schwarzschild limit of the QNM spectrum. In this limit, $t \to 0$ as $x_H$ is held fixed. So, as $t$ becomes progressively smaller, $x_A \to \infty$ and eventually merges with the future null infinity  $\mathscr{I}^+$. This intuitive picture was formally established in Sec.~\ref{sec:limitVaidyaHyperboloidal} where we showed that the limiting geometry of the SR metric reduces to that of the Schwarzschild BH in this scenario. To illustrate this point, we had introduced a radial coordinate $y = (1 -\sigma)/t$ which can be written as
\begin{equation}
	z \equiv \dfrac{x}{x_H} = 1 + y(1-t).
\end{equation}
Clearly, the event horizon is fixed at $y_H=0$, the acceleration horizon varies as $y_A=1/t$. Since the location of the acceleration horizon now varies parametrically with $t$, we can therefore use the radial coordinate $y$ to numerically probe the Schwarzschild limit as well. The coordinate $\sigma$ is inadequate for the present purpose since it keeps the two horizons at fixed coordinate locations, allowing us to only vary the separation between the two. The characteristic length is now set by $x_H$ so the QNMs in the Schwarzschild limit, $\Omega^{\mathrm{Sch}}$, scale as
\begin{equation}
	\tilde \Omega^{\mathrm{Sch}} = x_H \tilde \Omega \overset{t \to0}{\sim} \tilde \Omega, \label{eq:Sch_Lim_Normalization}
\end{equation}
where we used $x_H = 1+t$ to write the last approximation. We also use mesh refinement to cluster more points near the event horizon by setting $x_B=-1$ and $\kappa=7$.

\begin{figure}[t]
	\centering
	\begin{minipage}{\columnwidth}
		\centering
		\includegraphics[width=\linewidth]{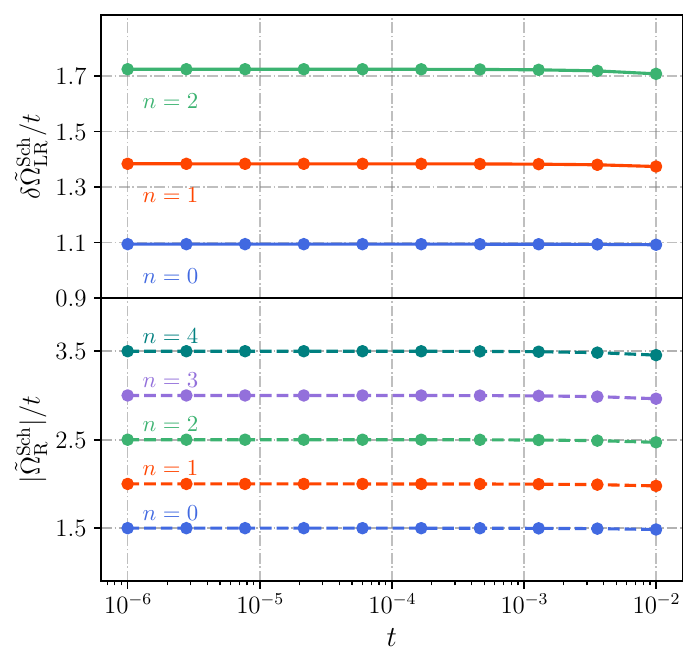}
	\end{minipage}
	\caption{\textbf{Parametric behavior of QNMs in the Schwarzschild Limit} for gravitational perturbations ($s=2$) with $\ell=2$. In the \textbf{top panel}, we plot the scaled relative difference $\delta \tilde \Omega_\mathrm{LR}^{\mathrm{Sch}}/t$ between the three lowest-lying LR modes of the Vaidya BH and the corresponding QNMs of the Schwarzschild BH as a function of $t$. In the \textbf{bottom panel}, we plot the scaled absolute values of the five lowest-lying Rindler modes $|\tilde \Omega_\mathrm{R}^{\mathrm{Sch}}|/t$ as a function of $t$. By scaling these quantities by $t$, we observe nearly constant values (horizontal lines), demonstrating that both $\delta \tilde \Omega_\mathrm{LR}^{\mathrm{Sch}}$ and $\tilde \Omega_\mathrm{R}^{\mathrm{Sch}}$ depend linearly on $t$ as $t \to 0$. The index $n$ denotes the overtone number for both LR and Rindler modes, with $n=0$ being the corresponding lowest-lying mode. The modes shown here are convergent with a relative difference $\delta \tilde \Omega_n < 10^{-8}$.}
	\label{fig:schwarzschild_limit_scaling_lr_modes}
\end{figure}

Let us now discuss our numerical results. In Fig.~\ref{fig:schwarzschild_limit_spectrum_vs_sch}, we show the QNM spectrum for gravitational perturbations with $\ell =2$ in the Schwarzschild limit for $t = 10^{-5}$, corresponding to a weak accretion/radiation rate of $M'\sim 2.5 \times 10^{-6}$. The static modes of the Vaidya BH are shown as translucent purple circles. The spectrum is superimposed on the usual QNM spectrum of the Schwarzschild BH~\footnote{Note that the Schwarzschild spectrum was computed using hyperboloidal coordinates in the minimal gauge~\cite{Jaramillo:2020tuu} on a standard CGL grid of size $N = 600$, and the data has been taken from~\cite{SubhodeepThesis}.} indicated by red dots. The light-ring modes completely overlap showing that the part of the spectrum governing the ringdown signal is (practically) identical. The relative difference between the fundamental modes of the Schwarzschild BH and Vaidya BH is of the order $t \sim 10^{-6}$. However, as anticipated, there is a glaring difference. The spectral calculation of the Schwarzschild eigenvalue problem reveals that a number of non-converging eigenvalues are present on the imaginary axis (indicated by green circles in Fig. \ref{fig:schwarzschild_limit_spectrum_vs_sch}), and the number and position of these modes change drastically on changing the grid resolution \cite{Jaramillo:2020tuu}. While they may be dismissed as spurious eigenvalues, it is worthwhile to interpret them as the branch-cut in the Schwarzschild Green's function, especially since the modes begin to approximate a continuum as we increase the grid resolution~\cite{Jaramillo:2020tuu}. Moreover, within the hyperboloidal framework, these branch-cut mode can be used to reconstruct the late-time tail in the time domain using a suitable integral transformation~\cite{Ansorg:2016ztf}. However, if we look at the spectrum of the Vaidya BH (or rather the SR metric), we find a number of convergent modes present on the imaginary axis. These Purely Imaginary modes are again of the order $t \sim 10^{-5}$. We quantify these observations with the help of the results presented in Fig.~\ref{fig:schwarzschild_limit_scaling_lr_modes} where we study the behavior of the modes as we vary $t$.
In order to study the behavior of the light ring modes, we define the relative difference  $\delta \tilde \Omega^{\rm Sch}_{\rm LR}$ between the LR modes of the SR metric and the Schwarzschild BH as,
\begin{equation}
	\delta \tilde \Omega^{\rm Sch}_{\rm LR}(t)  = \abs{1 - \dfrac{\tilde \Omega^{\rm Sch}_{\rm LR}(t)}{\Omega(0)}},
\end{equation}
where $\Omega^{\rm Sch}_{\rm LR}$ denotes the LR modes calculated in the Schwarzschild limit and $\Omega(0)$ denotes the usual LR modes the Schwarzschild BH, normalized with respect its event horizon. From the top panel of Fig.~\ref{fig:schwarzschild_limit_scaling_lr_modes}, we note that as $t \to 0$,
\begin{equation}
	\delta \tilde \Omega^{\rm Sch}_{\rm LR}(t) \propto t, \label{eqn:lr_scaling_schw_limit}
\end{equation}
for the three lowest-lying LR modes, meaning that the LR modes smoothly go over to the usual Schwarzschild modes in the Schwarzschild limit. The purely imaginary Rindler modes do not have a Schwarzschild counterpart as we have already discussed, and from the bottom panel of Fig.\ref{fig:schwarzschild_limit_scaling_lr_modes} we see that
\begin{equation}
	\lim_{t \to 0}   |\tilde \Omega^{\rm Sch}_{\rm R}(t)|  = t \dfrac{n+3}{2}, \label{eq:rindler_numerical_limit}
\end{equation}
where $\Omega^{\rm Sch}_{\rm R}$ denotes the Rindler modes calculated in the Schwarzschild limit.
\begin{figure}[t]
	\centering
	\begin{minipage}{\columnwidth}
		\centering
		\includegraphics[width=\linewidth]{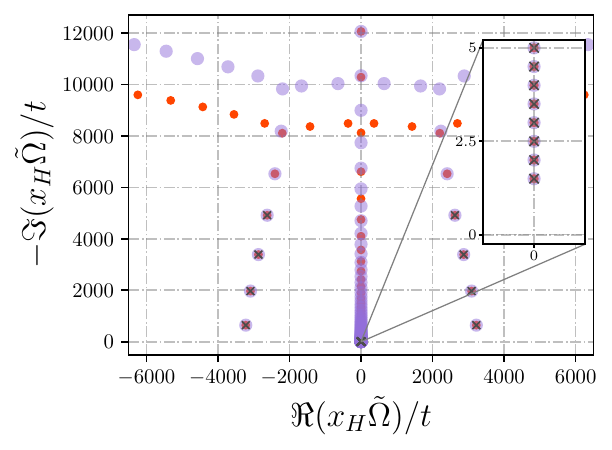}

	\end{minipage}

	\caption{\textbf{Static QNM Spectrum of the Vaidya BH in the Schwarzschild Limit} for scalar perturbations ($s=0$) with $\ell = 2$. We choose $t\equiv x_H/x_A = 0.0003$ and set $x_0 =x_H$, resulting in a very small accretion rate ($M' \sim 0.000074955$), and show the spectrum for $N = 200$ (filled orange circles) and $N= 250$ (translucent purple circles). The modes marked by crosses have a relative difference $\delta \tilde \Omega_n<10^{-8}$. In the \textbf{Inset}, we zoom into the region around $\tilde \Omega=0$ and find several convergent Rindler modes. Note that we have scaled the modes with $t$. Remarkably, the Rindler modes of the scalar spectrum (approximately) coincide with those found in the gravitational spectrum after we scale them by $t$. These results were obtained using mesh refinement.}
	\label{fig:schwarzschild_limit_spectrum_scalar_comparison}
\end{figure}
The above numerical result shows that the purely imaginary modes collapse to zero in the Schwarzschild limit, and it clearly matches with the analytical approximations given by Eqs.~\eqref{eq:PI_spectrum} and \eqref{eq:PI_spectrum_Heun_leading}, for $\ell =2$. The results are qualitatively similar for scalar and electromagnetic perturbations. To establish this fact, along with a demonstration of the robustness of our numerical results, in Fig.~\ref{fig:schwarzschild_limit_spectrum_scalar_comparison}, we show the $\ell = 2$ spectrum of scalar perturbations for $t = 3 \times10^{-4}$. Note that we have scaled the spectrum by $t$ and show the raw numerical spectrum for $N=200, 250$. We show this to reiterate that the raw spectrum contains spurious eigenvalues which form a horizontal branch. The numerical filter given by Eq.~\eqref{eq:numerical_filter} specifically eliminates these  non-convergent eigenvalues. It is also worth noting the distribution of the convergent Rindler modes shown in the inset and compare them with those in Fig.~\ref{fig:schwarzschild_limit_spectrum_vs_sch}. Clearly, both the scalar and gravitational Rindler modes satisfy Eq.~\eqref{eq:rindler_numerical_limit} showing that these modes do not strongly depend on the value of spin $s$ as $t \to 0$, a fact borne out by our analytical approximations as well.

\subsubsection{QNMs in the Rindler or Small BH  Limit}

We now want to examine the QNM spectrum in the limit $x_H \to0$ as $x_A$ is held fixed. Note that, in this limit, $t \to 0$ as well and we have shown in Sec.~\ref{sec:limitVaidyaHyperboloidal} that the limiting geometry is described by the Rindler metric. To study this limit, we use the radial coordinate,
\begin{equation}
	z \equiv \dfrac{x}{x_A} = 1 - \sigma(1-t).
\end{equation}

\begin{figure}[t]
	\centering
	\begin{minipage}{\columnwidth}
		\centering
		\includegraphics[width=\linewidth]{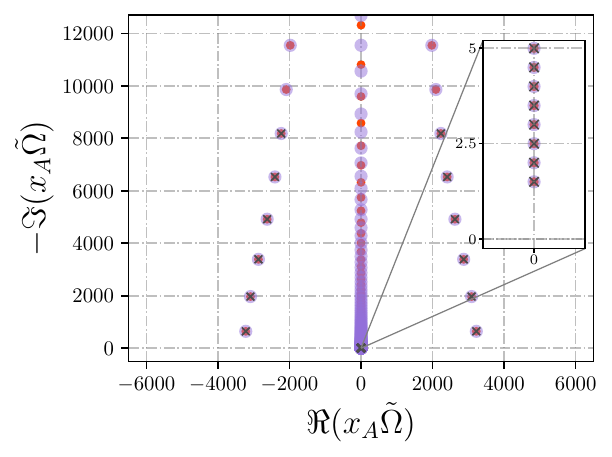}
	\end{minipage}
	\caption{\textbf{Static QNM Spectrum of the Vaidya BH in the Rindler Limit} for scalar perturbations ($s=0$) with $\ell = 2$. We choose $t\equiv x_H/x_A = 0.0003$ and set $x_0 =x_A$, resulting in a small accretion rate ($M' \sim 0.000074955$), and show the spectrum for $N = 200$ (filled orange circles) and $N= 250$ (translucent purple circles). The modes marked by crosses have a relative difference $\delta \tilde \Omega_n<10^{-8}$. In the \textbf{Inset}, we zoom into the region around $\tilde \Omega=0$ and find several convergent Rindler modes with $\mathfrak{I}(\tilde\Omega)<0$ of the order $\mathcal{O}(1)$. We find that the spectrum in the Rindler limit is related to that obtained in the Schwarzschild limit, $\tilde \Omega^{\mathrm{Rin}} \simeq \tilde \Omega^{\mathrm{Sch}}/t$. These results were obtained using mesh refinement.}
	\label{fig:rindler_limit_spectrum}
\end{figure}

\begin{figure*}[t]
	\centering
	\begin{minipage}{0.48\textwidth}
		\centering
		\includegraphics[width=\linewidth]{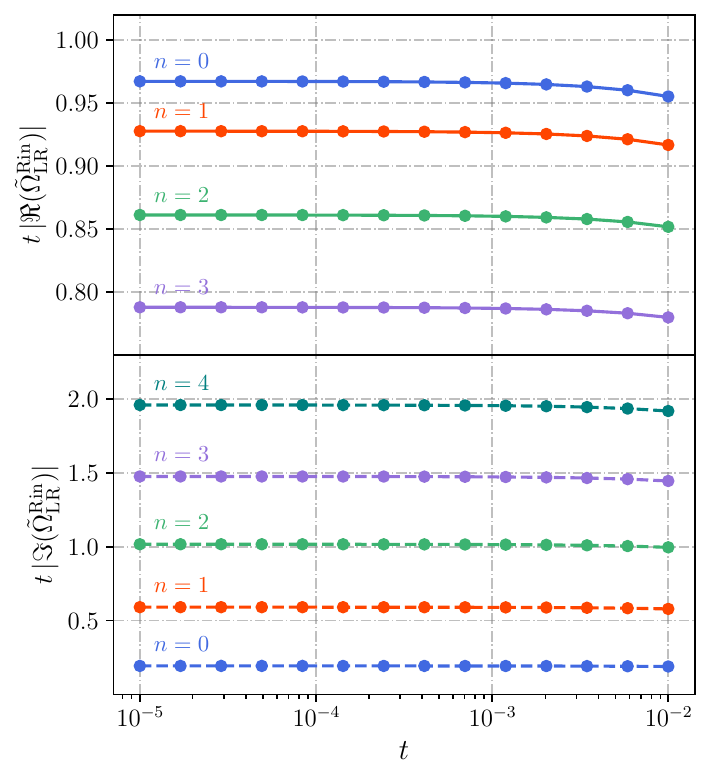}
		\label{fig:left_plot}
	\end{minipage}
	\hfill
	\begin{minipage}{0.48\textwidth}
		\centering
		\includegraphics[width=\linewidth]{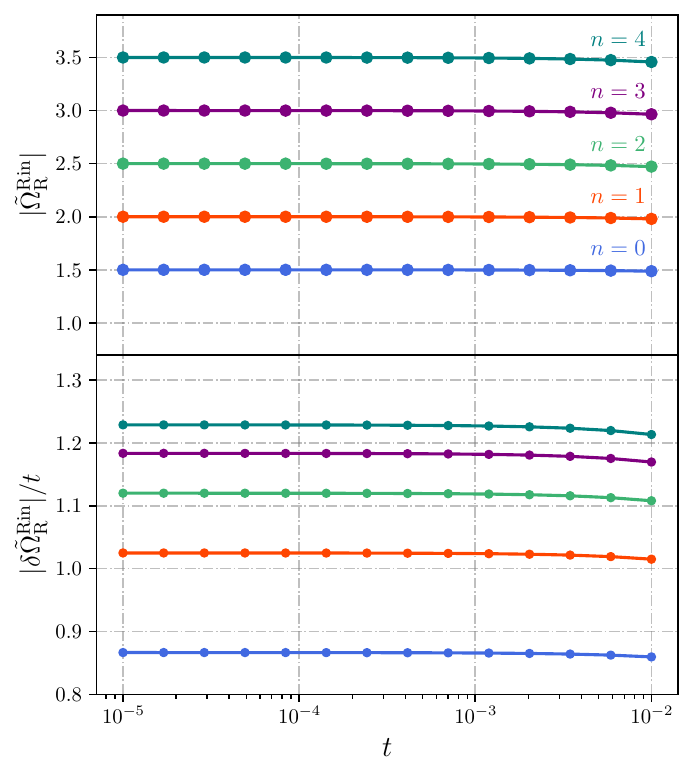}
		\label{fig:right_plot}
	\end{minipage}
	\caption{\textbf{Parametric behavior of QNMs in the Rindler Limit} for scalar perturbations ($s=0$) with $\ell =2$. In the \textbf{left panel}, we show the scaling of the Light Ring (LR) modes by plotting their real part $|\Re(\tilde \Omega^{\rm Rin}_{\rm LR})|$ (\textbf{top left}) and imaginary part $|\Im(\tilde \Omega^{\rm Rin}_{\rm LR})|$ (\textbf{bottom left}) as a function of $t$. We scale both real and imaginary parts by $t$, yielding nearly constant values in this limit, showing that they blow up as $\mathcal{O}(1/t)$ as $t \to 0$. The \textbf{right panel} focuses on the purely imaginary Rindler modes, showing their scaled absolute values $| \tilde \Omega^{\rm Rin}_{\rm R}|$ (\textbf{top right}) forming horizontal lines that approach the finite analytical limit $\Omega_A(0) = (n+\ell+1)/2$, and plotting the scaled relative difference $|\delta \tilde \Omega^{\rm Rin}_{\rm R}|/t$ (\textbf{bottom right}), which also becomes almost horizontal as $t \to 0$, demonstrating that the leading-order corrections vanish linearly with $t$ in this limit. The index $n$ denotes the overtone number for both LR and Rindler modes, with $n=0$ being the corresponding lowest-lying mode. The modes shown here are convergent with a relative difference $\delta \tilde \Omega_n < 10^{-8}$.}
	\label{fig:rindler_limit_rindler_modes}
\end{figure*}

Now, the acceleration horizon is fixed at $z_A=1$ and the event horizon varies as $z_H = t$. We therefore use the $z$ coordinate to probe this Rindler limit as the location of the event horizon now varies with $t$. The characteristic length scale is now set by $x_A$ and therefore the QNM spectrum in the Rindler limit $ \tilde \Omega^{\rm Rin}$ scale as
\begin{equation}
	\tilde\Omega^{\rm Rin} = x_A \tilde \Omega \overset{t \to0}{\sim} t^{-1} \tilde \Omega,  \label{eq:Rin_Lim_Normalization}
\end{equation}
where we used $x_A = (1+t)/t$ in writing the last approximation. Note that, using Eqs.~\eqref{eq:Sch_Lim_Normalization} and \eqref{eq:Rin_Lim_Normalization}, the spectrum in the Rindler limit can be theoretically related to the one obtained in the Schwarzschild limit as
\begin{equation}
	{\tilde \Omega^{\rm Rin}} =\dfrac{\tilde \Omega^{\rm Sch}}{t}. \label{eq:rin_sch_qnm_relation}
\end{equation}
We see that the above relationship indeed holds for our numerically computed spectra in Fig.~\ref{fig:rindler_limit_spectrum}, providing a strong consistency check for the numerical implementation. Here, we implement mesh refinement by clustering more points near the event horizon by setting $x_B = -1$ but take $\kappa = 1 - \log_{10}t$ \cite{Zhou:2025xta}.

In Fig.~\ref{fig:rindler_limit_spectrum}, we show the $\ell =2$ scalar QNM spectrum in the Rindler limit for $t=3 \times 10^{-4}$ for $N =200, 250$. A direct comparison with Fig.~\ref{fig:schwarzschild_limit_spectrum_scalar_comparison} shows that the two spectra are identical as far as the converged modes (indicated by black cross-marks) are concerned, thereby validating Eq.~\eqref{eq:rin_sch_qnm_relation}. In fact, the purely imaginary Rindler modes lying near the origin are of $\mathcal{O}(1)$ and our scaling violently blows up the LR modes. This indicates that a gauge choice adapted to the acceleration horizon is naturally better suited for studying the Rindler modes.

The blow-up of the LR modes in the Rindler limit becomes more evident in the left panel of Fig.~\ref{fig:rindler_limit_rindler_modes}: we see that in sharp contrast to the Schwarzschild limit (Eq.~\eqref{eqn:lr_scaling_schw_limit}), the LR modes now scale as $\mathcal{O}(1/t)$. This behavior is analogous to what is observed in the Schwarzschild-de Sitter case~\cite{Zhou:2025xta} and our study shows that this behavior may be a generic feature of spacetimes with multiple horizons. In~\cite{Zhou:2025xta} the authors reinterpreted this behavior in terms of the instability of the QNM spectrum: the Light Ring modes of the Schwarzschild BH are \emph{stable} under small deformations to the background since the shift in the spectrum is proportional to the strength of the deformation itself. While the deformation in the SdS case is introduced by the cosmological constant, the deformations in the present scenario can be ascribed to the BH environment since we are dealing with an accreting/radiating BH. On the other hand, in the Rindler limit, the result can be interpreted as an example of spectral instability: the introduction of a small BH in an accelerating spacetime modifies the causal structure of the spacetime at $x=0$, leading to the appearance of a new family of modes, in addition to the purely imaginary modes. This latter interpretation conveniently follows the SdS analogy but one must be cautious about the interpretation of these Rindler modes. While one can easily establish the existence of QNMs of pure de Sitter spacetime, the same is not true for a purely accelerating spacetime. To the best of our knowledge, the existence and interpretation of QNMs of purely accelerating spacetimes is an open question. However, from the top right panel of Fig.~\ref{fig:rindler_limit_rindler_modes}, we see that in the Rindler limit, the Rindler modes tend to almost finite values with subleading corrections depending on $t$, viz.,
\begin{equation}
	|\tilde \Omega^{\rm Rin}_{\rm R}| \simeq \dfrac{n+\ell+1}{2} +\mathcal{O}(t^\gamma),
\end{equation}
where $\ell=2$ and $\gamma>0$. To see the first leading $t$ dependence explicitly, we turn to the relative difference and see from the bottom right panel of Fig.~\ref{fig:rindler_limit_rindler_modes},
\begin{equation}
	\delta \tilde \Omega^{\rm Rin}_{\rm R}(t)  = \abs{1 - \dfrac{\tilde \Omega^{\rm Rin}_{\rm R}(t)}{\Omega_A(0)}} \overset{t \to 0} \sim t,
\end{equation}
where $\Omega_A(0)=({n+\ell+1})/{2}$. Here $\Omega_A(0)$ is interpreted as the limit, $\lim_{t \to0} \tilde\Omega/t$, of the analytical approximations derived here earlier (Eqs.~\eqref{eq:PI_spectrum} and \eqref{eq:PI_spectrum_Heun_leading}). In~\cite{Destounis:2020pjk,Xiong:2023usm, Chen:2024rov} the authors obtained the same expression for QNMs of accelerating BHs after taking the Minkowski limit (also see~\cite{Zhou:2024axx}). We therefore see that the \emph{Rindler spectrum} is stable under small deformations to the Rindler geometry due to the appearance of a small BH near $x=0$. Again, we draw inspiration from the SdS analogy while interpreting this behavior~\cite{Zhou:2025xta}.

In Appendix \ref{appsec:numerical_nariai_limit_QNMs}, we discuss the nearly-extremal limit, that is, the case when $\varepsilon \equiv x_A-x_H \to 0$. We show that the spectrum begins to resemble that of the P\"{o}schl-Teller potential~\cite{Poschl:1933zz}, just like the spectrum of the SdS BH in the Nariai limit \cite{Cardoso:2003sw,Sarkar:2023rhp}. This brings us to the conclusion of our analysis of the numerical QNM spectrum across various limiting regimes. Overall, the spectrum exhibits pronounced structural similarities to that of the SdS geometry. Furthermore, exploring extreme values of the parameter $t$ and recovering the expected scaling relations has served as a useful validation of the robustness of our numerical implementation.

\begin{figure}[t]
	\centering
	\begin{minipage}{\columnwidth}
		\centering
		\includegraphics[width=0.95\linewidth]{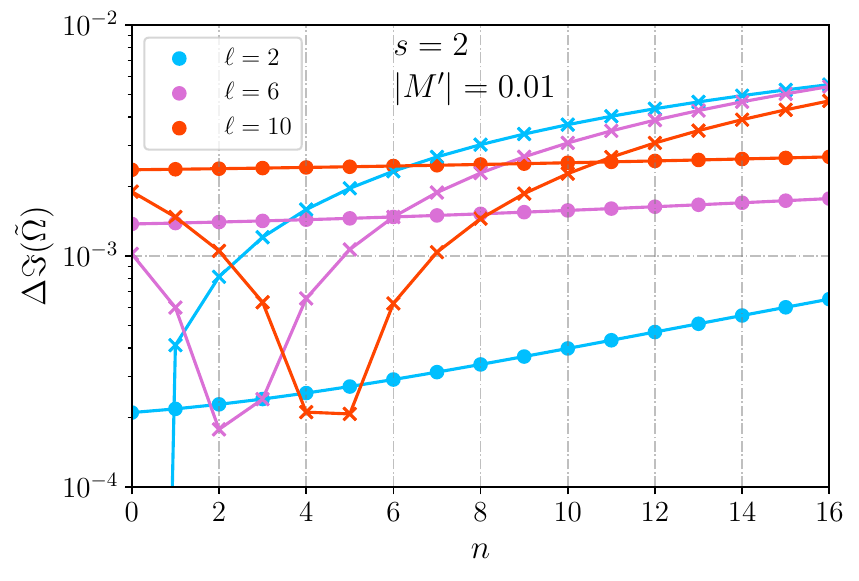}
	\end{minipage}
	\caption{\textbf{Frequency residual between numerical and analytic predictions} in the hypergeometric equation (cross points) and Heun equation (round points) approaches. The gravitational $|M'| = 0.01$ case is presented for multipole numbers $\ell = 2$ (orange), $\ell = 6$ (purple), and $\ell = 10$ (blue). Notice that we choose to connect the data points with full lines for representation purpose, despite the discreteness of QNMs.}
	\label{fig:RindlerLimit_Comparison}
\end{figure}
We conclude this subsection with a comparison between the QNM frequencies estimated analytically in Sec.~\ref{sec:QNMviaHeun} and the numerical frequencies computed here for different values of the multipole index $\ell$.
In Fig.~\ref{fig:RindlerLimit_Comparison}, we show the difference between the numerical prediction for the imaginary part of the mode frequency, and the analytic prediction obtained with the hypergeometric equation (cross points) and Heun equation approach, up to fifth order (round points). We recall that, as previously observed, the prediction obtained with the hypergeometric equation approach is equivalent to the first-order prediction given by the Heun equation. We only present the case in which $|M'|=0.01$, although different (small) values of the mass evolution rate produce qualitatively similar results. Different values of the multipole number are indicated with different colors. While in the $\ell = 2$ case, the fifth-order Heun prediction performs much better than the first-order/hypergeometric one, for $n \geq 1$, for higher multipoles, we observe that this only holds true for very large overtone numbers. While counterintuitive, this result can be explained as an accidental crossing of the $\tilde\Omega(n)$ curves predicted numerically and analytically~\footnote{Since the residual is defined as the absolute value of the difference between the numerical and analytic predictions, a sign-changing crossing of the corresponding signed difference curve is reflected as a sharp local minimum. Without the absolute value, the difference curve would instead pass continuously through zero and become negative beyond the intersection point, indicating that the agreement actually deteriorates for higher overtones, consistent with the trend observed for the fifth-order Heun prediction.}. This phenomenon does not happen, on the other hand, when we compare the numerical prediction with the fifth-order Heun equation one. We finally mention that all analytic predictions tend to perform better for lower multipoles.

\subsection{Mode Accumulation and Nonuniqueness of the Schwarzschild Branch-cut}\label{sec:mode_accumulation}
In \cite{Zhou:2025xta}, the authors sought to establish a connection between the Schwarzschild branch-cut and the PI de Sitter modes of the SdS BH by counting the QNMs contained in a small strip of the imaginary axis. This approach was inspired by the Weyl law for BH QNMs \cite{Jaramillo:2022zvf} and provided support in favour of recent analytical studies \cite{Arnaudo:2025kit}. We adopt a similar numerical approach to establish an analogous connection between the PI modes of the LMV spacetime and the Schwarzschild branch-cut, and discuss its potential impact on the BH spectroscopy program.

We begin by defining the following expression for the density $d$ of PI QNM along a finite strip $\delta \tilde \Omega$ of the imaginary axis,
\begin{equation}
	d(t) = \dfrac{\mathfrak{M}}{\delta \tilde \Omega}, \label{eq:qnm_density}
\end{equation}
where $\mathfrak{M}$ is a positive integer, representing the number of modes inside the strip $\delta \tilde \Omega$~\footnote{Note that for axial gravitational perturbations, $\tilde \Omega = \Omega +2i M'$, and all the physical modes of all spins of the LMV spacetime acquire a constant conformal shift given by Eq.~\eqref{eq:conformal_shift}, viz., $\Omega_*=\Omega+2i\chi M'$. Since the shift is constant for a particular value of $t$ or $M'$, the density of purely imaginary QNMs remain unaffected ($\delta \tilde \Omega = \delta \Omega_*$).}. We wish to calculate the density $d$ as a function of the BH parameter $t$ in the Schwarzschild limit ($t \to 0$ with $x_H$ kept fixed). In other words, we wish to find out how the distribution of the PI modes changes as the acceleration horizon $x_A$ merges with the future null infinity  $\mathscr{I}^+$. We adopt the following two strategies.

\begin{figure}[t]
	\centering

	\begin{minipage}{\columnwidth}
		\centering
		\includegraphics[width=\linewidth]{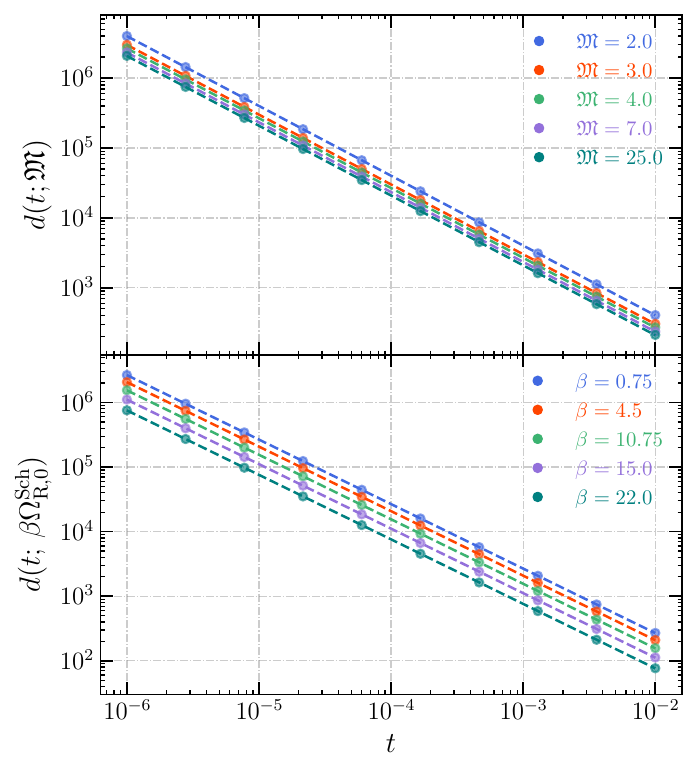}
	\end{minipage}
	\caption{\textbf{Accumulation of purely imaginary QNMs $(s=\ell=2)$ into the Schwarzschild branch-cut.} In the \textbf{top panel}, we plot the Purely Imaginary QNM mode density $d$ as a function of the parameter $t$ for a fixed number of modes $\mathfrak{M}$. In the \textbf{bottom panel}, we plot the density $d$ as a function of $t$ for fixed frequency strip widths, parameterized by $\beta$. Both panels show that the density diverges according to the power law $d \propto 1/t$ as $t \to 0$, illustrating the infinite accumulation of discrete modes at the origin in the Schwarzschild limit.  The modes shown here are convergent with a relative  difference $\delta \tilde \Omega_n < 10^{-8}$.}

	\label{fig:mode_accumulation}
\end{figure}

We first fix the number of modes $\mathfrak{M}$, and then, for a fixed value of $t$, we compute the width of the strip occupied by these modes. We start counting from the lowest-lying PI mode for simplicity, and get
\begin{equation}
	\delta \tilde \Omega =|\tilde \Omega^{\rm Sch}_{{\rm R},\mathfrak{M}-1} - \tilde \Omega^{\rm Sch}_{\mathrm{R},0} |.
\end{equation}
We calculate the density using Eq.~\eqref{eq:qnm_density} for different values of $t$ and then repeat the process for different values of $\mathfrak{M}$. We reuse the data generated for Fig.~\ref{fig:schwarzschild_limit_scaling_lr_modes} for this calculation, noting that we were able to recover at least $25$ PI modes for all the values of $t \in \left[10^{-6}, 10^{-2}\right]$ under consideration. The results are shown in the top panel of Fig.~\ref{fig:mode_accumulation}.

The second approach involves fixing $\delta \tilde \Omega$ instead of $\mathfrak{M}$. Using a positive real number $\beta$, we fix the width of the strip as
\begin{equation}
	\delta \tilde \Omega = |\tilde \Omega_{\rm max} -  \tilde \Omega^{\rm Sch}_{\mathrm{R},0}|,
\end{equation}
where
\begin{equation}
	\tilde \Omega_{\rm max} =\tilde \Omega^{\rm Sch}_{\mathrm{R},0}+\beta \tilde \Omega^{\rm Sch}_{\mathrm{R},0}.
\end{equation}
We again measure the width from the lowest-lying imaginary mode. We then count the number of modes $\mathfrak{M}$ lying inside the above strip to evaluate the density of a fixed $t$. As before, we obtain the density for different values of $t$ for a fixed $\beta$, and then repeat the process for different widths. The results are shown in the bottom panel of Fig.~\ref{fig:mode_accumulation}.

We see that the density diverges as $t \to 0$. Fitting a general power-law $d = A t^{k}$ to the simulated data\footnote{We estimate $A$ and $k$ by the method of least squares on the transformed relation $\ln d = \ln A + k \ln t$ \cite{bevington2003data,pine2019introduction}.}, we find that $k \simeq-1$ (see Table \ref{tab:power_law_fits_with_errors}) which shows that the density diverges as $d \propto 1/t$ when $t \to 0$.
Furthermore, we have seen in Fig. \ref{fig:schwarzschild_limit_scaling_lr_modes} that the PI modes approach $\tilde \Omega =0$ linearly, indicating that the end points of the strip $\delta \tilde \Omega$ collapse to the origin. Therefore, in the Schwarzschild limit, there is an infinite accumulation of PI modes at $\tilde \Omega =0$. This accumulation of discrete QNMs near the origin provides numerical evidence that the Schwarzschild branch-cut can emerge from the discrete PI spectrum of the LMV spacetime.

\begin{table}[t]
	\centering
	\caption{\textbf{Calculated parameters for the power-law describing the purely imaginary QNM density} ($d = A \cdot t^k$) for $s=\ell=2$, including standard errors $\sigma_k, \sigma_A$ in $k$ and $A$ respectively, and the coefficient of determination $R^2$ (obtained from the correlation coefficient $R$). The \textbf{upper section} shows the fits calculated for a varying, fixed number of modes $\mathfrak{M}$, while the \textbf{lower section} shows fits for varying frequency intervals parameterized by $\beta$. Across all values tested, the exponent is consistently $k \simeq -1$, confirming the inverse linear divergence ($d \propto 1/t$) of the mode density as $t \to 0$.}
	\vspace{0.2cm}

	\footnotesize
	\renewcommand{\arraystretch}{1.2}
	\setlength{\tabcolsep}{4.6pt}

	\begin{tabular}{lccc}
		\toprule
		\textbf{$\mathfrak{M}$} & \textbf{$k \pm \sigma_k$} & \textbf{$A \pm \sigma_A$} & \textbf{$R^2$} \\
		\midrule
		$2.0$                   & $-0.9989$ $\pm$ $0.0004$  & $4.0497$ $\pm$ $0.0148$   & $1.0000$       \\
		$3.0$                   & $-0.9989$ $\pm$ $0.0004$  & $3.0372$ $\pm$ $0.0111$   & $1.0000$       \\
		$4.0$                   & $-0.9989$ $\pm$ $0.0004$  & $2.6997$ $\pm$ $0.0099$   & $1.0000$       \\
		$7.0$                   & $-0.9989$ $\pm$ $0.0004$  & $2.3622$ $\pm$ $0.0086$   & $1.0000$       \\
		$25.0$                  & $-0.9989$ $\pm$ $0.0004$  & $2.1088$ $\pm$ $0.0075$   & $1.0000$       \\
		\midrule
		\addlinespace[0.5em]
		\textbf{$\beta$}        & \textbf{$k \pm \sigma_k$} & \textbf{$A \pm \sigma_A$} & \textbf{$R^2$} \\
		\midrule
		$0.75$                  & $-0.9993$ $\pm$ $0.0003$  & $2.6887$ $\pm$ $0.0066$   & $1.0000$       \\
		$4.50$                  & $-0.9993$ $\pm$ $0.0003$  & $2.0912$ $\pm$ $0.0051$   & $1.0000$       \\
		$10.75$                 & $-0.9993$ $\pm$ $0.0003$  & $1.5632$ $\pm$ $0.0038$   & $1.0000$       \\
		$15.00$                 & $-0.9993$ $\pm$ $0.0003$  & $1.1203$ $\pm$ $0.0027$   & $1.0000$       \\
		$22.00$                 & $-0.9993$ $\pm$ $0.0003$  & $0.7638$ $\pm$ $0.0019$   & $1.0000$       \\
		\bottomrule
	\end{tabular}
	\label{tab:power_law_fits_with_errors}
\end{table}

The above behavior is similar to that found in SdS BHs \cite{Zhou:2025xta}. In SdS BHs, PI modes are associated with the presence of a cosmological constant ($\Lambda$) and the cosmological horizon.
However, in the Vaidya spacetime they arise due to the change in the BH mass function encoded by $M'$. In both cases, these small deformations of the Schwarzschild geometry act as a regulator, in the sense that for small but nonzero $\Lambda$ or $M'$, the non-converging modes representing the branch-cut of the Schwarzschild problem are replaced by a discrete set of convergent PI modes. We then recover the Schwarzschild branch-cut in the limit where these modes accumulate on the imaginary axis at the origin as $\Lambda \to 0$ or $M' \to 0$. It is worth noting that the axial gravitational de Sitter modes of the SdS BHs can be approximated as \cite{Lopez-Ortega:2006aal,Lopez-Ortega:2012xvr,Cardoso:2017soq}
\begin{equation}
	\Omega_{\mathrm {dS}} \simeq - i \kappa_c(n+\ell+1),
\end{equation}
for gravitational perturbations, $\kappa_c$ being the surface gravity associated with the cosmological horizon\footnote{An analogous formula exists for accelerated BHs with  $\kappa_c$ replaced by the surface gravity of the accelerated horizon but with the same level-spacing~\cite{Destounis:2020pjk}.}. This bears a striking resemblance to the approximation obtained in Eq.~\eqref{eq:PI_spectrum} when we notice that the surface gravity $\kappa_A$ associated with the horizon $x_A$ of the SR metric is approximately $\kappa_A \to 2 |M'|$ as $t \to 0$. This shows that the late-time Price tail can be represented as the limit of a discrete spectra in more than one \emph{inequivalent or nonunique} way. In other words, we may obtain the same analytic structure of the Schwarzschild branch-cut from different discrete sets of modes, regardless of how we deform the Schwarzschild problem. From the point of view of BH spectroscopy, this might suggest that the late-time tail does not admit a unique modal interpretation. We will come back to this discussion at the end of the next section.

\section{Time-domain approach}
\label{sec:time_domain}

\begin{figure*}
	\centering
	\includegraphics[width=0.7\textwidth]{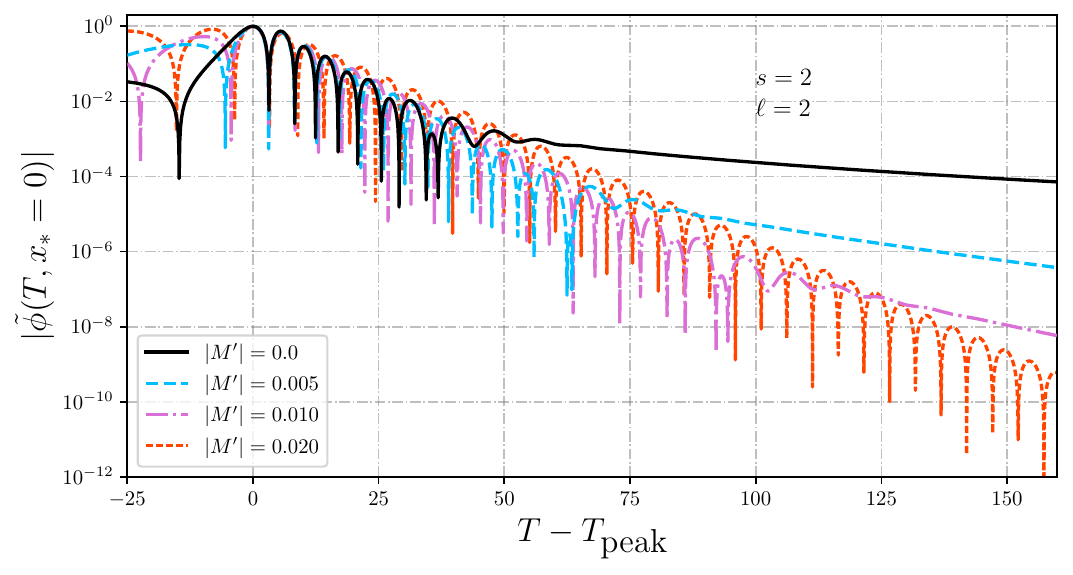}
	\caption{\textbf{Time-domain signal for gravitational axial quadrupole perturbations:} The result is shown for different values of the mass-evolution rate. The decay-time of the dominant QNM increases as $|M'|$ increases. It can also be observed that for $|M'|>0.01$ no late-time tail appears.}
	\label{fig:Vaidya_vs_Schwarzschild}
\end{figure*}

So far, we restricted our analysis to the frequency domain and obtained robust predictions for PI mode frequencies. However, at this stage we are still not able to tell whether the aforementioned PI modes are actually excited and visible in the time domain. The degree of excitation depends in general on the initial data in the wave equation Eq.~\eqref{eq:wave-equation}.  But even assuming that these modes are turned on in a given configuration, it is not guaranteed that they show up in the time-domain signal, because they could in principle be subdominant with respect to the LR modes at any time. This question can be addressed with a proper time-domain analysis \footnote{A possible frequency-domain  method for assessing the modes amplitude hierarchy involves calculating the QNM amplitudes~\cite{Ansorg:2016ztf, Ammon:2016fru} using the hyperboloidal foliation developed here. We hope to report on this complementary analysis in the near future.}, which will be the focus of this last section. Nevertheless, let us first make a few preliminary remarks. Assume now that all the PI modes are excited, but with a smaller amplitude with respect to the dominant LR mode with $(\ell,n)=(2,0)$. As shown in ~\cite{Capuano:2024qhv}, the damping of the LR modes can be approximated in the eikonal limit, and for small $|M'|$, with the analytic expression
\begin{equation}
	\Im(\tilde\Omega) \simeq -\frac{2}{\sqrt{3}}\left(n+\frac{1}{2}\right)\left(1-4 |M'|\right)\,.
	\label{eq:eikonal_QNMs}
\end{equation}
As it is evident, LR modes tend to the finite Schwarzschild frequencies in the $M'\to 0$ limit, while the PI ones pile up to $\Im(\tilde\Omega)\to 0$. Moreover, for small $M'$, the LR QNMs exhibit the positive linear scaling $\sim 8(n+1/2)/\sqrt{3} |M'|$, in contrast with the PI ones, behaving as $\sim -2(n+\ell +1)|M'|$. Hence, we expect that below a threshold value of the mass evolution rate, the PI modes are less damped, and thus more long-lived. Comparing Eq.~\eqref{eq:PI_spectrum} with Eq.~\eqref{eq:eikonal_QNMs}, for the fundamental $(\ell,n)=(2,0)$ mode, one can observe that this happens for $|M'|\lesssim 0.01$. This condition is generally reduced to smaller values of the mass evolution rate for higher $\ell,n$. Thus, for $\ell = 2$, and $|M'|\sim 0.01$, we expect the fundamental PI mode to become visible in the time-domain signal after an initial \emph{ringing-dominated stage}, producing a sort of exponential tail. For smaller values of $ |M'|$, higher PI overtones are supposed to appear between this stage and the final tail from the fundamental one.
This picture is not extremely rigorous, since it relies on a set of approximations, particularly the extrapolation of the eikonal regime prediction to small $\ell$. However, it gives a qualitative picture of the time-domain ringdown signal that one should expect on a LMV background.

We will now check this argument against a numerical time-domain simulation.
We evolve the perturbation equation Eq.~\eqref{eq:wave-equation} using the method of lines~\cite{Guzman2023}. The spatial derivatives are discretized with a fourth-order finite difference scheme on a uniform grid in $x_*$, and the resulting equation is integrated in time with a fourth-order Runge-Kutta method~\cite{Press:2007ipz}. The test of fourth-order convergence of the code is presented in Appendix~\ref{appendix:convergence}.
At the two ends of the computational domain we impose absorbing QNM boundary conditions, namely
\begin{equation}
	\begin{split}
		&(\partial_T-\partial_{x_*})\tilde\phi=0\,\,
		\text{for } x_*\to -\infty\,,\\
		&(\partial_T+\partial_{x_*})\tilde\phi=0
		\,\,\text{for } x_*\to +\infty.
	\end{split}
\end{equation}
These conditions (also called outgoing wave boundary conditions~\cite{Guzman2023}) are implemented numerically through one-sided fourth-order finite-difference stencils, consistently with the fourth-order bulk discretization.
However, the ends of the computational domain have been chosen far enough away from the region of interest that the choice of the boundary conditions is practically irrelevant.

\begin{figure*}
	\centering
	\includegraphics[width=0.7\textwidth]{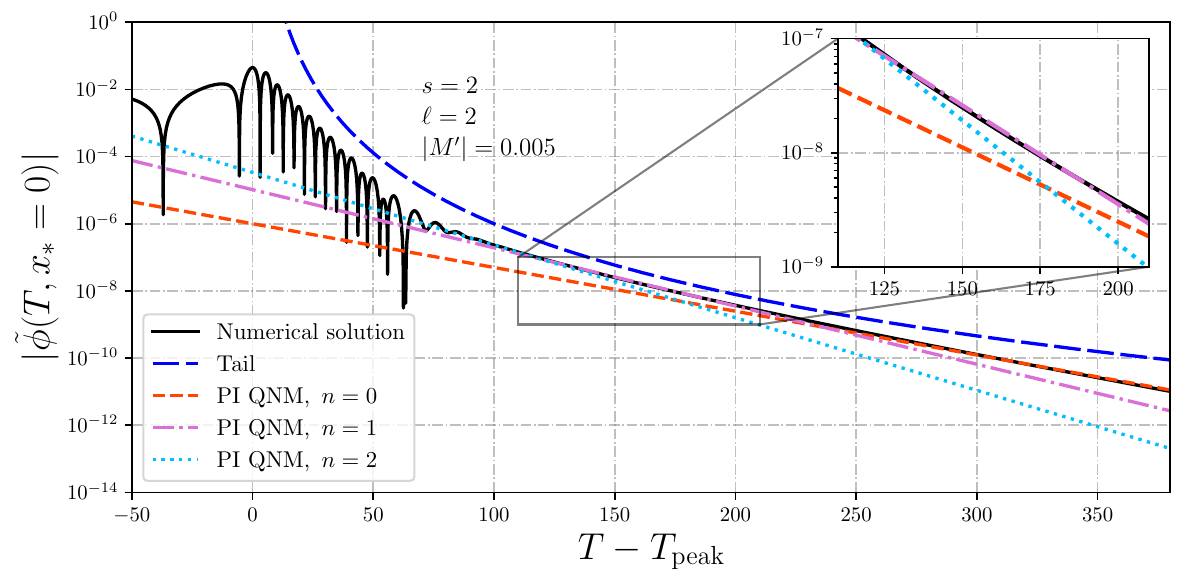}
	\caption{\textbf{Late-time signal for small mass-evolution rate:} The numerical ringdown signal for $s = \ell = 2$, and $|M'|=0.005$ is compared to the frequency-domain predictions for PI QNMs. The fundamental mode ($n = 0$), as well as the overtone modes $n = 1,2$ can be observed at different stages, after the LR mode-dominated part of the signal.}
	\label{fig:TD_signal_Vaidya}
\end{figure*}

Finally, the initial data consist of a Gaussian pulse centered away from the potential barrier, i.e.
\begin{equation}
	\tilde\phi(T=0,x_*)=A_0\exp\left(-\frac{(x_*-x_{*,0})^2}{2\sigma_0^2}\right)\,.
\end{equation}
We check the robustness of the results against different choices of $A_0$, $x_{*,0}$, and $\sigma_0$. In this section, we arbitrarily chose to present results for $A_0=0.4$, $x_{*,0}=5$, $\sigma_0=1$. The aforementioned numerical scheme has been implemented in \texttt{Python}. In Fig.~\ref{fig:Vaidya_vs_Schwarzschild}, the numerical solution extracted at $x_*=0$ is presented for $\ell = 2$ gravitational perturbations, for different values of the mass-evolution rate, down to the Schwarzschild case $|M'|=0$ (full black line). As it is clear from the picture, the first part of the signal right after the peak is characterized by the standard QNM profile. The damping of the QNMs decreases as $|M'|$ increases, which is consistent with the frequency-domain analysis of~\cite{Capuano:2024qhv}, and with the approximate relation reported in Eq.~\eqref{eq:eikonal_QNMs}. On the other hand, one can notice that the late-time tail starts dominating over LR modes at earlier stages as $|M'|$ is reduced. In contrast, the signal corresponding to $|M'| = 0.02$ does not exhibit any tail and is dominated by the loudest LR mode at any time. This is consistent with the discussion at the beginning of the present section, in which we predicted from the frequency-domain computation that an exponential tail given by PI QNMs is generally present for a mass-evolution rate below a certain threshold, given by $|M'|\simeq 0.01$. Moreover, as pointed out in Sec~\ref{sec:mode_accumulation}, in the $|M'|\to 0$ limit, PI modes accumulate on the imaginary axis and produce the characteristic branch-cut that is present in the Schwarzschild space-time~\cite{Leaver:1986gd}. In the time-domain, the branch-cut produces a power-law contribution over time, scaling as $T^{-(2\ell+3)}$~\cite{Price_tails}. This effect is generally known as Price tail. Let us look more closely into the small $|M'|$ limit.
In Fig.~\ref{fig:TD_signal_Vaidya}, the numerical result (full black line) for a gravitational $\ell = 2$ perturbation, and the same initial data as the ones employed previously, for $|M'|=0.005$ is shown. The orange, pink, and light-blue lines represent the predicted slopes for PI QNMs, for $n=0,1,2$ respectively. As it is evident, these modes dominate at different stages of the late-time ringdown signal. In particular, this result suggests that higher overtones are excited with increasingly higher amplitudes, in such a way that they dominate in finite time windows after the ringing stage, until the fundamental $n = 0$ mode shows up. The resulting curve from the aforementioned mode hierarchy approximates with higher accuracy the Schwarzschild power-law tail (dashed blue line), as the mass-evolution rate is reduced.

We conclude this section, stressing the fact that the numerical solutions  presented here do not represent the full physical signal of a ringing LMV BH. In particular, as for the LR modes, also the PI mode frequencies should account for the time-dependent scaling of Eq.~\eqref{eq:time_dependent_omega}, which affects the tail shape. Schematically the late-time signal in physical coordinates will behave as
\begin{equation}
	\tilde\phi(w,r) \propto \exp\left(-\Im(\Omega^{\rm dom}_*) \frac{w}{2 M(w)}\right)\,,
\end{equation}
where $\Omega^{\rm dom}_*$ is the static frequency of the dominant Rindler mode. This peculiar time-domain feature makes the case of an evolving BH mass, with constant rate $|M'|$, different, in principle, from the SdS and the accelerating BH \footnote{A similar mechanism is expected to arise in accelerating BH spacetimes (e.g. C-metrics~\cite{Destounis:2020pjk}), or more generally in any setup admitting an additional outer horizon.} cases, in which an exponential tail dominates the late-time ringdown signal. However, astrophysically relevant scenarios, in which the BH mass evolves, are not expected to be fully captured in our mathematical framework, which strongly relies on the $|M'| = \text{const.}$ assumption.
We leave a more detailed analysis in the time domain, also for general and more physical mass evolution rates, to a future study.

\section{Conclusions}
\label{sec:conclusions}

Understanding the ringdown phase of binary BH coalescence is a central goal of GW physics, as it provides a direct probe of the underlying spacetime geometry and the nature of compact objects and their environment. While most studies model the post-merger remnant as a stationary BH, realistic astrophysical scenarios are generally dynamical, and deviations from stationarity may leave observable imprints on the QNM spectrum. In this work, we have explored this scenario, in the context of a dynamical BH spacetime described by a linear-mass Vaidya metric, building on the formalism developed in~\cite{Capuano:2024qhv}.

On the frequency-domain side, we first

showed that the wave equation can be mapped to a Heun form with four regular singular points. Exploiting recent results on connection coefficients, we derived a quantization condition for the QNM frequencies, analogous to that known for the SdS geometry. In the regime of weak mass accretion/radiation rate, we obtained explicit expressions as a series expansion in $|M'|$.

This result was independently confirmed through a numerical computation using hyperboloidal slices, based on a spectral method, showing excellent agreement with the analytical predictions. In particular, we verified the existence and robustness of the PI branch across a range of parameters, and for different kinds of perturbing (scalar, electromagnetic and axial gravitational) fields.

A key outcome of our analysis is that, in contrast to stationary BHs, a certain class of dynamical spacetimes can support QNM families that are not directly tied to null geodesics at the LR. This is consistent with analogous findings in the context of SdS~\cite{Zhidenko:2003wq,Konoplya:2022xid,Cardoso:2017soq,Arnaudo:2025kit}, and accelerating~\cite{Destounis:2020pjk,Chen:2024rov,Lei:2023mqx,Mi:2025fbt} BH spacetimes.

Finally, we evolved the perturbation equation numerically in the time-domain using a fourth-order method-of-lines scheme. Our simulations show that PI modes are generally visible in the late-time signal in the form of exponential tails. Furthermore, we showed that in the Schwarzschild limit, the PI modes accumulate near the origin of the imaginary axis in the complex frequency plane, reconstructing the branch-cut of the Schwarzschild Green function. In this sense, the exponential tails observed in the dynamical case can be interpreted as the discrete counterpart of the continuous spectrum responsible for the late-time power-law decay in stationary spacetimes.  Our findings also point towards a potentially nonunique modal interpretation of the Schwarzschild branch-cut, an observation that requires further investigation.

From a phenomenological perspective, the presence of non-LR modes and exponential tails may impact BH spectroscopy, in scenarios where the remnant object is not well described by a stationary background. This could be relevant, for instance, in highly dynamical environments or in situations where accretion or radiation effects are non-negligible. In particular, while a dynamical evolution of the BH mass is expected to occur in the early stages of BH ringdown and may affect even the early stages right after merger~\cite{Redondo-Yuste:2023ipg}, the excitation of PI modes could potentially affect the late-time signal as well.

Future work could extend this analysis to more general dynamical spacetimes, as well as investigate the detectability of these effects in GW observations.

\bigskip
\section*{Acknowledgments}

We warmly thank Vitor Cardoso for valuable discussion and feedback on a preliminary version of the draft. We also acknowledge useful discussions with David Hilditch, Srijit Bhattacharjee, Sreelakshmi M. D., Sumanta Chakraborty,  Matteo Della Rocca, Alessandro Tanzini and Giulio Bonelli. S.S. acknowledges membership in the Hyperboloidal Research Network~\cite{HyperboloidalNetwork} and thanks the collaboration for its support and for providing opportunities for discussion.
S.S. and E.B. were supported in part by the International Centre for Theoretical Sciences (ICTS), Bengaluru, India, for participation in the program \emph{The Future of Gravitational-Wave Astronomy 2025} (code: ICTS/FGWA2025/10) \cite{ICTS_FGWA_2025}; this work arose in part from discussions at the program and they also thank the members of the Astrophysical Relativity Group
and the staff at ICTS for their warm hospitality during the program.
The work by L.C. has been partially supported by the MUR FIS2 Advanced Grant ET-NOW (CUP: B53C25001080001) and by the INFN TEONGRAV initiative.
The research of G.P.-V. is partly supported by the INFN Iniziativa Specifica ST\&FI.
S.S. and D.K. acknowledge funding from the Institutions of Eminence initiative of the Government of India through the Centre for Industrial Consultancy and Sponsored Research (IC\&SR) of the Indian Institute of Technology Madras for the project titled Centre for Strings, Gravitation and Cosmology (Project No.: SB22231259PHE-TWO008479).
E.B. and T.L. acknowledge support from the European Union’s Horizon ERC Synergy Grant ``Making Sense of the Unexpected in the Gravitational-Wave Sky'' (Grant No. GWSky-101167314)
A.K  thanks the Fundação para a Ciência e Tecnologia (FCT), Portugal, for the financial support to the Center for Astrophysics and Gravitation (CENTRA/IST/ULisboa) through grant No. UID/PRR/00099/2025 and grant No. UID/00099/2025, as well as to the FCT project ``Gravitational waves as a new probe of fundamental physics and astrophysics'' grant agreement 2023.07357.CEECIND/CP2830/CT0003.

\appendix

\section{QNMs in the Nearly-Extremal Limit}
\label{appendix:next_limit}

In this appendix, we collect results related to the nearly-extremal limit of the SR metric, namely the regime in which $|M'| \to 1/16$ while $M_0 \neq 0$. In this case, the black hole possesses two distinct horizons that are arbitrarily close to each other. We show that the limiting geometry approaches a Nariai-like solution, while the QNM spectrum matches that of the P\"{o}schl-Teller potential, further strengthening the analogy with the SdS spacetime. Charged and/or rotating black holes generally possess an inner horizon in addition to the event horizon. Likewise, a charged LMV metric also contains an inner horizon. In much of the literature, the term \emph{nearly-extremal limit} typically refers to the situation in which the inner and event horizons coincide. In the present context, however, the nearly-extremal limit instead refers to the regime where the event horizon and the acceleration horizon approach each other. For de Sitter black holes, this is commonly known as the Nariai limit.

\subsection{Nearly-extremal limit of the SR metric}
\label{appsec:nearly_ext_limit_of_metric}
The nearly extremal limit corresponds to the case when the two horizons are very close to each other, that is, when $|M'| \to 1/16$ but $M_0 \neq 0$. So let us introduce a parameter $\varepsilon = x_A - x_H$ such that $\varepsilon \to 0 $ as $|M'| \to 1/16$. Then in the nearly extremal limit, we can rewrite Eq.~\eqref{eq:vaidya_metric_static} as
\begin{equation}
	\d \tilde s^2 = 8 \left[ -\sigma(1-\sigma) \d\tilde{T}^2 + \frac{\d\sigma^2}{\sigma(1-\sigma)} \right] + 4 \d\Omega^2_{S{^2}} \label{eqn:near_extremal_metric}
\end{equation}

In writing the above line element, we have used the fact that $f(x(\sigma)) \sim\varepsilon^2\sigma(1-\sigma)/8$ in the extremal limit. To ensure that the metric is well-behaved in the limit, we introduced $\d \tilde{T} = \varepsilon \d T/8$ which amounts to setting a characteristic length scale using $\varepsilon$. Note that, under the linear rescaling given by Eq.~\eqref{eqn:x_sigma_linear}, $\d x = - \varepsilon \d \sigma$ but the $g_{\sigma \sigma}$ component of the above metric remains finite. Finally we use $\tilde\rho = 2 \sigma -1$ and $\d\tilde{T} \to 2\d \tilde{T}$ to write,

\begin{equation}
	\d \tilde{s}^2 \sim {8}\left[-(1-\tilde\rho ^2)\d \tilde{T}^2+\dfrac{\d \tilde\rho ^2}{(1-\tilde\rho ^2)}+ \dfrac{1}{2} \d \Omega^2_{S{^2}}\right]. \label{eqn:nariai_like_metric}
\end{equation}

The above metric can be described as the topological product $\d S_2 \times S^2$ where $\d S_2$ is a 2-dimensional de-Sitter ($\d S$) spacetime and $S^2$ is a 2-sphere of constant radius. We note that the geometry is Nariai-like since the curvature radii of $\d S_2$ and $S^2$ are different. In the canonical Nariai spacetime, the two curvature radii are equal~\cite{Bini:2025wef}. Now, in the nearly extremal limit, the metric in hyperboloidal coordinates Eq.~\eqref{eqn:hyperboloidal_metric} reduces to,
\begin{equation}
	{\d}\tilde{s}^2 \sim -f{\d}\tau^2 - 2\varepsilon(2\sigma-1){\d}\tau{\d}\sigma + 32 {\d}\sigma^2 + 4{\d}\Omega^2_{S^2} \label{eqn:intermediate_hyperboloidal}
\end{equation}
Notice that this metric is manifestly regular at both the acceleration ($\sigma=0$) and event ($\sigma=1$) horizons, since $f \to 0$ at the boundaries. To interpret the causal structure, recall that $\sigma$ maps the domain monotonically such that the physical radius decreases as $\sigma$ increases. Consequently, a physically outgoing ray corresponds to ${\d}\sigma/ \d \tau < 0$, while an ingoing ray corresponds to ${\d}\sigma/ \d \tau > 0$. Evaluating the radial null geodesic equation (${\rm d}\tilde{s}^2 = 0$) at the boundaries yields coordinate velocities of ${\rm d}\sigma/{\rm d}\tau = -\varepsilon/16$ at $\sigma=0$ and $+\varepsilon/16$ at $\sigma=1$.  So it is outgoing at the acceleration horizon and ingoing at the event horizon, as desired.
We can transform the metric given by Eq.~\eqref{eqn:intermediate_hyperboloidal} into Eq.~\eqref{eqn:near_extremal_metric} by using
\begin{equation}
	\d \bar\tau \equiv {\varepsilon}  \d \tau  =   8{\d}\tilde{T} + \frac{8(1-2\sigma)}{\sigma(1-\sigma)}{\d}\sigma \,.
\end{equation}
The above transformation is the nearly extremal approximation of the height function given by Eq.~\eqref{eqn:height_func} and is well-behaved once we use $\varepsilon$ to set the characteristic length scale. Therefore, we establish that the limiting geometry of the SR metric for $t \to 1$ is a Nariai-like spacetime. This exercise also demonstrates that the coordinate system that we have constructed using our choice of $h(\sigma)$ and $g(\sigma)$ is naturally well-suited to probe the nearly-extremal limit. Note that although we take $\varepsilon \to 0$, the ratio of the two horizons $t = x_H/x_A$ approaches the limit $t \to 1$. The two horizons remain well-resolved and stay at fixed coordinate locations in this limit. In this sense, our construction is analogous to the one employed in~\cite{Zhou:2025xta} to probe the Nariai limit of the Schwarzschild-de Sitter geometry.

\subsection{Analytic solution}\label{sec:HeunNariai}
As pointed out in Sec~\ref{sec:Vaidya_geometry}, gravitational axial and scalar QNMs in the Nariai regime are typically affected by instabilities, respectively in the mass-accreting and mass-radiating case, once the physical perturbation frequency is reconstructed via Eq.~\eqref{eq:conformal_shift} and Eq.~\eqref{eq:time_dependent_omega}. With this in mind, we will however focus here on the static QNM frequencies, i.e. we will restrict our analysis to the nearly-extremal limit of the SR geometry, and not on the full LMV, with the primary purpose of confirming/excluding the existence of non-LR modes in the strong-accretion/radiation regime. For $|M'|\to 1/16$, the two horizons get close, i.e. $x_H\simeq x_A$.
In this limit, the expansion parameter $t=x_H/x_A$ introduced in Sec.~\ref{sec:QNMviaHeun} approaches one. This means that the frame introduced earlier is not the correct one to study this limit. Therefore, we introduce the new variable $\Tilde{z}=1-z=(x_A-x)/x_A$, which maps the points $\{0,x_H,x_A,\infty\}$ to $\{1,\tilde{t},0,\infty\}$. The new expansion parameter, $\tilde{t}=1-t=(x_A-x_H)/x_A$, approaches zero in this limit. The rescaling of the radial function remains $R(x)=\psi(x)/\sqrt{f(x)}$. The resulting differential equation is of the same form as Eq.~\eqref{eq:HeunNormalForm}
\begin{equation}
	(\partial_{\tilde{z}}^2+\tilde V)\psi=0,
\end{equation}
with the same form of the potential
\begin{equation}
	\tilde V=\frac{1}{\tilde{z}^2(\tilde{z}-1)^2(\tilde{z}-\tilde t)^2}\sum_{i=0}^4\tilde{V}_i\tilde{z}^i.
\end{equation}
The coefficients of the potential, however, are not the same. They can be found in appendix~\ref{appendix:Vaidya_potential}. Comparing again with the normal form of the Heun equation in appendix~\ref{appendix:NSHeun}, we find the following dictionaries
\begin{align}
	\tilde a_0      & =\tilde\theta^{(0)}\frac{i\tilde{\Omega}}{4|M'|\,(1-x_H/x_A)},\nonumber   \\
	\tilde a_1      & =\tilde\theta^{(1)}\sqrt{1-\frac{1-s^2}{4|M'| \,x_H\, x_A}}, \nonumber    \\
	\tilde a_\infty & =\tilde\theta^{(\infty)}\frac{i\tilde{\Omega}}{4|M'|},\nonumber           \\
	\tilde a_t      & =\tilde\theta^{(t)}\frac{i\,\tilde{\Omega}\ x_H/x_A}{4|M'|\,(1-x_H/x_A)},
\end{align}
and
\begin{align}
	\tilde u & =\frac{\ell(\ell+1)}{4\abs{M'}x_Hx_A^2}-\frac{x_A}{2x_H}+\frac{1-s^2}{4\abs{M'}x_A^2x_H^2} \nonumber \\&-\frac{\tilde\Omega^2\ x_H/x_A}{8\abs{M'}^2(1-x_H/x_A)^2x_A^2}.
\end{align}
The $\tilde\theta^{(i)}$ are independent sign choices that we take to be $+1$. Note that the parameters $\tilde a_i$ are the same as the $a_i$ in Sec.~\ref{sec:QNMviaHeun} up to the exchange $(0\leftrightarrow1)$. The behavior of the solutions around the singular points is again
\begin{equation}
	\psi^{(\tilde z_i)}_\theta\sim(\tilde z-\tilde z_i)^{\frac{1}{2}+\theta\tilde a_i}(1+\mathcal{O}(\tilde z-\tilde z_i)).
\end{equation}
The event horizon and acceleration horizon now correspond to $\tilde z=\tilde t$ and $\tilde z=0$ respectively. So we are interested in the connection coefficients between the solutions around these two singular points. It turns out that these coefficients are much simpler in this case and they are given by~\cite{Bonelli:2022ten}
\begin{equation}
	\psi^{(0)}_\theta=\sum_{\theta'=\pm}\tilde t^{\theta \tilde a_0-\theta'\tilde a_t}e^{(\frac{\theta}{2}\partial_{\tilde a_0}-\frac{\theta'}{2}\partial_{\tilde a_t})F}\mathcal{M}_{\theta\theta'}(\tilde a_0,\tilde a_t;\tilde a)\psi^{(\tilde t)}_{\theta'},
\end{equation}
where
\begin{equation}
	\mathcal{M}_{\theta\theta'}(\tilde a_0,\tilde a_t;\tilde a)=\frac{\Gamma(1+2\theta \tilde a_0)\Gamma(-2\theta'\tilde a_t)}{\prod_{\sigma=\pm}\Gamma(\frac{1}{2}+\theta\tilde a_0-\theta'\tilde a_t+\sigma \tilde a)}.
\end{equation}
In an analogous study to the one carried out in the weak accretion/radiation limit, we find that ingoing boundary conditions at the event horizon correspond to $\psi^{(\tilde t)}_-$ and outgoing boundary conditions at the acceleration horizon correspond to $\psi^{(0)}_-$. Therefore, our QNM condition is $\mathcal{M}_{-+}(\tilde a_0,\tilde a_t;\tilde a)=0$. The only way for this condition to be satisfied is if we hit a pole of one of the $\Gamma$ functions in the denominator. This gives the conditions
\begin{equation}
	\frac{1}{2}-\tilde a_0-\tilde a_t\pm \tilde a=-n,\quad n\in\mathbb Z_{\geq0}.
\end{equation}
Now the parameter for the instanton expansion is $\tilde t=(x_A-x_H)/x_A=8\sqrt{1/16-\abs{M'}}+\mathcal{O}(1/16-\abs{M'})$.
Therefore, the instanton expansion is equivalent to an expansion in $y=\sqrt{1/16-\abs{M'}}$. Expanding $\tilde a_0$, $\tilde a_t$ and $\tilde a$ to leading order in $y$, we find that
\begin{equation}
	\tilde\Omega=\tilde\Omega^{(1)}y+\mathcal{O}(y^2),
\end{equation}
with
\begin{equation}
	\tilde\Omega^{(1)}=-i\left(\frac{1}{2}+n\right)\pm\frac{1}{2}\sqrt{3+8\ell(\ell+1)-4s^2}.
\end{equation}
If we now expand all the parameters to next-to-leading order, we find
\begin{equation}
	\tilde\Omega=\tilde\Omega^{(1)}y+\tilde\Omega^{(2)}y^2+\mathcal{O}(y^3),
\end{equation}
with the new correction given by
\begin{equation}
	\tilde\Omega^{(2)}=\pm4\frac{1+2\ell(\ell+1)-s^2}{\sqrt{3+8\ell(\ell+1)-4s^2}}.
\end{equation}
As mentioned earlier in the $\abs{M'}\to0$ limit, in order to obtain the higher order corrections to the QNMs we simply have to include more instantons in our computations. The explicit expressions can be found in appendix~\ref{appendix:QNMsHeunNariai}.

As done in the weak accretion/radiation limit, we also show here, that Eq.~\eqref{eq:wave-equation} can be solved in terms of hypergeometric functions by removing one of the four singular points.

In the nearly-extremal limit, the tortoise coordinate reads
\begin{equation}
	x_*\simeq \frac{x_H}{4|M'|(x_A-x_H)}\ln\left|\frac{x-x_H}{x_A-x}\right|\,,
\end{equation}
and the perturbation equation becomes
\begin{equation}
	\left[\frac{{\rm d}^2}{{\rm d}x_*^2}+\left(\tilde{\Omega}^2-\frac{V_0}{\cosh^2(q x_*)}\right)\right]R(x_*)=0\,,
	\label{eq:master_eq_nearly_ext}
\end{equation}
with
\begin{equation}
	\begin{split}
		&V_0 =\frac{(x_A-x_H)^2}{x_H^4}|M'|\left(\ell(\ell+1)x_H+1-s^2\right)>0\,,\\
		&q=2|M'|\left(\frac{x_H-x_A}{x_H}\right)<0\,.
	\end{split}
\end{equation}
The potential in Eq.~\eqref{eq:master_eq_nearly_ext} is the well-known P\"{o}schl-Teller potential~\cite{Poschl:1933zz}, for which the QNMs are given by the expression

\begin{equation}
	\tilde\Omega = i q \left(n+\frac{1}{2}\right)\pm\sqrt{V_0-\frac{q^2}{4}}\,.
	\label{eq:QNMs_nearly_ext}
\end{equation}
Notice that, if the argument of the square root is smaller than zero, we are going to have PI modes. This condition can be expressed as
\begin{equation}
	|M'|>\frac{\ell(\ell+1)x_H+1-s^2}{ x_H^2}\,.
	\label{eq:threshold_imaginary_mods}
\end{equation}
However, for physical values of the mass-evolution rate, this condition is never met. Hence, in the nearly extremal regime, we will only observe standard damped-sinusoidal LR modes.

\subsection{Numerical solution}
\label{appsec:numerical_nariai_limit_QNMs}
\begin{figure}[t]
	\includegraphics[width=0.8\columnwidth]{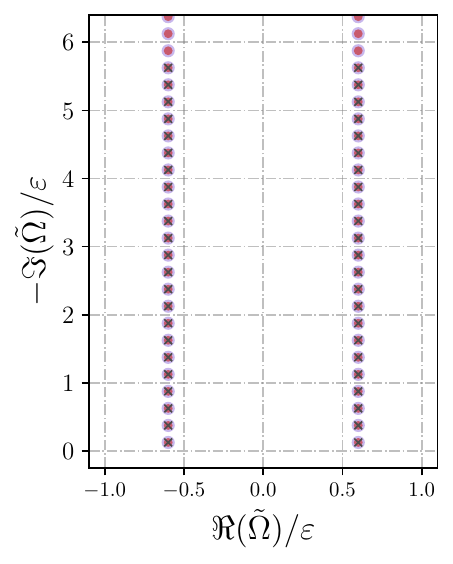}
	\caption{\textbf{QNM Spectrum in the Nariai Limit} for gravitational perturbations ($s=2$) with $\ell=2$. The spectrum is computed in the nearly-extremal limit with $\varepsilon=10^{-100}$, and we show the spectrum for $N = 200$ (filled orange circles) and $N = 250$ (translucent purple circles). The modes marked by crosses have a relative difference $\delta \tilde \Omega_n < 10^{-8}$. The numerical frequencies are scaled by $\varepsilon^{-1}$ and appear as two parallel vertical lines, symmetrical about the imaginary axis, bearing a strong visual resemblance to the analytical P\"{o}schl-Teller (PT) potential spectrum.}
	\label{fig:qnm_spectrum_nariai_limit}
\end{figure}
In this appendix, we have shown that in the nearly-extremal limit, the spacetime resembled a Nariai-like geometry. To establish this, we had introduced a parameter $\varepsilon = x_A-x_H$ to keep track of the separation between the two horizons which were kept fixed at $\sigma_H=1$ and $\sigma_A=0$. Specifically, this means that we set $x_H = 1 - \varepsilon/2$ and $x_A = 1 + \varepsilon/2$, such that,
\begin{equation}
	x  = 1 + \varepsilon\left(\dfrac{1}{2}-\sigma\right)\,.
\end{equation}
Just as our choice of $x(\sigma)$ was found to be quite suitable for studying the Nariai limit theoretically, we find that it is quite useful for computing the QNM spectrum as well. In Fig.~\ref{fig:qnm_spectrum_nariai_limit}, we show the spectrum of gravitational perturbations with $\ell=2$ computed at the extremal limit ($\varepsilon =10^{-100}$ with the internal precision set to $10^{-160}$). We see that the spectrum resembles that of the P\"{o}schl-Teller potential, given by Eq.~\eqref{eq:QNMs_nearly_ext}. We find a remarkable agreement between the numerical and analytical result, especially after scaling both frequencies by $\varepsilon$. The scaling is necessary since we had to scale the time coordinate as $T \to \varepsilon T$ in Sec.~\ref{sec:limitVaidyaHyperboloidal} to ensure the metric is well-behaved in the near-extremal limit. So we define the QNMs in the nearly-extremal limit as
\begin{equation}
	\tilde \Omega^{\rm NE} = \varepsilon^{-1} \tilde\Omega\,.
\end{equation}

\begin{figure}[t]
	\centering
	\includegraphics[width=\columnwidth]{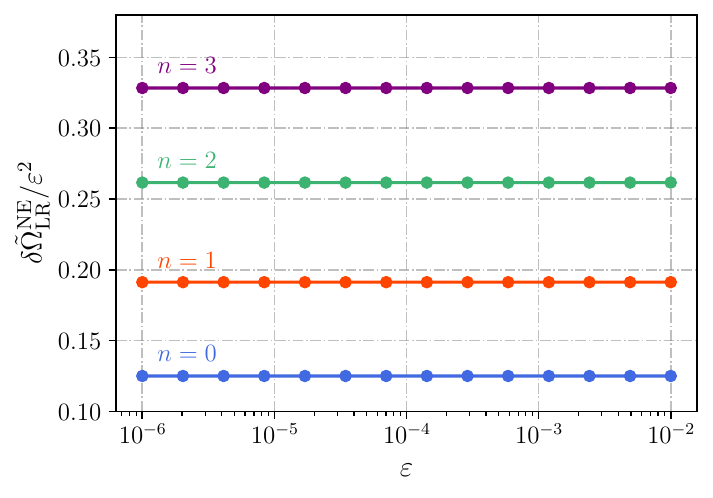}
	\includegraphics[width=\columnwidth]{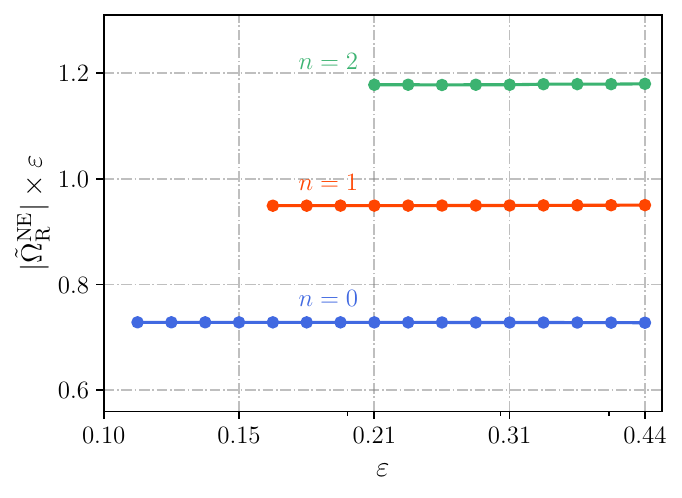}
	\caption{\textbf{Parametric behavior of QNMs in the nearly-extremal limit.} In the \textbf{top panel}, we show the scaling behavior of the Light Ring (LR) modes by plotting their scaled relative difference $\delta\tilde{\Omega}^{\rm NE}_{\rm LR}/\varepsilon^2$ as a function of $\varepsilon$. The horizontal lines confirm that the relative difference scales quadratically as $\mathcal{O}(\varepsilon^2)$ as $\varepsilon \to 0$. In the \textbf{bottom panel}, we plot the scaled absolute magnitude of the purely imaginary Rindler modes $|\tilde{\Omega}^{\rm NE}_{\rm R}| \times \varepsilon$. The constant values demonstrate that these modes diverge as $\mathcal{O}(\varepsilon^{-1})$ in the nearly-extremal limit. The index $n$ denotes the overtone number for each respective family of modes. All modes shown here are convergent with a relative difference $\delta \tilde \Omega_n < 10^{-8}$.}
	\label{fig:nearly_extremal_combined}
\end{figure}

To make the match with the analytical result explicit, we plug in the values of $x_H, x_A$ and $M'$ in terms of $\varepsilon$ into Eq.~\eqref{eq:QNMs_nearly_ext}, and get the following in the limit $\varepsilon \to 0$,
\begin{equation}
	\tilde \Omega_{\rm PT}(0) \equiv \dfrac{\tilde \Omega}{\varepsilon} = -\dfrac{i}{4} \left(n+\dfrac{1}{2}\right) \pm \dfrac{1}{4} \sqrt{V^{\rm PT}_0-\dfrac{1}{4}}\,,
\end{equation}
where $V^{\rm PT}_0 = 2[\ell(\ell+1)+1-s^2]$ and $\Omega_{\rm PT}(0)$ denotes the QNMs of the PT potential. Then, as shown in top panel of Fig.~\ref{fig:nearly_extremal_combined}, the Light Ring modes in the nearly-extremal limit converge to the P\"{o}schl-Teller values, with their relative difference scaling as
\begin{equation}
	\delta \tilde \Omega^{\rm NE}_{\rm LR}(t)  = \abs{1 - \dfrac{\tilde \Omega^{\rm NE}_{\rm LR}(t)}{\Omega_{\rm PT}(0)}} \overset{\varepsilon \to 0}\sim \varepsilon^2.
\end{equation}
We also see from the bottom panel of Fig.~\ref{fig:nearly_extremal_combined} that the purely imaginary modes diverge as
\begin{equation}
	|\tilde \Omega^{\rm NE}_{\rm R}| \sim \varepsilon^{-1}.
\end{equation}

\begin{figure}[t]
	\includegraphics[width=\columnwidth]{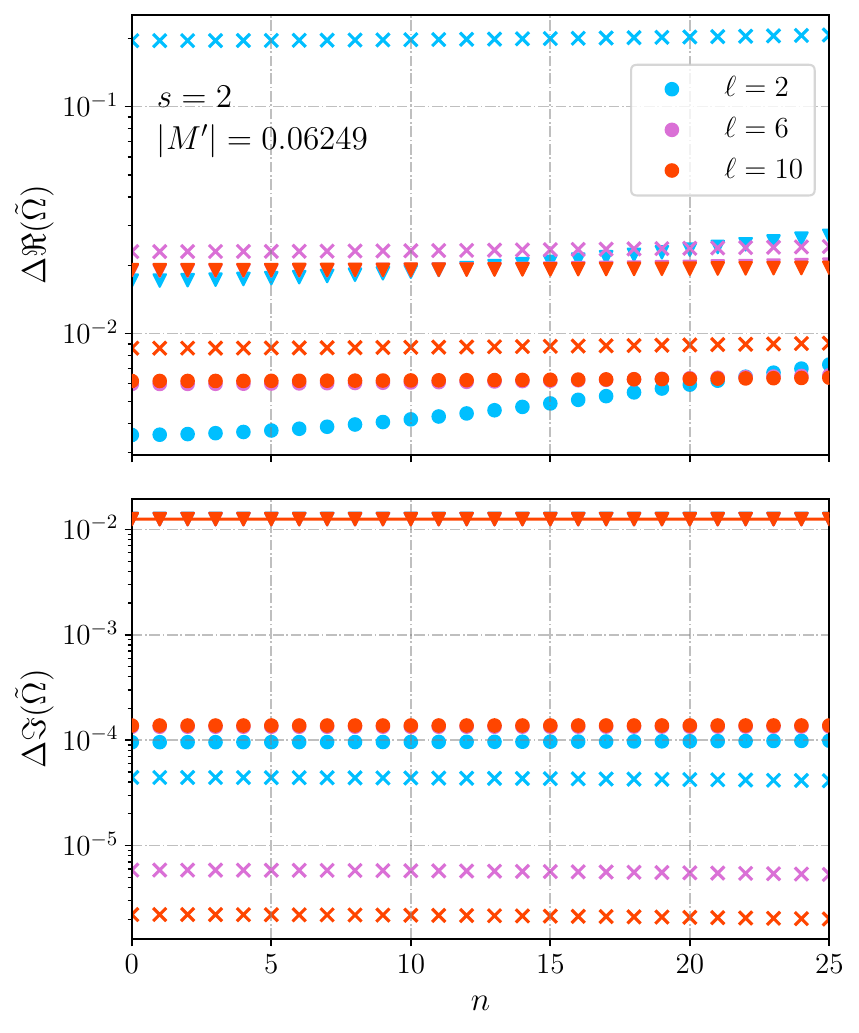}
	\caption{\textbf{QNM frequency residuals in the nearly-extremal limit:} in the case of the hypergeometric equation (triangle points) and Heun equation (round points) and eikonal approximation (cross points). The gravitational $|M'| = 0.06249$ case is presented for multipole indices $\ell = 2$ (blue), $\ell = 6$ (purple), and $\ell = 10$ (orange).
	}
	\label{fig:QNMs_nearly_ext}
\end{figure}

To validate the analytic computation that we performed in the nearly-extremal limit, we compute the QNMs for the quadrupole gravitational $|M'|=0.06249$ case. Remarkably, we did not find PI modes in this limit, consistently with our analytical computations. The comparison of LR modes is shown in Fig.~\ref{fig:QNMs_nearly_ext}. The upper panel shows the residual between numerical and analytic predictions for the real part of the QNM frequency, while the lower panel shows the same residual for the imaginary part. The full circles indicate the residual with the Heun equation prediction, the triangles the one corresponding to the hypergeometric equation prediction (i.e. Eq.~\eqref{eq:QNMs_nearly_ext}). Finally, the cross points represent the residual between numerical prediction and the one obtained through the eikonal approximation in~\cite{Capuano:2024qhv}, given by
\begin{equation}
	\begin{split}
		&\mathfrak{R}( \tilde{\Omega})= \sqrt{\frac{V_{\rm eik}(x_M)}{\ell^2}}\left(\ell+\frac{1}{2}\right)+\mathcal{O}(\ell^{-1}),\\
		&\mathfrak{I}(\tilde{\Omega})=-\left.\frac{{\rm d}x}{{\rm d}x_*}\sqrt{\frac{V_{\rm eik}''(x)}{2V_{\rm eik}(x)}}\right|_{x_M}\left(n+\frac{1}{2}\right)+\mathcal{O}(\ell^{-1})\,,
	\end{split}
	\label{eikonal_frequencies}
\end{equation}
where the eikonal-limit potential reads
\begin{equation}
	V_{\rm eik}(x)=\frac{\ell^2}{x^2}f(x)\,.
\end{equation}

The different colors indicate different multipole numbers. It can be observed that the eikonal approximation, as expected, always performs better for higher multipoles. In the prediction of the imaginary part of QNM frequencies, it also performs better than the other theoretical estimates for every $\ell$. On the other hand, the real part appears to be best estimated by the fourth-order Heun equation, though we note that the accuracy of both the Heun and hypergeometric predictions for the real part noticeably degrades as the overtone number $n$ increases. Moreover, both the predictions obtained with the Heun and hypergeometric equations exhibit a smaller sensitivity to the multipole number. In particular, the residuals for $\ell = 6$ and $\ell = 10$ coincide for both the real and imaginary part of the QNM frequency.

\section{Obtaining the height function via the algebraic approach}\label{appendix:height_function_algebraic_approach}

In Section~\ref{sec:Hyperboloidal_coord_geometric_approach} of the main text, we discussed how the height function can be obtained using purely geometrical arguments, independent of the wave equation. However, since the hyperboloidal foliation constructed via the height function respects the QNM boundary conditions, we saw in Section~\ref{sec:wave_eq_hyperboloidal} that the wavefunction remains regular when the wave equation is expressed in these coordinates, an insight first noted in \cite{Zenginoglu:2011jz}. We previously claimed that the height function is closely related to the choice of ansatz for the wave equation used in computing QNMs via the method of continued fractions. We shall now demonstrate this connection explicitly.

We start by noting that when computing the QNMs using the method of continued fractions \cite{Capuano:2024qhv}, one performs the following separation of variables in Eq.~\eqref{eq:wave-equation}:
\begin{equation}
	\tilde{\phi}(T,x_*) = e^{- i \Omega T} {R(x_*)},\label{phi_decomp}
\end{equation}
resulting in Eq.~\eqref{eq:VaidyaFourierDomain}, the wave equation in the Fourier domain. We then choose the following ansatz:
\begin{equation}
	R(x)= |x-x_A|^{-i\Omega \eta x_A}(x-x_H)^{-i\Omega  \eta x_H}S(x), \label{eq:leaver_ansatz}
\end{equation}
to put the wave equation in a form amenable to numerical analysis. The choice of the prefactor in the above ansatz encodes the QNM boundary conditions. It factors out the exponentially diverging behavior at the two boundary points so that the resultant equation, when written in terms of $S(x)$, is regular. It is therefore already evident that this ansatz achieves the same goal as the hyperboloidal foliation.
We can now substitute Eq.~\eqref{eq:leaver_ansatz} into Eq.~\eqref{phi_decomp} to write:
\begin{equation}
	\tilde{\phi} = e^{-i\Omega\tau}{S(x)}.
\end{equation}
The expression in the exponential helps us to identify a new temporal coordinate:
\begin{equation}
	\begin{split}
		\tau &= T + \eta x_A \ln|x-x_A| + \eta x_H \ln(x-x_H) \\
		&= T + x_* + 2 x_A \eta \ln|x-x_A| \\
		&= V + 2 x_A \eta \ln|x-x_A| \\
		&= T + h(x),
	\end{split}
\end{equation}
where
\begin{equation}
	h(x) = \eta x_H \ln(x-x_H) + \eta x_A \ln|x-x_A|.
\end{equation}
The above expression matches the height function obtained in the main text (cf.~Eq.~\eqref{eqn:height_func}) up to an additive constant, $\eta ( x_A + x_H) \ln(x_A - x_H)$, and we identify $\tau$ with the now familiar hyperboloidal coordinate. So by demonstrating that the ansatz used in the method of continued fractions is essentially identical to choosing a hyperboloidal foliation, we have shown that the algebraic manipulations traditionally required to ensure regularity of the wave equation are elegantly bypassed by the geometric approach, which provides a more fundamental and direct path to the same result.

\section{Vaidya potential in normal form of the Heun equation}\label{appendix:Vaidya_potential}

As mentioned in section~\ref{sec:QNMviaHeun}, making the coordinate transformation $z=x/x_A$, the differential equation for the radial part of the perturbation can be brought to the form
\begin{equation}
	(\partial_z^2+\bar V)\psi=0,
\end{equation}
with
\begin{equation}
	\bar V=\frac{1}{z^2(z-1)^2(z-t)^2}\sum_{i=0}^4\bar V_iz^i.
\end{equation}
The explicit formulae for the coefficients of the potential are given by
\begin{equation}
	\begin{aligned}
		\bar V_0 & =  -\frac{3(x_H/x_A)^2}{4}+\frac{x_H(1-s^2)}{4|M'|x_A^3},                                                                                   \\
		\bar V_1 & =\frac{x_H}{x_A}+\left(\frac{x_H}{x_A}\right)^2-\frac{\ell(\ell+1)x_H}{4|M'|x_A^2}-\frac{1-s^2}{4|M'| x_A^2}\left(1+\frac{x_H}{x_A}\right), \\
		\bar V_2 & =-\frac{3x_H}{2x_A}-\frac{\ell(\ell+1)}{4|M'| x_A}\left(1+\frac{x_H}{x_A}\right)+\frac{1-s^2}{4|M'|^2x_A^2},                                \\
		\bar V_3 & =\frac{\ell(\ell+1)}{4|M'|x_A},                                                                                                             \\
		\bar V_4 & =\frac{1}{4}+\frac{\Tilde{\Omega}^2}{16|M'|^2}.
	\end{aligned}
\end{equation}
Similarly, making the change of coordinates $\tilde{z}=(x_A-x)/x_A$, the differential equation can be transformed into
\begin{equation}
	(\partial_{\tilde{z}}^2+\tilde V)\psi=0,
\end{equation}
with
\begin{equation}
	\tilde V=\frac{1}{\tilde{z}^2(\tilde{z}-1)^2(\tilde{z}-\tilde t)^2}\sum_{i=0}^4\tilde{V}_i\tilde{z}^i.
\end{equation}
The coefficients of this potential are given by
\begin{equation}
	\begin{aligned}
		\tilde V_0 & = \frac{1}{4}\left(1-\frac{x_H}{x_A}\right)^2+\frac{\tilde{\Omega}^2}{16\abs{M'}^2},                                         \\
		\tilde V_1 & =-\frac{\ell(\ell+1)}{4\abs{M'}x_A}\left(1-\frac{x_H}{x_A}\right)-\frac{1-s^2}{4\abs{M'}x_A^2}\left(1-\frac{x_H}{x_A}\right) \\&-\frac{\tilde{\Omega}^2}{4\abs{M'}^2}-\left(1-\frac{x_H}{x_A}\right)^2,\\
		\tilde V_2 & =\frac{\ell(\ell+1)}{2\abs{M'}x_A}\left(1-\frac{x_H}{2x_A}\right)+\frac{1-s^2}{4\abs{M'}x_A^2}                               \\&+\frac{3\tilde{\Omega}^2}{8\abs{M'}^2}+\frac{3}{2}\left(1-\frac{x_H}{x_A}\right),\\
		\tilde V_3 & =-\frac{\ell(\ell+1)}{4\abs{M'}x_A}-\frac{\tilde{\Omega}^2}{4\abs{M'}^2}-1,                                                  \\
		\tilde V_4 & =\frac{1}{4}+\frac{\tilde{\Omega}^2}{16\abs{M'}^2}.
	\end{aligned}
\end{equation}

\section{Heun equation from gauge theory}\label{appendix:NSHeun}

The connection coefficients introduced in Eq.~\eqref{eq:connectionformula} are written in terms of parameters describing states in Liouville CFT~\cite{ZamolodchikovZamolodchikov1991,Teschner_2001}. Via the AGT correspondence~\cite{Alday_2010} (see~\cite{Le_Floch_2022} for a review), this can be described by means of an $\mathcal{N}=2$ $SU(2)$ gauge theory with $N_f=4$ fundamental hypermultiplets. For an in depth discussion of the derivation, we refer the reader to the original work~\cite{Bonelli:2022ten} in which the machinery was developed. Here, we limit ourselves to giving a brief description of the steps that one needs to follow in practice in order to implement the results.
The Heun equation in normal form can be written as
\begin{equation}
	\begin{aligned}
		\Big[\partial_z^2+\frac{u}{z(z-t)}-\frac{\frac{1}{2}-a_1^2-a_t^2-a_0^2+a_\infty^2+u}{z(z-1)} \\
			+\frac{\frac{1}{4}-a_1^2}{(z-1)^2}+\frac{\frac{1}{4}-a_t^2}{(z-t)^2}+\frac{\frac{1}{4}-a_0^2}{z^2}\Big]\psi=0.
	\end{aligned}
\end{equation}
This can be expressed in the following way
\begin{equation}
	(\partial_z^2+V^\mathrm{Heun})\psi=0,
\end{equation}
where the potential has the structure
\begin{equation}
	V^\mathrm{Heun}=\frac{1}{z^2(z-1)^2(z-t)^2}\sum_{i=0}^4V_i^\mathrm{Heun}z^i.
\end{equation}
The coefficients of the potential take the following shape
\begin{equation}
	\begin{aligned}
		V_0^\mathrm{Heun} & = t^2\left(\frac{1}{4}-a_0^2\right),                                        \\
		V_1^\mathrm{Heun} & =-\frac{t}{2}+t^2\left(a_\infty^2+a_0^2-a_1^2-a_t^2\right)+2ta_0^2+ut(t-1), \\
		V_2^\mathrm{Heun} & =\frac{1}{2}+2a_1^2t-a_\infty^2t(t+2)-a_0^2(2t+1)                           \\&+a_t^2(t^2+2t-1)+u(1-t^2),\\
		V_3^\mathrm{Heun} & =-\frac{1}{2}+a_0^2-a_1^2+a_\infty^2(1+2t)+a_t^2(1-2t)+u(t-1),              \\
		V_4^\mathrm{Heun} & =\frac{1}{4}-a_\infty^2.
	\end{aligned} \label{eq:HeunCoeffsa}
\end{equation}
We obtain the dictionaries used in section~\ref{sec:QNMviaHeun} by imposing equality between the coefficients in Eq.~\eqref{eq:HeunCoeffsa} and the ones in Appendix~\ref{appendix:Vaidya_potential} and solving the resulting system of equations. As already shown in the formulae in section~\ref{sec:QNMviaHeun}, the connection coefficients are written in terms of the instanton part of the NS free energy $F(t)$~\cite{NEKRASOV_2010}, which can be computed as a series in the gauge coupling $t$ via some combinatorial formula~\cite{FLUME_2003}. As an example, here we show the expansion at leading order
\begin{equation}
	F(t)=\frac{\left(\frac{1}{4}-a^2-a_1^2+a_\infty^2\right)\left(\frac{1}{4}-a^2-a_t^2+a_0^2\right)}{2\left(\frac{1}{4}-a^2\right)}t+\mathcal{O}(t^2).
\end{equation}
The connection formulae also depend on the parameter $a$. The computation of this parameter is carried out by first fixing the order of the instanton expansion of $F$, and then inverting the Matone relation~\cite{Matone_1995} order-by-order in $t$
\begin{equation}
	u=-\frac{1}{4}-a^2+a_t^2+a_0^2+t\partial_tF.
\end{equation}
In practice, we solve this equation iteratively as a series in $\abs{M'}$ in the limit $\abs{M'}\to0$, and as a series in $\sqrt{1/16-\abs{M'}}$ in the case $\abs{M'}\to1/16$.

\section{Higher order correction to the QNMs obtained using the Heun equation}
In this appendix we show the explicit expressions for the higher order corrections to the QNMs computed up to 5 instantons. We do this in both the weak and strong accretion/radiation limits, discussed in section~\ref{sec:QNMviaHeun} and in appendix~\ref{appendix:next_limit} respectively.

\subsection{Weak Accretion/Radiation Limit}
\label{appendix:QNMsHeunSchwarzschild}

As explained in section~\ref{sec:QNMviaHeun}, in the small $\abs{M'}$ limit the QNMs are computed as a series in $\abs{M'}$
\begin{equation}
	\tilde\Omega=\sum_{k\geq1}\tilde\Omega^{(k)}\abs{M'}^k.
\end{equation}

Here we show the first few terms in the series:

\begin{widetext}
	\begin{equation}
		\begin{aligned}
			\tilde{\Omega}^{(1)} & =-2i\left(n+\ell+1\right),                                                                                                                 \\
			\tilde{\Omega}^{(2)} & =-4i\frac{(\ell-n)(2\ell+1)-2s^2}{2\ell +1},                                                                                               \\
			\tilde\Omega^{(3)}   & =-\frac{8i}{(1+2\ell)^3} \left[ (1+2\ell)^2 (9\ell^2 + 7\ell + 2 + 3n^2 - 6s^2) + n(1+2\ell)(1-2\ell-8\ell^2) - 4s^4 \right],              \\
			\tilde\Omega^{(4)}   & =-\frac{16i}{(1+2\ell)^5} \Big[ \big( -40 - 63\ell + 311\ell^2 + 1712\ell^3 + 3256\ell^4 + 2864\ell^5 + 944\ell^6 \big)                    \\
			                     & + 2n(1+2\ell)\big( 39 + 268\ell + 364\ell^2 + 240\ell^3 + 140\ell^4 + 16\ell^5 \big) + n^2(1+2\ell)^2 \big( 37 + 100\ell + 100\ell^2 \big) \\
			                     & + 2(1-s^2)\big( 57 + 190\ell + 190\ell^2 + 232\ell^2(1+\ell)^2 - 6n(1+2\ell) - 6n^2(1+2\ell)^2 \big)                                       \\
			                     & - 4(1-s^2)^2 (17 + 20\ell + 20\ell^2) + 16(1-s^2)^3 \Big],                                                                                 \\
			\tilde\Omega^{(5)}   & =-\frac{32i}{(1+2\ell)^7} \Bigg[
			\Big( 28464\ell^8 + 64\ell^7(1871 + 184n) + 8\ell^6(26493 + 6732n + 1580n^2) + 24\ell^5(8601 + 4156n + 1580n^2)                                                   \\
			                     & + \ell^4(119283 + 100832n + 49168n^2 - 288n^3 - 144n^4) + \ell^3(40238 + 61184n + 35136n^2 - 288n^3 - 144n^4)                              \\
			                     & +2\ell^2(3259 + 11367n + 7395n^2 - 108n^3 - 54n^4) + \ell(-105 + 4902n + 3542n^2 - 144n^3 - 72n^4)                                         \\
			                     & - 224 + 481n + 380n^2 - 18n^3 - 9n^4 \Big)                                                                                                 \\
			                     & + (1-s^2) \Big( 11072\ell^6 + 33216\ell^5 + (44672 - 1984n(1+n))\ell^4 + (34048 - 3968n(1+n))\ell^3 + (16308 - 3552n(1+n))\ell^2           \\
			                     & + (4852 - 1568n(1+n))\ell + 990 - 268n(1+n) \Big)                                                                                          \\
			                     & + (1-s^2)^2 \Big( -1824\ell^4 - 3648\ell^3 + 24(12n^2 + 12n - 167)\ell^2 + 24(12n^2 + 12n - 91)\ell + 24(3n^2 + 3n - 38) \Big)             \\
			                     & + (1-s^2)^3 \Big( 448\ell^2 + 448\ell + 432 \Big) - 80(1-s^2)^4 \Bigg]
		\end{aligned}
	\end{equation}
\end{widetext}

\subsection{Strong Accretion/Radiation Limit}\label{appendix:QNMsHeunNariai}

In a similar fashion, in appendix~\ref{sec:HeunNariai} we explained how in the $\abs{M'}\to1/16$ limit the QNMs are computed as a series in $y=\sqrt{1/16-\abs{M'}}$
\begin{equation}
	\tilde\Omega=\sum_{k\geq 1}\tilde\Omega^{(k)}y^k.
\end{equation}

Here we show the first terms of the series:

\begin{widetext}
	\begin{equation}
		\begin{aligned}
			\tilde\Omega^{(1)} & =-i\left(\frac{1}{2}+n\right)\pm\frac{1}{2}\sqrt{3+8\ell(\ell+1)-4s^2},                                                                                                       \\
			\tilde\Omega^{(2)} & =\pm4\frac{1+2\ell(\ell+1)-s^2}{\sqrt{3+8\ell(\ell+1)-4s^2}},                                                                                                                 \\
			\tilde\Omega^{(3)} & =\mp\frac{16 \left[ (1 + 2\ell + 2\ell^2)^2 + n(1 + n)(3 + 8\ell + 8\ell^2 - 4s^2) - s^2 \right]}{\left(3 + 8\ell + 8\ell^2 - 4s^2\right)^{3/2}},                             \\
			\tilde\Omega^{(4)} & =\pm\frac{32(1 + 2\ell + 2\ell^2 - 2s^2) \left[ (1 + 2\ell + 2\ell^2)^2 + n(1 + n)(3 + 8\ell + 8\ell^2 - 4s^2) - s^2 \right]}{\left(3 + 8\ell + 8\ell^2 - 4s^2\right)^{5/2}}, \\
			\tilde\Omega^{(5)} & =\mp\frac{256 \left[ (1+2\ell+2\ell^2)^2 + n(1+n)(3 + 8\ell + 8\ell^2 - 4s^2) - s^2 \right] }{\left(3 + 8\ell + 8\ell^2 - 4s^2\right)^{7/2}}                                  \\&\phantom{=\mp}\left[ 5(1+2\ell+2\ell^2)^2 + n(1+n)(3 + 8\ell + 8\ell^2 - 4s^2) - 12s^2(1+2\ell+2\ell^2) + 16s^4 - s^2 \right].
		\end{aligned}
	\end{equation}
\end{widetext}

\section{Explicit expression for the wave operator in hyperboloidal coordinates}
\label{appendix:wave_operator_hyperboloidal_explicit_expressions}
In the main text we have shown how the QNM problem is reduced to an eigenvalue problem given by Eq.~\eqref{eqn:qnm_eigenvalue} with the hyperboloidal coordinates providing a way to incorporate the boundary conditions in a geometric manner.
For the benefit of the reader, we note that for our choice of height function and radial rescaling, that is, using Eqs.~\eqref{eqn:height_func} and \eqref{compact_func_tortoise}, the exact expressions for the quantities in Eqs.~\eqref{eqn:L_L1_L2} and \eqref{eqn:aux_func_L1_2} are as follows,
\begin{equation}
	p(\sigma) =  \dfrac{\left(1-\sigma\right)\sigma}{\eta\left(x_A\left(1-\sigma\right)+x_H\sigma\right)},
\end{equation}
and
\begin{align}
	w(\sigma)          & = \dfrac{4 x_A x_H \eta}{x_A - x_A \sigma + x_H \sigma},\nonumber                              \\
	q_{\ell,s}(\sigma) & =    \dfrac{ \left(x_A-x_H\right)Q_{\ell,s}}{\left(x_A-x_A\sigma+x_H\sigma\right)^3},\nonumber \\
	Q_{\ell,s}(\sigma) & = \bar{s} + \ell(\ell+1)\left(x_A-x_A\sigma+x_H\sigma\right),\nonumber                         \\
	\gamma{(\sigma)}   & =\dfrac{x_A-\left(x_A+x_H\right)\sigma}{x_A-x_A\sigma+x_H\sigma},
\end{align}
with $\bar s = {(1-s^2)}$. Note that $p(x)$ vanishes at the boundaries as we had claimed in the main text.
Finally, we also write,
\begin{align}
	L_1 & = f_2(\sigma) \partial_\sigma^2 + f_1(\sigma) \partial_\sigma + f_0{(\sigma)}, \nonumber \\
	L_2 & = g_1(\sigma) \partial_\sigma + g_0(\sigma),
\end{align}
with
\begin{align}
	f_0(\sigma) & = \dfrac{-\left(x_A-x_H\right) Q_{\ell,s}}{4 x_A x_H \eta \left(-x_A+x_A\sigma-x_H\sigma\right)^2}, \nonumber                     \\
	f_1(\sigma) & = -\dfrac{\left(x_A-2 x_A\sigma+x_A\sigma^2-x_H\sigma^2\right)}{4 x_A x_H \eta^2 \left(-x_A+x_A\sigma-x_H\sigma\right)},\nonumber \\
	f_2(\sigma) & =\dfrac{\left(1-\sigma\right)\sigma}{4 x_A x_H \eta^2},\nonumber                                                                  \\
	g_0(\sigma) & =\dfrac{1}{2 \eta \left(-x_A+x_A\sigma-x_H\sigma\right)},\nonumber                                                                \\
	g_1(\sigma) & =-\dfrac{-x_A+x_A\sigma+x_H\sigma}{2 x_A x_H \eta}.
\end{align}
Note that the highest-order coefficient $f_2(\sigma)$ of the Sturm-Liouville operator vanishes precisely at the boundaries $\sigma=0$ and $\sigma=1$. This renders the $L_1$ operator singular, allowing the physical boundary conditions to be satisfied simply by requiring the eigenfunctions to be regular. Because $L_1$ contains the principal part (the highest-order spatial derivative) of the full differential operator $L$, the singular structure at the boundaries is entirely governed by $L_1$. Consequently, the lower-order terms in $L$ do not alter the singularity, meaning the requirement of regularity naturally extends to the full operator.
\section{Chebyshev spectral method with mesh refinement}
\label{app:chebyshev_spectral_method_details}

In this appendix, we discuss some of the salient aspects of the Chebyshev Spectral Method and mesh refinement.
We define the standard Chebyshev-Gauss-Lobatto (CGL) collocation grid on the interval $[a,b]$ by defining a mapping from the canonical CGL nodes on $[-1,1]$ to the physical interval. The canonical nodes are
\begin{equation}
	\xi_k = \cos\left(\frac{k\pi}{N}\right), \qquad k = 0, 1, \ldots, N,
\end{equation}
where $N$ is the number of subintervals (with $N+1$ grid points). These grid points are mapped to the physical interval $[a,b]$ using the affine transformation\footnote{Note that $\xi_0 =1$ and $\xi_N =-1$; the nodes are ordered in reverse, and consequently $X_0=b$ and $X_N=a$.},
\begin{equation}
	X_k = \frac{a + b}{2} + \frac{b - a}{2}\,\xi_k.
	\label{eq:cgl_standard}
\end{equation}

Note that the grid $\{X_k\}$ is nonuniform, clustering near the edges of the grid. Now, following~\cite{Zhou:2025xta}, we introduce a mesh refinement scheme to further concentrate the grid points near a boundary by refining the above map with
\begin{equation}
	\sigma_k = \frac{a + b}{2} + \frac{b - a}{2}\,\mathcal{F}(\xi_k;\,\kappa,\,x_B),
	\label{eq:sigma_cgl_amr}
\end{equation}
where
\begin{equation}
	\mathcal{F}(\xi_k;\, \kappa,\, x_B) = x_B\left(1 - \frac{2\sinh\!\left[\kappa(1 - x_B\,\xi_k)\right]}{\sinh(2\kappa)}\right),
	\label{eq:amrfunc}
\end{equation}
and $\kappa > 0$ controls the concentration of points along the boundary specified by $x_B$. Note that $x_B=1$ clusters near the right boundary located\footnote{It is worth mentioning that we usually take $\sigma_H=a=1$ and $\sigma_A=b=0$.} at $\sigma =b$, and $x_B=-1$ clusters near the left boundary located at $\sigma =a$.

Having defined the collocation grid $\{X_k\}_{k=0}^N$, we approximate a function $f$ as a truncated expansion in Chebyshev polynomials of the first kind, $T_m(\xi) = \cos(m\arccos\xi)$, for $m = 0, 1, \ldots, N$. This choice is natural since the CGL nodes coincide with the extrema of $T_N(\xi)$, yielding spectrally accurate interpolation for sufficiently smooth functions while mitigating the Runge phenomenon. Differential operators acting on $f$ are then represented as matrices acting on the nodal values $\mathbf{f} = (f(X_0), \ldots, f(X_N))^T$. We can now construct the Chebyshev spectral differentiation matrix $\mathbf{D}$ for a grid $\{X_k\}^N_{k=0}$, where
\begin{equation}
	(\mathbf{D}\,\mathbf{f})_k \approx \frac{df}{dX}\bigg|_{X_k},
\end{equation}
and the entries follow the formula~\cite{trefethen2000spectral}:
\begin{equation}
	D_{ij} = \frac{c_i}{c_j}\frac{(-1)^{i+j}}{X_i - X_j},
	\qquad i,j=0,\ldots,N,\;\; i \neq j,
\end{equation}
\begin{equation}
	D_{ii} = -\sum_{\substack{j=0 \\ j \neq i}}^N D_{ij},
\end{equation}

\begin{figure*}[t]
	\centering
	\includegraphics[width=0.32\textwidth]{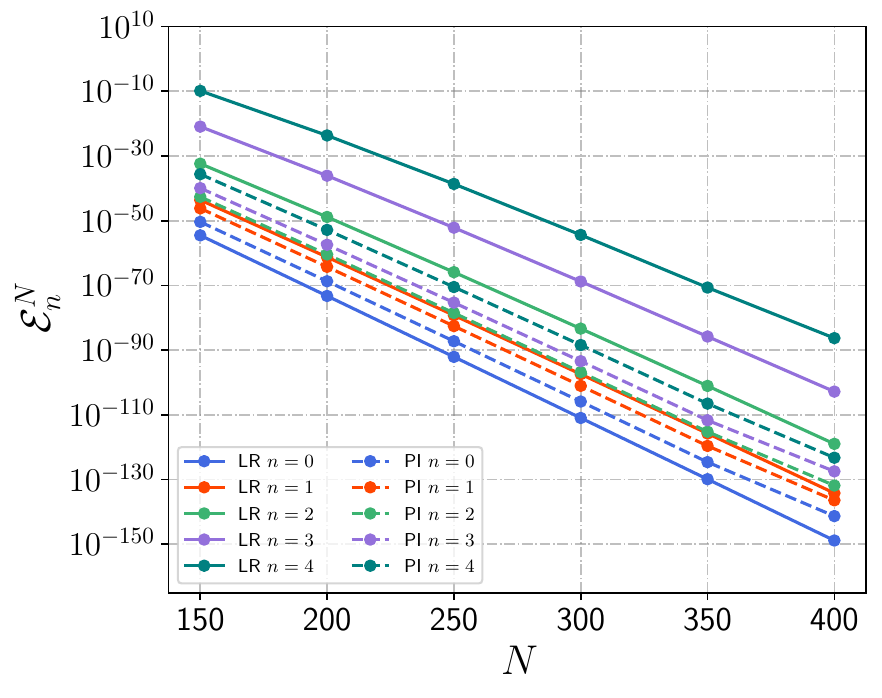}
	\includegraphics[width=0.32\textwidth]{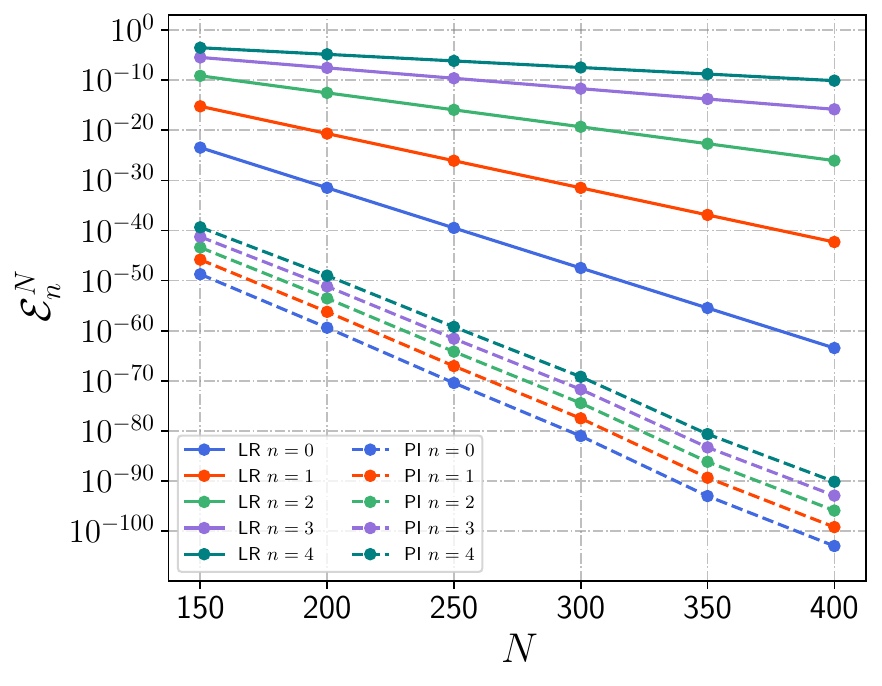}
	\includegraphics[width=0.32\textwidth]{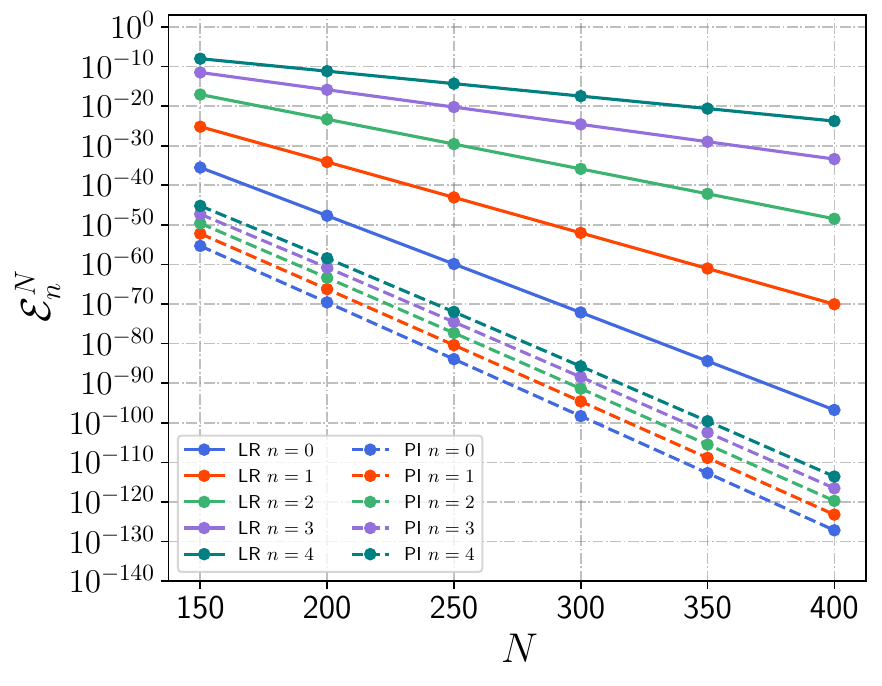}
	\caption{\textbf{Convergence tests for the frequency-domain code.}
		The solid lines represent the LR modes, while the dashed lines correspond to the PI modes. \textbf{Left}: convergence of the gravitational QNMs for $\ell = 0$ and $|M'| = 1/32$ without mesh refinement. \textbf{Middle}: convergence of the $\ell = 2$ gravitational QNMs in the Schwarzschild limit for $t = 10^{-5}$ with mesh refinement. \textbf{Right}: convergence of the $\ell = 2$ scalar QNMs in the Rindler limit for $t = 0.0003$ with mesh refinement. The linear behavior in the semilog plots indicates exponential convergence.}
	\label{fig:conv_test_freq_all}
\end{figure*}

with $c_0 = c_N = 2$ and $c_i = 1$ otherwise. The second-order differentiation matrix is obtained by squaring: $\mathbf{D}^{(2)} = \mathbf{D}\cdot\mathbf{D}$. The differentiation matrix $\mathbf{D}$ computes $d/dX$; to obtain derivatives in the coordinate $\sigma$, we compute the Jacobian
\begin{align}
	J(\xi) & = \frac{d\sigma}{dX} = \frac{d\sigma}{d\xi}\cdot\frac{d\xi}{dX} \nonumber              \\
	       & = 2\,x_B^2\,\kappa\,\cosh\!\left[\kappa - x_B\kappa\,\xi\right]\,\text{csch}(2\kappa),
	\label{eq:Jfinal}
\end{align}
and write the first derivative as
\begin{equation}
	\frac{d}{d\sigma} = \frac{1}{J}\frac{d}{dX}.
\end{equation}
The second derivative is given by
\begin{align}
	\frac{d^2}{d\sigma^2} f
	 & = \frac{1}{J}\frac{d}{dX}\!\left(\frac{1}{J}\frac{df}{dX}\right) \nonumber \\
	 & = \frac{1}{J^2}\frac{d^2f}{dX^2} - \frac{J'}{J^3}\frac{df}{dX},
	\label{eq:d2sigma}
\end{align}
where
\begin{align}
	J' & \equiv \frac{dJ}{dX} = \frac{dJ}{d\xi}\cdot\frac{d\xi}{dX} \nonumber                                   \\
	   & =\frac{-4\,x_B^3\,\kappa^2\,\text{csch}(2\kappa)\,\sinh\!\left[\kappa - x_B\kappa\,\xi\right]}{b - a}.
\end{align}
Using the above expressions, we can obtain $\mathbf{D}_\sigma$ and $\mathbf{D}^{(2)}_\sigma$ from $\mathbf{D}$, and with these finite $(N+1) \times (N+1)$-dimensional approximations of the derivative operator at hand, we can write the scattering operator $L$ obtained in the previous subsection as a $2(N+1) \times 2(N+1)$ matrix whose eigenvalues correspond to the QNM frequencies (provided they survive the convergence test).

\section{Convergence tests}
\label{appendix:convergence}

In this last appendix, we provide convergence tests for our numerical results, both in the frequency and the time domain.
\subsection{Frequency domain}

In Section \ref{sec:numerical_results}, we presented the numerical results of our computation of the QNMs in the hyperboloidal framework using the Chebyshev spectral method. We also mentioned that we had adopted a method to filter spurious eigenvalues and estimate the accuracy of the results. However, in order to establish the exponential convergence of the QNMs, which is characteristic of spectral methods, we perform a convergence test in the following manner: We fix the internal precision to a high value ($10 \times$\texttt{MachinePrecision}) and compute the spectrum for a large grid size ($N_\mathrm{Ref} = 450$), which we treat as the set of reference values. We then repeat the computation for $N = 150$ to $400$ in steps of $dN = 50$ and compare these values against the corresponding reference values. We specifically consider the five lowest-lying LR modes and PI modes. We define the numerical error as
\begin{equation}
	\mathcal{E}^N_n = \abs{1 - \dfrac{\tilde \Omega^N_n }{\tilde \Omega^{N_{\mathrm{Ref}}}_n } },
\end{equation}
where $N$ denotes the grid size and $n$ denotes the overtone number. We then plot $\mathcal{E}^N_n$ as a function of $N$ to examine how rapidly the modes converge with increasing grid resolution.

We present the results of our convergence tests for the different scenarios considered in the main text in Fig. \ref{fig:conv_test_freq_all}. The solid lines represent $\mathcal{E}^N_n$ for the LR modes, while the dashed lines correspond to the PI modes. The convergence of the modes reported in the bottom panel of Fig. \ref{fig:combined_qnm_spectrum}, namely the gravitational QNMs for $\ell = 0, |M'| = 1/32$, is shown in the left panel of Fig. \ref{fig:conv_test_freq_all}. Meanwhile, the convergence of the $\ell = 2$ gravitational modes in the Schwarzschild limit for $t = 10^{-5}$ (Fig. \ref{fig:schwarzschild_limit_spectrum_vs_sch}) is shown in the middle panel of Fig. \ref{fig:conv_test_freq_all}. Lastly, the right panel of Fig. \ref{fig:conv_test_freq_all} shows the convergence of the $\ell = 2$ scalar QNMs in the Rindler limit for $t = 0.0003$ (Fig. \ref{fig:rindler_limit_spectrum}). In the last two cases, we have used mesh refinement, as mentioned in the main text. In all three cases, we observe straight lines in the semilog plot. The linear behavior in the semilog plot implies that the numerical error decreases exponentially with increasing grid resolution.

\subsection{Time domain}
\begin{figure}[t]
	\centering
	\includegraphics[width=0.49\textwidth]{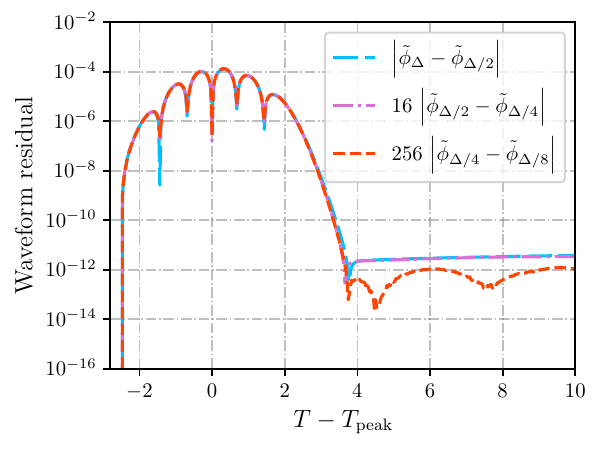}
	\caption{\textbf{Convergence test for the time-domain code:} the curves represent the difference between the numerical signal computed with different grid sizes, and rescaled with the expected numerical factor for our fourth-order scheme. The curves show excellent agreement.}
	\label{fig:conv_test_td}
\end{figure}

To verify the numerical accuracy of our time-domain solver, we performed a convergence test using four different resolutions, i.e. $\Delta$, $\Delta/2$, $\Delta/4$ and $\Delta/8$, with $\Delta = 0.25$.
We set $|M'| = 0.01$ and consider the gravitational quadrupole case.
For each resolution, we extract the waveform at a fixed observer location. We then computed the differences between consecutive resolutions, rescaling them according to the expected fourth-order convergence.
The waveform residuals as a function of $T-T_{\rm peak}$ are shown in Fig.~\ref{fig:conv_test_td}, from which it is evident that the rescaled differences nearly overlap, confirming the expected fourth-order accuracy of the code in the physically relevant regime.
To quantify the convergence, we computed the $L_2$-norm of the differences between two successive resolutions $i,j$, over a time window $[T_{\rm start},T_{\rm end}]$, which in our discrete case reads
\begin{equation}
	\|\tilde\phi_i - \tilde\phi_{j}\| \simeq
	\sqrt{\sum_k \big( \tilde\phi_i(t_k) - \tilde\phi_{j}(t_k) \big)^2 }\,,
\end{equation}
with $k$ running over all the points such that $T_k\in [T_{\rm start},T_{\rm end}]$.
Then, the convergence order can be numerically estimated as
\begin{equation}
	p_{\rm num}=\log_b\left(\frac{\|\tilde\phi_\Delta-\tilde\phi_{\Delta/2}\|}{\|\tilde\phi_{\Delta/2}-\tilde\phi_{\Delta/4}\|}\right)\,,
\end{equation}
where the basis $b$ of the logarithm is given by the refinement factor among successive grids, which is $b = 2$ in our case. Numerically, we obtain $p_{\rm num}\simeq 3.996$, in excellent agreement with the expected fourth-order accuracy of the spatial discretization and time integration scheme.

This test demonstrates that the solver correctly captures the waveform evolution with the anticipated numerical order, providing confidence in the reliability of the computed ringdown signals.


\end{document}